%% file: dissertation.tex
\documentclass[12pt,PhD]{ucdavisthesis} 

\input{tex/environment}
\input{tex/packages}
\input{tex/macros}

\input{tex/info}
\abstract{\input{tex/abstract}}
\acknowledgments{\input{tex/acknowledgments.tex}}

\begin{document}
\makeintropages{}
\captionsetup{justification=centering}

\chapter{Introduction}
\label{ch:introduction}
\input{tex/introduction.tex}

\chapter{Background}
\label{ch:background}
\input{tex/background.tex}

\chapter{A Survey of GPU Load-Balancing Techniques for Irregular Applications}
\label{ch:survey}
\input{ch_survey/survey}

\chapter{A Programming Model for GPU Load Balancing}
\label{ch:loadbalance}

\input{ch_loadbalance/loadbalance}
\chapter[Work-centric Parallel Decomposition for GEMM on the GPU]{Work-centric Parallel Decomposition for Dense Matrix-Matrix Multiplication on the GPU}
\label{ch:streamk}
\input{ch_streamk_v2/streamk_v2}

\input{tex/conclusion.tex}

\bibliography{bib/all,dissertation}

\end{document}

%% file: tex/environment.tex
\usepackage{times} 

\usepackage[hidelinks]{hyperref}
\usepackage[square,comma,numbers,sort&compress]{natbib} 
\renewcommand{\cite}{\citep}

%% file: tex/packages.tex
\usepackage{url}
\usepackage{amsmath,amssymb,amsfonts}
\usepackage{algpseudocode,algorithm,algorithmicx}

\usepackage{graphicx}

\usepackage{textcomp}
\usepackage[table]{xcolor}
\usepackage{comment}
\usepackage{tikz}

\usepackage{caption}
\usepackage{subcaption}

\usepackage{booktabs}
\usepackage{array,tabularx}

\usepackage{multirow}
\usepackage{makecell}

\usepackage[most]{tcolorbox}

\usepackage{inconsolata}

\usepackage[subdued,defaultmathsizes]{mathastext}

\usepackage{alltt}
\usepackage{listings}
\usepackage{textcomp} 
\usepackage{color} 
\usepackage{verbatim}
\usepackage{ifthen}

\usepackage[chapter,frozencache=true]{minted}
\usepackage{float,ragged2e}

\usepackage[export]{adjustbox}

\usepackage{lscape}

\usepackage{natbib}
\usepackage[nottoc,numbib]{tocbibind}

%% file: tex/macros.tex
\Mathastext[inconsolata]
\MTlettershape{it}
\Mathastext[inconsolatait]

\AtBeginEnvironment{tabular}{\singlespacing}
\AtBeginEnvironment{tabularx}{\singlespacing}
\AtBeginEnvironment{algorithmic}{\singlespacing\MTversion{inconsolata}}
\AtEndEnvironment{algorithmic}{\MTversion{inconsolatait}}
\AtBeginEnvironment{listing}{\singlespacing\MTversion{inconsolata}}
\AtEndEnvironment{listing}{\MTversion{inconsolatait}}

\newcolumntype{b}{X}
\newcolumntype{m}{>{\hsize=.70\hsize}X}
\newcolumntype{s}{>{\hsize=.35\hsize}X}

\graphicspath{{ch_streamk_v2/}, {ch_essentials/}, {ch_loadbalance/}, {ch_survey/}}

\newcommand{\cpp}{\texorpdfstring{{C\nolinebreak[4]\hspace{-.05em}\raisebox{.10ex}{{++}}}}{C++}}

\algnewcommand\algorithmicforeach{\textbf{for each}}
\algdef{S}[FOR]{ForEach}[1]{\algorithmicforeach\ #1\ \algorithmicdo}

\algdef{SE}{Fork}{EndFork}[1]{\textbf{fork} \(\mbox{#1}\) \textbf{do}}{\textbf{join}}%

\newcommand{\pluseq}{\mathrel{+}=}

\newcommand{\shared}{\textbf{\_shared\_} }

\definecolor{comment}{HTML}{6a9955}
\newcommand{\algcomment}   [1]{{{\color{comment}$\rhd$ #1}}}

\definecolor{darkred}{rgb} {0.5,0.0,0.0}
\definecolor{darkgreen}{rgb} {0.0,0.5,0.0}
\definecolor{darkblue}{rgb} {0.0,0.0,0.5}
\definecolor{orange}{rgb} {1.0,0.65,0.0}

\definecolor{commentgreen}{RGB}{2,112,10}
\definecolor{eminence}{RGB}{108,48,130}
\definecolor{weborange}{RGB}{255,165,0}
\definecolor{frenchplum}{RGB}{129,20,83}
\definecolor{brightgreen}{RGB}{170,255,0}

\definecolor{pastelpink}{RGB}{250,190,233}
\definecolor{pastelyellow}{RGB}{250,246,186}
\definecolor{pastelpurple}{RGB}{221,199,255}
\definecolor{pastelblue}{RGB}{183,239,255}
\definecolor{purple}{RGB}{159, 90, 253}
\definecolor{comment}{HTML}{6a9955}
\definecolor{lilac}{RGB}{108,48,130}

\newcommand{\final}{1}
\newcommand{\todo}   [1]{{{\color{frenchplum}\colorbox{frenchplum!30}{todo} #1}}}
\ifthenelse{\equal{\final}{1}}{\renewcommand{\todo}[1]{}}{}

\newcommand{\mosama}   [1]{{{\color{red}(mosama) #1}}}
\ifthenelse{\equal{\final}{1}}{\renewcommand{\mosama}[1]{}}{}

\newcommand{\jwapman}   [1]{{{\color{weborange}(jwapman) #1}}}
\ifthenelse{\equal{\final}{1}}{\renewcommand{\jwapman}[1]{}}{}

\newcommand{\jowens}   [1]{{{\color{blue}(jowens) #1}}}
\ifthenelse{\equal{\final}{1}}{\renewcommand{\jowens}[1]{}}{}

\newcommand{\toluwa}   [1]{{{\color{magenta}(toluwa) #1}}}
\ifthenelse{\equal{\final}{1}}{\renewcommand{\toluwa}[1]{}}{}

\newcommand{\matt}   [1]{{{\color{purple}(matt) #1}}}
\ifthenelse{\equal{\final}{1}}{\renewcommand{\matt}[1]{}}{}

\newcommand{\serban}   [1]{{{\color{red}(serban) #1}}}
\ifthenelse{\equal{\final}{1}}{\renewcommand{\serban}[1]{}}{}
\DeclareFontFamily{OT1}{pzc}{}
\DeclareFontShape{OT1}{pzc}{m}{it}{<-> s * [1.10] pzcmi7t}{}
\DeclareMathAlphabet{\mathpzc}{OT1}{pzc}{m}{it}

\tcbset{textmarker/.style={%
        enhanced,
        parbox=false,boxrule=0mm,boxsep=0mm,arc=0mm,
        outer arc=0mm,left=4mm,right=3mm,top=7pt,bottom=7pt,
        toptitle=1mm,bottomtitle=1mm,oversize}}

\newtcolorbox{evaluation_box}{textmarker,
    borderline west={6pt}{0pt}{pastelyellow},
    colback=pastelyellow!10!white}

\newtcolorbox{definition_box}{textmarker,
borderline west={6pt}{0pt}{pastelpurple},
colback=pastelpurple!10!white,text width=\columnwidth}

\newtcolorbox{characterization_box}{textmarker,
    borderline west={6pt}{0pt}{pastelpink},
    colback=pastelpink!10!white,text width=\columnwidth}

\newtcolorbox[auto counter,number within=chapter]{sidebar_box}[2][]{textmarker,
    borderline west={6pt}{0pt}{pastelblue},
    colback=pastelblue!10!white,
    text width=\columnwidth,
    halign=justify,
    title=\textcolor{black}{\textbf{Sidebar~\thetcbcounter}~#2},
    title code={
      \path[fill=pastelblue!10!white] (title.south west) rectangle (title.north east);
      \path[draw=pastelblue,solid,line width=0.75mm]
      ([xshift=0mm]title.south west) -- ([xshift=0mm]title.south east);
      },
    nameref={#2},
    #1
}

\newcommand{\characterization}[1]{
  \begin{characterization_box} \textbf{Characterization\quad}\newline 
    #1 
  \end{characterization_box}
}

\newcommand{\evaluation}[1]{
  \begin{evaluation_box} \textbf{Evaluation\quad}\newline 
    #1 
  \end{evaluation_box}
}

\newcommand{\definition}[1]{
  \begin{definition_box} \textbf{Definition\quad}\newline 
    #1 
  \end{definition_box}
}

\newcommand{\taxonomy}[1]{\textbf{#1}}
\newcommand{\schedule}[1]{\emph{#1}}

%% file: tex/info.tex
\title{GPU Load Balancing}

\author{Muhammad Osama}

\officialmajor{Electrical and Computer Engineering}

\graduateprogram{Electrical and Computer Engineering}

\degreeyear{2022}

\degreemonth{December}

\committee
  {John D.\ Owens}
  {Venkatesh Akella}
  {Jason Lowe-Power}
  {}
  {}

\copyrightyear{2022}

\dedication{\textsl{To my parents, Shahid and Fozia, my siblings, Umair and Sahar,\\ and my fianc\'ee Melissa.}}

%% file: tex/abstract.tex
Fine-grained workload and resource balancing is the key to high performance for regular and irregular computations on the GPUs. In this dissertation, we conduct an extensive survey of existing load-balancing techniques to build an abstraction that addresses the difficulty of scheduling computations on the GPU.

We propose a GPU fine-grained load-balancing abstraction that decouples load balancing from work processing and aims to support both static and dynamic schedules with a programmable interface to implement new load-balancing schedules. Prior to our work, the only way to unleash the GPU's potential on irregular problems has been to workload-balance through application-specific, tightly coupled load-balancing techniques.
With our open-source framework for load-balancing, we hope to improve programmers' productivity when developing irregular-parallel algorithms on the GPU, and also improve the overall performance characteristics for such applications by allowing a quick path to experimentation with a variety of existing load-balancing techniques. Consequently, we also hope that by separating the concerns of load-balancing from work processing within our abstraction, managing and extending existing code to future architectures becomes easier.

Using our insights from load-balancing irregular workloads, we build \emph{Stream-K}, a work-centric parallelization of matrix multiplication (GEMM) and related computations in dense linear algebra. Whereas contemporary decompositions are primarily tile-based, our method operates by partitioning an even share of the aggregate inner loop iterations among physical processing elements. This provides a near-perfect utilization of computing resources, regardless of how efficiently the output tiling for any given problem quantizes across the underlying processing elements. On GPU processors, our \emph{Stream-K} parallelization of GEMM produces a peak speedup of up to 14$\times$ and 6.7$\times$, and an average performance response that is both higher and more consistent across 32,824 GEMM problem geometries than state-of-the-art math libraries such as CUTLASS and cuBLAS\@. Furthermore, we achieve this performance from a \emph{single} tile size configuration per floating-point precision, whereas today's math libraries employ complex kernel-selection heuristics to select from a large ensemble of kernel variants.

%% file: tex/acknowledgments.tex
I owe my deepest gratitude to my advisor John D. Owens. He gave me the opportunity and the freedom to seek exciting research and become who I am today. His enthusiasm for research and admiration of his students made the last seven years of my life truly exciting and endearing. He has helped me find my own style of research, writing and mentoring, and these are the best gifts anyone can ever hope to receive. Thank you John, you taught me how to look for and attempt to solve challenging problems, and I am honored to have been a part of your exceptional research family.

Over the years, I also had the pleasure of learning and discussing research with remarkable individuals. Through early part of my Ph.D., Yuechao Pan and Yangzihao Wang mentored me and taught me much of what I know about parallel programming, graphical processing units (GPUs) and graph analytics. During the later half I had the pleasure of learning from Serban D. Porumbescu, a terrific post graduate of our research group. Since joining our group, Serban has been a key resource and has helped me not just with the academic but also personal and professional matters. He is also responsible for helping me edit many of my research works, and I will forever be grateful for the time that he has spent making my graduate school experience truly extraordinary.

Among the folks in the industry, I was lucky enough to learn CUDA from the CUDA architect himself, Stephen Jones! Stephen has always been a great friend to the group, but has been especially kind to me and taught me the value of good software design and engineering. From Duane Merrill and Cris Cecka I learned everything about low-level optimizations and writing high-performance GEMM kernels. I also had the chance to work with Michael Garland, with whom I enjoyed fruitful discussions that led to many significant design decisions within my research.

I truly enjoyed working with my co-authors and collaborators: Jonathan Wapman, Carl Yang, Minh Truong, Chenshan Yuan, Leyuan Wang, Weitang Liu, Andrew Davidson, Yuduo Wu, Andy T. Riffel, Aydin Bulu{\c{c}}, Toluwanimi Odemuyiwa, Cameron Shinn, Collin Michael McCarthy, and Saurav Muralidharan. I also would like to thank my students Minh Truong, Jonathan Wapman, Daniel Loran, Cameron Shinn and Annie Robison, who I had the pleasure of mentoring as they found their own place within the research group. It was a privilege. 

Among the terrific people already mentioned, I would like to add the rest of my outstanding colleagues, with whom I have spent large parts of the last seven years: Muhammad Awad, Kerry Seitz, Afton Geil, Agnieszka \L{}upi\'nska, Saman Ashkiani, Vehbi E\c{s}ref Bayrktar, Ahmed H. Mahmoud, Yuxin Chen, Zhongyi Lin, Radoyeh Shojaei, Chuck Rozhon, Matthew Drescher, Shalini Venkataraman, Nima Johari and Teja Aluru. I am fortunate to have experienced the influence of so many talented and hardworking people in my life.

I would also like to extend my gratitude to Prof. Venkatesh Akella, Prof. Soheil Ghiasi, Jason Lowe-Power and Sean Treichler for dedicating their time and attention to be a part of my qualification exam committee, and also Prof. Akella and Jason for further providing me with feedback and guidance on my dissertation. Additionally, I would like to thank Sean and Aamer Jaleel for dedicating an hour out of their busy schedule every week for months to give feedback on mine and my colleagues' research works. I also wish to thank the past and current UC Davis ECE staff, who always work hard to remove frictions from the path of students and professors, and enable us to focus on our research. Among the staff, I especially thank Sacksith Ekkaphanh, Phil Young, Fredrick P. Singh, Jennifer Y. Torres, Carole Bustamante, and Michelle T. Walker.

Graduate school would be way too stressful without friends, and I have been lucky enough to have a lot of wonderful friends that helped me get through some of the most stressful times with lots (and lots) of distractions. The RPG, anime and manga theory crafting with Aaron Feaster and Thea Tesoro, the Dungeon \& Dragons sessions and the countless hours\footnote{Please don't tell John.} spent playing Destiny and Overwatch with Nate Bryant, Jesse Navarro, Christian Perry, William ``Billy'' Lucas, Krestine Whaley, Jace Locario and Tiffany Simonet, and the League of Legends games with Megan Vieira. My late friend Pierre ``Link'' Schreurs, dearly missed, who's absurdity and jokes always made us laugh till we had tears in our eyes. Thank you all for being the best of friends.

I extend my love and gratitude to my fianc\'ee, Melissa Vieira. Despite the distance that separated us during these years, you have supported me with your love and joy. I am lucky to have you and truly excited to spend the rest of my life with you.


My research would not have been possible without the financial support from the National Science Foundation (SI2-SSE), the Defense Advanced Research Projects Agency (XDATA, HIVE and Symphony), the United States Geological Survey Agency (USG23), equipment donations from NVIDIA and their funding of an NVIDIA AI Lab at UC Davis, and a 2022 UC Davis ECE Dissertation Fellowship.

And finally, to my family---I am what I am today because of you. I thank my parents, Shahid and Fozia, who have worked tirelessly throughout their lives, foregoing personal pleasure to raise me and my sibilings with all the love, care and support in the world, and who always taught us to learn and unravel the secrets of this world. I thank my sister, Sahar, who continues to be the symbol of patience and courage, which I needed to chase after my dreams. I thank my brother, Umair, who ever since I was a kid, has been my role model---your hardwork, passion and persevarence has led me down a path of excellence. I thank my sister-in-law and brother-in-law, Maria and Rameez, for their support and love. I thank my niece and nephews, Hadiya, Hashir, Musab, and Mahad, who are my sweet little bundles of joy and my source of happiness. Whatever I did to deserve you, it could not have been enough.
\newline \newline
Thank you.

%% file: tex/introduction.tex
Graphics Processing Units (GPUs), born out of the needs of graphics for games, dealing with vertices that make up triangles and subsequent fragments generated by the rasterizer, have evolved to be more general-purpose and programmable. GPUs now excel in many other domains beyond graphics, such as machine learning applications, graph analytics, sparse linear algebra, real-time interactive media, physical simulations, and bioinformatics. GPUs are used as a staple for accelerating data-parallel workloads within these domains, where parallelism is exploited by operating on multiple elements of data simultaneously. 

Data parallelism within these application domains can be classified based on the granularity of work, where granularity (or grain size) is a measure of the amount of work performed by each task~\cite{Hwang:1992:ACA}. \emph{Fine-grained} data parallelism is present when the granularity of work is much smaller than the entire problem. The problem is broken down into smaller components with small amounts of computations that map well to less sophisticated processing elements. Computations between fine-grained components have high communication overhead, and the impact of workload imbalance among processing elements is fairly large.
\emph{Coarse-grained} data parallelism is present when a program is broken down into a small number of large, computationally intensive tasks, and between the tasks, there is low communication overhead~\cite{Hwang:1992:ACA}. Due to their large number of cores and their throughput-oriented design, modern GPUs excel at processing fine-grained data-parallel tasks. NVIDIA's GA100 GPU, for example, boasts 128~Streaming Multiprocessors (SMs) with 8192 total cores and High Bandwidth Memory (HBM) to keep the cores fed with data, both promoting fine-grained parallelism~\cite{NVIDIA:2020:Ampere}.

\begin{figure}
    \centering
    \begin{subfigure}[t]{0.49\textwidth}
        \includegraphics[width=\columnwidth]{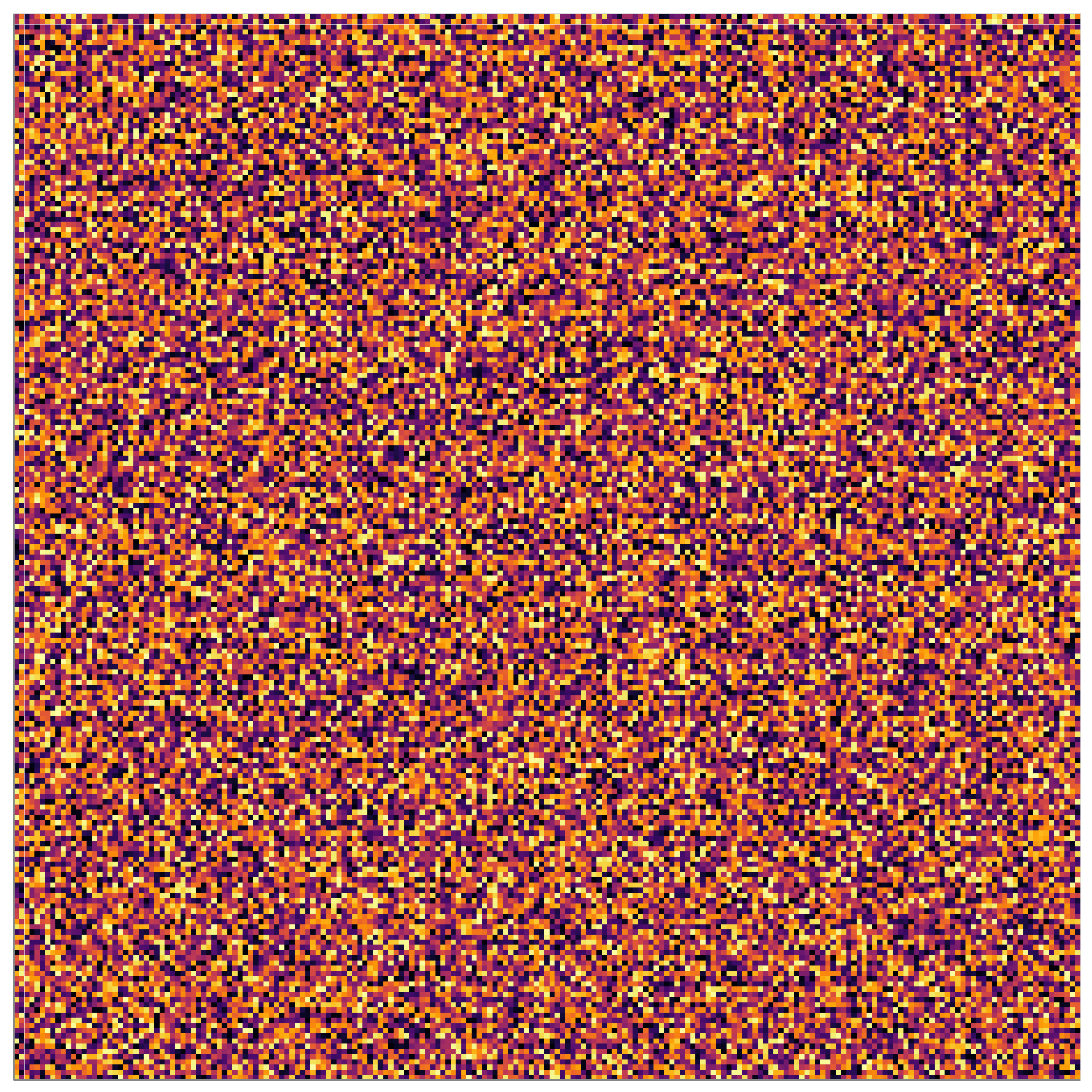}
        \caption{An example dense matrix of size 2048$\times$2048 used as an input to linear layer of a ConvNeXt architecture~\cite{Liu:2022:ACF}. In a dense matrix-matrix multiply (GEMM), all values of this matrix will be consumed, with neighboring threads each processing an identical number of elements.} \label{fig:reg_example}
    \end{subfigure}
    \hfill%
    \begin{subfigure}[t]{0.49\textwidth}
        \includegraphics[width=\columnwidth]{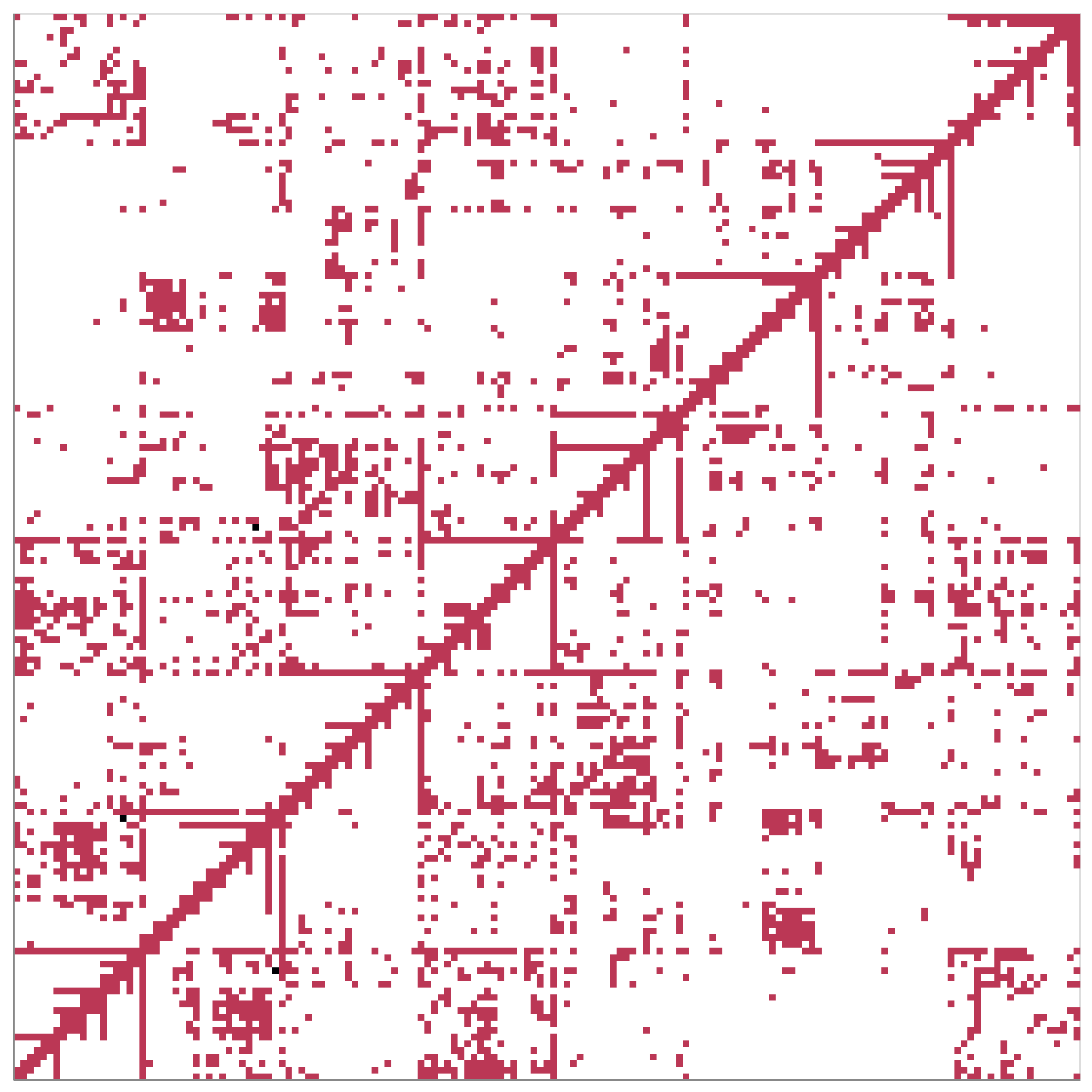}
        \caption{Xyce circuit simulation matrix (graph) of size 321,671$\times$321,671 with 1,316,085 nonzero values irregularly distributed~\cite{Davis:2011:TUO}. If this data is stored in a typical sparse-matrix format (e.g., CSR), only nonzero values will be consumed, and the varying amount of work per matrix row means that a straightforward mapping to a GPU results in neighboring threads consuming different amounts of work.} \label{fig:irr_example}
    \end{subfigure}
    \caption[Visualization of regular and irregular matrices]{Visualization of regular and irregular matrices.} \label{fig:regular_vs_irregular}
\end{figure}

In this dissertation we take a closer look at two categories of fine-grained parallel work, \emph{regular} and \emph{irregular} computations, and explore the fundamental underlying software technology that allows these computations to run efficiently on the GPU architecture: load balancing. In this context, a computation is defined to be regular when the neighboring processing elements have similar or identical workloads. In contrast, a computation is irregular when neighboring processing elements have varying amounts of work to process. Figure~\ref{fig:reg_example} shows an example of a dense matrix used in a General-Matrix Multiplication (GEMM) computation, a regular workload where the partition of the work can be statically and trivially determined, such that each processing element performs identical amounts of reads, writes, and multiply-accumulate instructions. In contrast, Figure~\ref{fig:irr_example} visualizes a graph represented as a sparse matrix used in sparse-linear algebra operations such as Sparse-Matrix Matrix Multiplication (SpMM), an irregular workload where each processing element has varying amounts of reads, writes, and nonzero values to process, and thus an inherent load imbalance within the dataset. 

To leverage fine-grained parallelism within regular and irregular computations on a GPU, we attempt to maximize the available memory bandwidth and compute throughput. For compute-bound, regular problems such as GEMM, the predominant approach is to find the optimal size of chunks of the complete workload, which efficiently utilizes the GPU memory hierarchy, and map that onto the GPU cores in oversubscribed ``waves'' of work. These waves get scheduled onto the GPUs as resources become available until all waves/work is completed. In practice, the last wave of scheduled work often underutilizes the GPU, causing millions of instructions to parallelize over only a small subset of the total available cores. In contrast to prior approaches, which built complicated heuristics selecting different chunk sizes for the waves of work to better utilize the device, this dissertation addresses the \emph{resource imbalance} problem for GEMM by instead balancing the total available work to a fixed, device-filling number of processing elements. Our approach is a generalization of previous parallel decompositions and removes the need for complicated heuristics for kernel selection for GEMM (and other regular problems alike).

For \emph{irregular} computations, however, the irregularity within the program loop-structure or the data itself varies tremendously, making it difficult to find \emph{one} parallel decomposition that balances all problems onto available GPU resources. Instead, there is an inherent tradeoff between the quality of the balance and the amount of extra work needed to compute the workload balance.
This dissertation addresses the \emph{workload balancing} problem for irregular workloads with a load-balancing abstraction, which separates workload mapping from work execution. We argue and demonstrate that this is the superior approach to building high-performance kernels for irregular applications as it promotes reuse of existing load-balancing techniques and facilitates improved portability and programmability.

\section{Contributions}
\label{sec:contributions}

This dissertation makes several contributions to the areas of load balancing regular and irregular computations on the GPU\@.

\begin{itemize}
    \item We conduct a survey of current load-balancing techniques used for sparse-irregular applications (Chapter~\ref{ch:survey}). 
    \item We present our GPU load-balancing abstraction that promotes the separation of concerns between workload mapping and work execution. Our GPU load-balancing framework reconstructs existing application-dependent techniques that address irregularity to be more general, portable, and programmable (Chapter~\ref{ch:loadbalance}).
    \item We provide a solution to the \emph{resource balancing} problem for dense-regular computations such as GEMM\@. Our approach provides improved performance compared to highly-optimized vendor library, cuBLAS, generalizes existing solutions and removes the need for complicated heuristics to select the right set of tile sizes and shapes when scheduling (Chapter~\ref{ch:streamk}).
\end{itemize}

%% file: tex/background.tex
\section{Graphics Processing Unit (GPU) Architecture}
\label{sec:cuda-hierarchy}

\begin{figure}
    \centering
    \begin{subfigure}[t]{0.45\textwidth}
        \includegraphics[width=\columnwidth]{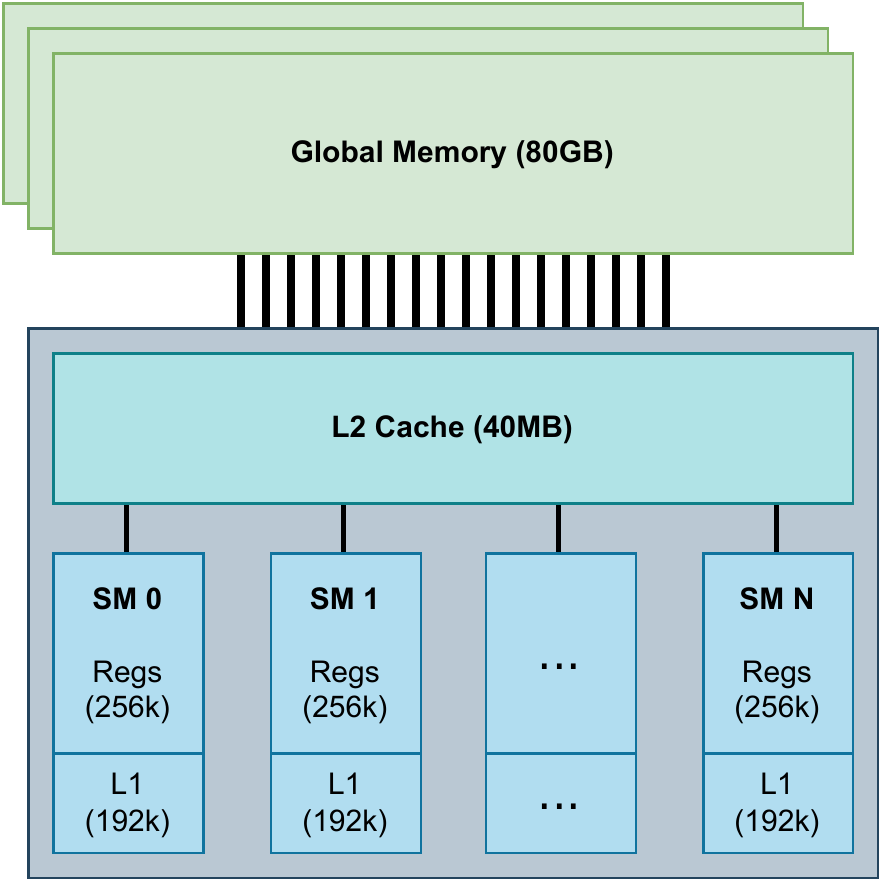}
        \caption[CUDA's Memory System]{\textbf{CUDA's Memory System} NVIDIA A100 sizes for global memory, L1- and L2-caches, registers used as an example.} \label{fig:em}
    \end{subfigure}
    \hfill%
    \begin{subfigure}[t]{0.54\textwidth}
        \includegraphics[width=\columnwidth]{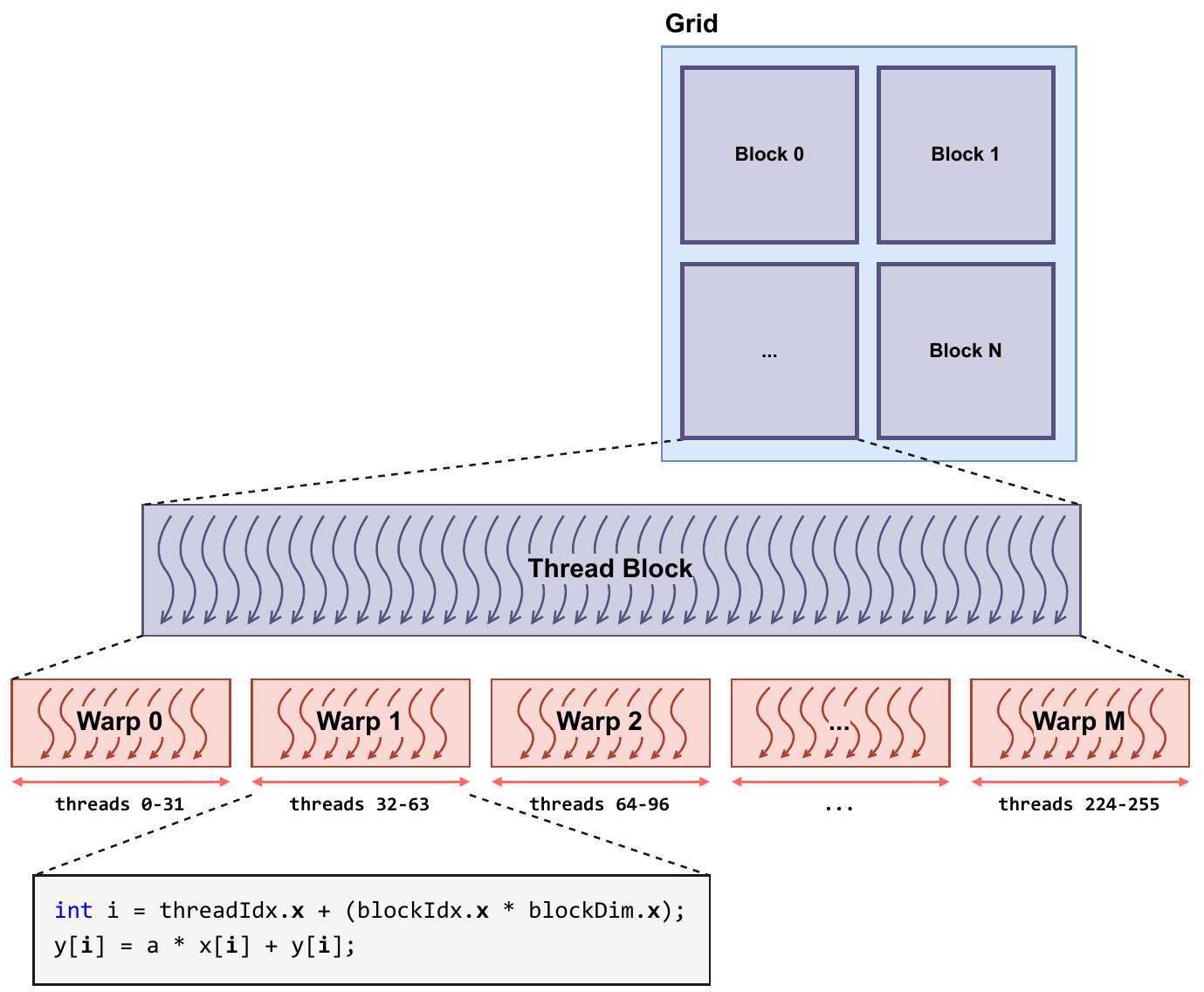}
        \caption[CUDA's Compute Hierarchy]{\textbf{CUDA's Compute Hierarchy} Shows how a kernel is mapped onto the individual threads of a CUDA grid.} \label{fig:compute_hierarchy} %
    \end{subfigure}
    \caption[Graphics Processing Unit (GPU) Architecture]{GPU architecture illustrated as CUDA's memory system and compute hierarchy.} \label{fig:gpu_arch}
\end{figure}

Modern GPU architectures feature two separate hierarchies, one targeted at the individual processing cores organized into a parallel compute hierarchy (Figure~\ref{fig:compute_hierarchy}) and the second targeted at the memory scopes such as L1 and L2 cache, programmable shared memory, and global memory, all organized into a memory hierarchy (Figure~\ref{fig:em}). Using NVIDIA's popular Compute Unified Device Architecture (CUDA) programming model for parallel computing, the compute and memory hierarchies are exposed for programmers to build general-purpose computing applications for GPUs. In this chapter, we provide a brief overview of CUDA's compute and memory hierarchies and highlight their significance for load balancing.

\subsection{CUDA's Compute Hierarchy}
\label{sec:cuda-compute}
The compute hierarchy within the CUDA programming model leverages multiple compute ``perspectives'' that get mapped onto physical streaming multiprocessor cores (SMs) on the GPU\@. The SMs feature a \emph{Single Instruction Multiple Thread (SIMT)} programming model to achieve multithreaded data-parallel execution~\cite{Choquette:2018:VPA}. CUDA further expands the compute hierarchy within the SMs using a software abstraction of threads, the smallest unit, grouped to create larger compute perspectives like a warp, a block, and more. Figure~\ref{fig:compute_hierarchy} illustrates how a CUDA grid, mapped onto the physical SM, is broken down into a number of compute perspectives. The finer-grained perspectives within the hierarchy allow us to do more flexible, higher-performance communication when compared to coarser-grained perspectives. The following list summarizes each of these perspectives and their purpose:

\begin{itemize}
  \item \textbf{Thread} The smallest processing unit. Each thread runs an instance of a GPU program called a \emph{kernel}.
  \item \textbf{Warp} 32~CUDA threads that run in lockstep. Threads in a warp are divergent-free, and run in a Single Instruction Multiple Data (SIMD) fashion.
  \item \textbf{Block} A block or Cooperative Thread Array (CTA) in a GPU is a group of threads that cooperate together to process a slice of data in an algorithm and map onto the same Streaming Multiprocessor (SM). Multiple blocks may run on the same SM concurrently.
  \item \textbf{Clusters} A thread block cluster is a collective of blocks, co-scheduled on adjacent SMs.
  \item \textbf{Grid} The collection of all blocks that run on all the SMs in a single device.
\end{itemize}

\subsection{CUDA's Memory Hierarchy}
\label{sec:cuda-memory}
The CUDA memory model is similar to modern CPU architectures. It features \emph{L1- and L2-cache}, with varying caching policies and traits depending on the GPU architecture. It also features the following list of programmable memory scopes:

\begin{itemize}
  \item \textbf{Global Memory} An off-chip memory shared by all SMs.
  \item \textbf{Shared Memory} Fast on-chip memory shared among different threads in a single block, but not between blocks. Shared memory is limited per SM and is often used to store data that is accessed across an entire block. On some architectures, shared memory shares the same pool of memory as the GPU's L1 cache and the programmer may specify how much of this pool to assign to cached vs.\ explicit data storage.
  \item \textbf{Registers} Any scalar variables declared within the scope of a kernel are by default stored within registers. Register data is only visible locally at the thread-level scope. It is the fastest memory in a GPU\@.
\end{itemize}

\subsection{Impact on Load-Balancing} 
High-performance implementations of fine-grained parallel workloads require programmers to design their kernels with the previously described compute and memory hierarchies in mind. Warps' SIMD execution model is hardware-efficient and is a great match for many workloads. However, it forces programmers to avoid any discrepancies in instructions between the neighboring threads within the same warp. Load-balancing techniques must ensure that threads within the same warp have approximately the same amounts of work to process as well as the same set of instructions. At a block level, NVIDIA's hardware scheduler also plays an important role in scheduling blocks to the underlying SM as resources become available. Work scheduled to the GPU is often oversubscribed to its resources, therefore, the blocks get scheduled out as they finish their set of work with new blocks assigned to keep the GPU fully utilized. Programmers often spend time balancing the workload for a given block (or various compute hierarchies) and then rely on the hardware scheduler or the data workload distribution itself to balance the resources of an SM to a given problem. We discuss this and other load-balancing techniques in detail in Chapter~\ref{sec:deeper-dive}.

Memory hierarchy also plays a significant role in optimizing kernels, where the most important consideration is minimizing the memory accesses required to access data at a thread-level granularity. This includes coalesced memory accesses, where multiple memory accesses are grouped into a single transaction. However, many irregular problems inherently result in uncoalesced loads and stores where the accesses become serialized; this is due to the sparse or random access nature of these workloads. For both regular and irregular workloads, memory hierarchy is utilized to reduce the number of off-chip (global memory) accesses needed by caching the accesses to L2/L1, registers, and the shared memory.

%% file: ch_survey/survey.tex
\input{ch_survey/chapters/introduction}
\input{ch_survey/chapters/related}
\input{ch_survey/chapters/taxonomy}
\input{ch_survey/chapters/implementations}
\input{ch_survey/chapters/primitives}
\input{ch_survey/chapters/summary}
\input{ch_survey/chapters/optimizations}
\input{ch_survey/chapters/conclusion}

%% file: ch_survey/chapters/introduction.tex
Graphical Processing Units (GPUs) excel at and are generally designed for regular fine-grained parallel problems, such as General Matrix Multiplication (GEMM)\@. In regular problems like GEMM, neighboring threads have similar or identical workloads and often achieve nearly 100\% of peak GPU theoretical performance.
What is much more challenging is an application with ample \emph{fine-grained} parallelism but \emph{irregular} parallelism. In such applications, neighboring threads running in a lockstep fashion will have different workloads---perhaps different amounts of work---making an efficient implementation on a highly parallel machine like a GPU a significant challenge.

Consider Sparse-Matrix Vector Multiplication (SpMV), a critical kernel within scientific computing for sparse linear algebra and sparse eigenvalue solvers~\cite{Filippone:2017:SMM}. A sparse matrix $\textbf{A}$ and a dense vector $x$ defined as inputs, SpMV computes the output vector  $y = \textbf{A}x$ and is an example of irregular fine-grained parallelism. Unlike in GEMM, the sparse matrix in SpMV can contain irregularity within the rows of the matrix: the rows of the matrix can have different numbers of non-zero entries. A simple mapping of one row to each GPU thread can expose this irregularity, where the neighboring threads may get different amounts of non-zeros to process, causing threads within the same warp\footnote{Reminder: A CUDA warp is a collection of 32~threads that execute instructions in lockstep. Threads in a warp are divergent-free, and run in a Single Instruction Multiple Data (SIMD) fashion.}
to wait on threads with large amounts of non-zeros. The imbalance created due to this irregularity---specifically, when the work is not equally distributed among the parallel actors, and consequently, some actors are idle while others do more work---is defined as the load imbalance problem. Current approaches to addressing the load-imbalance problem is by distributing the work evenly among threads using sophisticated load-balancing techniques.

In this survey, we look at these fine-grained GPU load-balancing techniques that are used to date to implement high-performance kernels for irregular applications such as graph algorithms (Breadth-First Search), sparse-linear algebra (Sparse-Matrix Multiplication), and many others. But this problem is not specific to GPUs alone. In the past several decades, researchers have proposed many solutions to load imbalance for SIMD and SIMT programming models alike~\cite{Blumofe:1995:CAE,Blelloch:1993:IPN,Liu:2013:ESM}. However, given the prevalence of GPUs, their unique compute and memory model (see Chapter~\ref{ch:background}), this work addresses the need for a broad understanding of the literature space. Our contributions are the following:

\begin{enumerate}
    \item A taxonomy of GPU load-balancing techniques to categorize the load-balancing algorithms (Section~\ref{sec:load-balancing-taxonomy}).
    \item A deeper dive on key techniques that appear in the literature (Section~\ref{sec:deeper-dive}).
    \item An understanding of implementation details and building-block algorithms required to implement these techniques (Section~\ref{sec:primitives}).
\end{enumerate}

The irregularly-parallel workloads we encountered in our survey predominantly focus on sparse linear algebra (with problems such as sparse-matrix dense-vector multiplication, sparse-matrix sparse-matrix multiplication), graph analytics (with algorithms such as breadth-first search, single-source shortest path, triangle counting, and general graph analytics programming abstractions), and computer graphics (ray traversal).

%% file: ch_survey/chapters/related.tex
\section{Related Works}
\label{sec:related-works}

To the best of our knowledge, only one other paper from 2017 conducts a survey of GPU load-balancing techniques~\cite{Busato:2017:APP}. Busato and Bombeiri explain some of the key techniques used at that time to implement load balancing on older GPU architectures, and provide and performance, power and energy analysis on a small subset of sparse datasets. Contemporary to Busato and Bobeiri's approach, our survey not only updates the load-balancing survey literature with new and old techniques, but also provides a deeper dive into their implementations and the underlying low-level primitives. We also build a fine-grained GPU load balancing taxonomy to characterize any current or future load-balancing algorithms.

\subsection{Compressed Sparse Data Structures}
\label{sec:csr}

Irregularly-parallel applications often operate on \emph{sparse} datasets, or datasets in which a significant portion of the data consists of ``zero'' values that make up the bulk of the data in the data structure, but that would be wasteful to store. Sparse data formats allow the programmer to both store the data more efficiently in memory by omitting explicit ``zero'' values as well as access the data more efficiently once it is in a sparse format without performing useless work. Since many sparse workloads are bandwidth-bound rather than compute-bound, the cost of moving data into the GPU's memory hierarchy is a key factor in determining the performance of a sparse workload. In fact, load balancing is often a function of making sure that all processors are \emph{reading or writing} equal amounts of data rather than performing arithmetic on equal amounts of data. Performant sparse data formats commonly group logically-adjacent work items in memory so that they can be read or written in a single memory transaction, so that elements within a group share data (such as row or column metadata), or so that the size of all groups in the data structure are equal. In this section we describe a few common compressed data structures that are used to represent these datasets.

First, consider a matrix where only a handful of the elements are non-zero. We can represent this matrix as a \emph{COO} (coordinate) format, which is a simple list of nonzero elements. Each element is stored as a tuple of row, column, and value. These tuples are often sorted first by row and then by column within each row, but this step is optional. The COO format works well when the programmer simply wants to split nonzeros into multiple groups.

The primary downsides of the COO format are that it uses more space than necessary due to explicitly-stored row and column indices for each nonzero, and it is computationally expensive to perform queries on the matrix, especially if it is unsorted by row and column. For example, determining the number of nonzeros in a row requires an iteration over the list, as do random accesses into the matrix. To solve this problem, we can use the Compressed Sparse Row (CSR) format, which is a more space-efficient format that stores nonzero elements in a compressed format. CSR uses a simple row-major list of nonzero column indices and values, as well as a prefix-sum that stores a running total of the sum of the nonzeros in all previous rows. Compared to COO, this eliminates the need for row index storage, allows kernels to quickly determine the number of nonzeros in each row with a single subtraction operation, and enables faster random or iterative accesses into a given row.

There are many other alternative data structures that typically require more preprocessing, as described in Table~II by Fillipone et al.~\cite{Filippone:2017:SMM}. In general, preprocessed formats are a form of static load balancing, where the programmer analyzes how the data will be accessed and uses this information to optimize the data structure, with an assumption that the cost of preprocessing the data format will pay off on net, either by amortizing the preprocessing cost over many computations or by significantly speeding up the computation. In this work we primarily focus on CSR and COO formats in part because they are the most common formats used in sparse linear algebra on GPUs, and in part because they represent two different ways of organizing data with opposing trade-offs. COO is a list of coordinate/value tuples, where the programmer can easily subdivide the data structure based on an equal split of nonzeros, but where the cost to access information about the row the nonzero is a part of is more expensive. In contrast, CSR is a compressed format allows the programmer to easily subdivide the data structure into an equal number of rows, but requires the programmer to pay an additional cost to load balance \emph{across} rows or access a random nonzero element.

%% file: ch_survey/chapters/taxonomy.tex
\section{Taxonomy of GPU Load Balancing}
\label{sec:load-balancing-taxonomy}
This section focuses on building a taxonomy of GPU load balancing. Our goal with this taxonomy is to (1) characterize existing load-balancing techniques found in GPU literature (see Section~\ref{sec:deeper-dive} for a deeper-dive on implementations, Section~\ref{sec:primitives} for a study on common low-level algorithmic primitives used for load-balancing, and~\ref{sec:summary} for a complete summary), and (2) provide an abstraction to classify future works searchability.

\begin{table}
    \centering
    \begin{tabular}{lllll}
    \toprule
    Configuration & Accuracy & Granularity & Communication & Topology \\
    \midrule
    Static & Exact & Hierarchical & Cooperative & Centralized \\
    Dynamic & Approximate & Flat & Non-Cooperative & Distributed \\
    \bottomrule
    \end{tabular}
    \caption[Taxonomy of Load-Balancing techniques for GPUs.]{Taxonomy of Load-Balancing techniques for GPUs, a layout elaborating how load-balancing algorithms can be characterized.}
    \label{tab:taxonomy}
\end{table}

\subsection{Configuration: Static vs.\ Dynamic}
\label{sec:static-dynamic}
Load-balancing algorithms fall into two distinct approaches; \taxonomy{static} and \taxonomy{dynamic}. Static algorithms use only the information known before the algorithm launches. This may include the number of work items or tiles in the dataset, the number of threads or SMs in a given GPU, or any other characteristics of a given dataset, algorithm, or GPU architecture that can be determined without actually running the algorithm. The work distribution does not change as the algorithm proceeds, and static load-balancing strategies do not account for any runtime changes in the state of the processors or the problem. Therefore, static load-balancing algorithms have no notion of the load on each processor during the execution period. Conversely, dynamic load-balancing algorithms account for the changing incoming or outgoing load on each execution unit by measuring and (re-)distributing that load at runtime. For example, some kernels may dynamically create new work during runtime, and a dynamic load-balancing strategy allows the kernel to immediately process this new work with idle processors rather than waiting for a later kernel launch to do so. Other load balancing strategies may rely on processors greedily requesting or donating work during runtime. Because dynamic approaches can redistribute work at runtime, they can potentially have better overall performance than static approaches, which do not. But this performance comes at the cost of additional runtime work. Thus dynamic approaches win only if the performance gains from better load balance exceed the additional cost of the runtime work to compute it. This is not always a win; Yang et al.'s sparse-matrix multiplication analysis~\cite{Yang:2018:DPF} and concludes that static approaches for this problem deliver better overall performance.

For load-balancing techniques that use both static and dynamic approaches, we use the classification of \taxonomy{Hybrid}~\cite{Chen:2022:AAT}. Hybrid algorithms may use a static technique for one segment of the problem (or one part of the compute hierarchy) and a dynamic technique for another.

\subsection{Accuracy: Exact vs.\ Approximate}
\label{section:subsection:acc}
The ideal load-balancing strategy aims to minimize runtime. In practice, such a strategy aims to balance work evenly across all compute elements (threads, warps, or blocks) so that each compute element completes its work at the same time and no units are ever idle waiting for others to finish. Optimal algorithms use the properties of the problem and the number of available execution units to determine the \taxonomy{exact} amount of work to assign to each unit to ensure all execution units are fully utilized at all times. These algorithms may achieve optimal load balancing for a given cost function.
In GPUs, the cost function often results in modeling the variance of work between neighbouring threads, and the goal of the load-balancing is to attempt to assign equal amount of work items per thread (an ``even-share'' schedule~\cite{Green:2012:GMP,Dalton:2015:OSM,Merrill:2016:MPS,CUB:2016,Baxter:2013:MPA,Steinberger:2017:GHL}, another cost function may model execution time, and a load-balancing schedule optimizes this cost function to ensure that each thread runs for approximately the same time). However there is often an overhead associated with load balancing techniques that attempt to achieve exact workload balance across all processing units. \taxonomy{Approximate} load-balancing algorithms, on the other hand, focus on metrics \emph{within} the problem or the dataset, or the attributes of the system (such as the number of threads) to attempt to distribute the work equally across processors. Approximate algorithms may achieve inconsistent load balancing depending on the metrics used, but can be tuned to reduce the overhead required to achieve a balanced workload. The resulting trade-off between the cost of load balancing and the quality of the workload balanced achieved is an important consideration when opting for an exact or approximate load-balancing algorithm.

\subsection{Granularity: Hierarchical vs.\ Flat}
\taxonomy{Hierarchical} schedules leverage the compute hierarchy available within NVIDIA GPUs by breaking the workload into coarse-grained chunks that map to the highest level of the compute hierarchy (for example the blocks) and then further break down the work into smaller granularities that proportionally map to smaller computation constructs such as warps and/or threads~\cite{Merrill:2016:MPS}. Section~\ref{sec:cuda-hierarchy} explores this concept in more detail.
\taxonomy{Flat} scheduling algorithms are designed to target the smallest computational unit in the hierarchy (threads) and are not mapped to the compute hierarchy available within GPU architectures~\cite{Merrill:2012:SGG,Davidson:2014:WPG,Wang:2017:GGG}. Hierarchical schedules often improve load balance by ensuring that a given work chunk maps to a group of computational units instead of a single (smallest) unit with proportional computation capabilities, and process the mapped work in parallel using the compute hierarchy within the group (CUDA's compute hierarchy explained in Chapter~\ref{ch:background}). For example, a row in a CSR matrix with few nonzeros may be most efficiently processed with a single thread, since processing this row with an entire block would leave many threads sitting idle when they could otherwise be processing other rows. As another example, to avoid significant amounts of atomic contention, a queue-based algorithm may assign a single thread to fetch a large chunk work on behalf of all threads in a block, but then further subdivide this work amongst the threads in the block. Beyond their increased complexity, the primary downside of hierarchical schedules is that the performance gains from leveraging coarser-granularity workers may not be enough to outweigh imperfect load balance within the coarser-granularity work chunks.

\subsection{Topology: Centralized vs.\ Distributed}
\taxonomy{Centralized} schedules focus on a distribution unit responsible for assigning work items to multiple processors. As an example, the GPU's (hardware) block scheduler distributes work to all SMs, assigning a new block to an SM when the SM is ready for new work. In contrast, \taxonomy{distributed} schedules rely on individual processors making independent, local decisions about when they need more work.

Consider a shared queue of work, accessible by all blocks. A centralized strategy for managing this queue has a single entity (either hardware or software) that is responsible for distributing work to processors. A distributed strategy instead delegates the responsibility for fetching new work to each processor: processors must actively fetch new work for themselves as needed, while competing with other processors to claim work from the queue. Centralized and Distributed schedules only apply to \emph{dynamic} schedules. In static schedules, each processor already knows exactly which work items it must process, so there is no concept of ``shared'' work that could go to one of many processors depending on runtime decisions.

\subsection{Communication: Cooperative vs.\ Non-Cooperative}
\taxonomy{Cooperative} schedules allow processing entities to communicate with each other, for example in work-stealing load-balancing algorithms where threads are able to steal work from other threads within a warp or another block~\cite{Blumofe:1999:SMC}. \taxonomy{Non-cooperative} algorithms require processors to operate autonomously within their own pool of work and with no ability to access pools of work that may be simultaneously accessed by \emph{other} processors. In a throughput-oriented architecture, such as a GPU, cooperation often requires expensive synchronization, such as locks or atomics on queues. However, cooperation may result in a better workload balance~\cite{Chen:2022:AAT}. Similar to Centralized and Distributed schedules, the concept of ``cooperation'' only applies to \emph{dynamic} schedules. Cooperation requires that processors are able to make decisions at runtime about whether they need more work, or whether they have additional work to give to another processor. \jowens{... if there was one or two high-level abstract example(s) that could be carried through this whole section for illustration purposes, I think that could be helpful. (e.g., the work-pool example in the centralized/distributed discussion.)  Nothing about its/their implementation(s), just about its/their operation.}

%% file: ch_survey/chapters/implementations.tex
\section{Implementations of Load Balancing on GPUs}
\label{sec:deeper-dive}
In this section we conduct a study of how load-balancing techniques (summarized in Table~\ref{tab:annotated-bib}) actually get implemented. We characterize these implementations using the taxonomy built in the previous section, and group each implementation under common patterns of scheduling work.

\paragraph{Terminology}
\label{sec:terminology}

For consistency, we will use the following terms to describe the partitions of \emph{work} within the different load-balancing techniques.

\begin{itemize}
  \item \textbf{Work Item:} A single unit of work that is to be scheduled onto the GPU; for example, a non-zero element of a sparse matrix.
  \item \textbf{Work Tile:} A collection of work items, for example, a single row of a sparse matrix.
  \item \textbf{Tile Set:} A collection of work tiles that together comprise the entire problem, for example, an entire sparse matrix.
\end{itemize}

The terminologies summarized above are later built into a complete abstraction for load-balancing on the GPU, and are explained in detail in Section~\ref{ch:loadbalance}. As an example to illustrate the terminology, consider Sparse-Matrix Dense-Vector Multiplication (SpMV). The irregular work in SpMV is within the sparse matrix, where the work items (the smallest units of work) are the non-zero elements, a work tile is a row of the sparse matrix, and a tile set is the entire sparse matrix. The work items are scheduled onto the GPU using a load-balancing technique.

The rest of this section describes different strategies for load-balancing work. We can choose to assign an equal number of work tiles to each thread (Section~\ref{sec:thread-mapped}) or to a group of threads (Section~\ref{sec:group-mapped}), or instead assign an equal number of work items to each thread (Section~\ref{sec:work-oriented}). We use several low-level algorithmic building blocks to construct these load-balancing techniques, such as \emph{parallel prefix-sum (scan)}, \emph{binary search}, \emph{sorting}, and more. We defer a detailed explanation of these building blocks to Section~\ref{sec:primitives}.

\subsection{Thread-Mapped}
\label{sec:thread-mapped}
\definition{Assign a fixed, constant number of work tiles to each thread. Resultant work items from each work tile are processed sequentially within the thread.}
\characterization{Static, Approximate, Flat.}
\evaluation{Works well for small, balanced work tiles; not recommended for tile sets where the variance between work tile size is large (e.g., scale-free graphs).}

Perhaps the most natural way to parallelize across work tiles is to assign each work tile to a GPU thread, and sequentially process the work items within the work tile.
For problems where the workload is inherently balanced, the additional overhead of a sophisticated load-balancing scheduling of work can often be detrimental to performance. In such cases, this simple thread-mapped assignment can prove to be a viable substitute. However, the resultant workload balance with this method is often at a coarser granularity.
As discussed in detail in Merrill et al.'s ``Merge-Based Parallel Sparse Matrix-Vector Multiplication''~\cite{Merrill:2016:MPS},
the performance for a thread-mapped assignment is typically not as good as other schedules if the problem is highly irregular or if there are large amounts of fine-grained parallelism left to explore. \jowens{This latter point deserves a sentence of explanation, because it's not clear to me that this is always the case. If there is enough per-thread coarse-grained parallelism to fill the machine, who cares if there's fine-grained parallelism unexposed within each thread?} Because each thread contains a fixed-size set of tiles and in a highly irregular dataset, each tile may contain a different amount of work, such datasets cause threads within the same warp to be waiting on threads with significantly more work, causing very low warp or block utilization and poor performance. Furthermore, since a thread is assigned an entire work tile, the work items within a tile---for example, the nonzeros within a sparse-matrix's row---are sequentially processed, and the fine-grained parallelism within the work tile is effectively serialized, leaving the device underutilized. \jowens{See comment above. I'd like to understand this better.} Algorithm~\ref{alg:thread-mapped} shows a simple example of thread-mapped assignment as a scheduling strategy where each thread is assigned a fixed, statically determined input and output elements to process~\cite{Merrill:2012:SGG,Wang:2017:GGG,Brahmakshatriya:2021:CGA}.

\begin{algorithm}
    \caption[SpMV using thread-mapped scheduling]{Simple pseudocode to illustrate thread-mapped scheduling where each thread is assigned to work on $N$ input and output elements of the \lstinline{saxpy} operation. \jowens{the \texttt{for} notation you're using here (and elsewhere) is weird} 
    \mosama{I tried some stuff, but I like this the most John. :(}}\label{alg:thread-mapped}
    \begin{algorithmic}
    \State $N \gets 4$ \algcomment{Four items per thread.}
    \Procedure{Thread-Mapped}{$\alpha, x, y$}
    \State $t_{id} \gets N \times \textbf{ThreadIndex}()$
    \State \algcomment{Loop can be perfectly unrolled.}
    \For{$k \gets 0$, $k{+}{+}$, \textbf{while} $k < N$}
        \State $y[t_{id} + k] \gets \alpha \cdot x[t_{id} + k] + y[t_{id} + k]$
    \EndFor
    \EndProcedure
    \end{algorithmic}
\end{algorithm}

\jowens{It would be nice to have a (second) equally simple example that has distinct tiles and threads within a tile. In the above example, tiles \emph{are} threads.}

\subsection{Group-Mapped}
\label{sec:group-mapped}
\definition{Assign an equal amount of work tiles to a group of threads (warp or block). Threads within each group process individual work items in parallel.}
\characterization{Static, Approximate, Hierarchical.}
\evaluation{Exposes fine-grained parallelism within a work tile, and is highly tunable to different problems (allowing the group size to be configured to fit a problem). Works well for fairly regular workloads where no one group takes too much GPU resources.}

In contrast to the coarse-grained thread-mapped approach, for irregular problems with ample amounts of fine-grained work items, consider assigning groups of threads to process a single work tile in parallel. For this approach, we can leverage the existing compute hierarchy available within CUDA's programming model: threads, warps, and blocks. A larger compute unit such as a block or a warp could potentially prove to be a better mapping for a work tile that has a large number of work items to process. This is because instead of serializing the work within a work tile, we can assign all threads within a block or a warp to process that work in parallel, promoting better utilization and load balancing within the thread group.

This strategy of assigning an entire tile of work to a group of threads (either a warp, a block, or a ``Cooperative Group''~\cite{NVIDIA:2016:CUDA}) of a GPU is called  group-mapped load balancing. Group-mapped is typically implemented by first assigning an equal number of work tiles to each group, $\frac{\text{Total Work Tiles}}{\text{Total Groups}}$. Then to effectively map the assigned tiles onto the threads within a group, each thread individually needs two sets of information: (1) The total number of work items it is processing (a range from work item $i$ to work item $j$) and (2) what work tile each work item belongs to. A parallel prefix sum gives us both pieces of information. Discussed in detail in Section~\ref{sec:primitives}, a prefix sum is a sequence of numbers that is the running total of an input sequence. In group-mapped's case, a prefix sum of work items per assigned work tiles is constructed, where the last element of the prefix sum sequence is the total number of work items the group needs to process. To solve (1), we simply divide the total number of work items assigned to the group by the number of threads within the group to get the ``range'' of work items any given thread is going to process. And for (2), when processing the work items within a loop, each thread performs a binary search within the prefix sum sequence to determine the work tile index the current work item corresponds to (see Algorithm~\ref{alg:group-mapped}, lines~8--10).

Using the running SpMV example, if a row of a sparse matrix is assigned to an entire warp, all 32 threads within a warp will collectively process each nonzero element within the row. Similarly, if a row is assigned to an entire block, all threads within the block will process the individual nonzero elements of the row~\cite{Davidson:2014:WPG}.

Group-mapped load balancing exposes the fine-grained parallelism available in each work tile to a group, allowing the work items to be processed in parallel, with the cost being the overhead of a parallel prefix sum per group (warp or block) and the binary search to find the work tile within the parallel prefix sum array. A group-mapped assignment also heavily relies on the underlying hardware scheduler's ability to schedule new thread groups as a group finishes processing a work tile. This method is also commonly known as warp-mapped or block-mapped scheduling, and is often used in conjunction with dynamic scheduling techniques such as a dynamic task queue~\cite{Chen:2022:AAT} to create a hybrid load-balancing strategy discussed later (see Section~\ref{sec:task-oriented-scheduling}). Generally, irregular problems with structured regular blocks within them \jowens{``structured regular blocks'' is pretty vague, be a little more precise please} map well to this load-balancing schedule. 

\begin{algorithm}
    \caption[SpMV using group-mapped scheduling]{An example group-mapped scheduled kernel for sparse-matrix dense-vector multiplication (\lstinline{spmv}), where each group is assigned an entire row and threads within the group process individual nonzero elements.}\label{alg:group-mapped}
    \begin{algorithmic}[1]
    \State \textbf{input:} \textbf{$A$}, CSR Matrix. \textbf{$x$}, Dense Vector.
    \State \textbf{output:} $y$, Dense Vector.
    \Procedure{Group-Mapped}{$A, x, y$}
    \State $items\_per\_tile \gets$ \textbf{Zeros}($GROUP\_SIZE$)
    \State $t_{id} \gets$ \textbf{ThreadIndex}()
    \State $gt_{id} \gets$ \textbf{ThreadIndexInGroup}()
    \State \algcomment{Each group populates an array with the number} 
    \State \algcomment{of work items per tile.}
    \State $items\_per\_tile[gt_{id}] \gets A.offsets[t_{id} + 1] - A.offsets[t_{id}]$
    \State $prefix\_sum\_array \gets$ \textbf{ParallelPrefixSum}($items\_per\_tile, t_{id}$)
    \State \algcomment{Last element of prefix-sum array corresponds}
    \State \algcomment{to total work items per group.}
    \State $ total\_items \gets prefix\_sum\_array[GROUP\_SIZE - 1]$
    \State \algcomment{Loop over total work, each thread processing individual work items.}
    \For{$k \gets t_{id}$, $k + GROUP\_SIZE$, \textbf{while} $k < total\_items$}
        \State \algcomment{Perform a binary-search to find the tile index.}
        \State $row \gets$ \textbf{BinarySearch}($prefix\_sum\_array, k$)
        \State $y[row] \gets A.values[k] \times x[A.indices[k]]$
    \EndFor
    \EndProcedure
    \end{algorithmic}
\end{algorithm}

\subsection{Work-Oriented}
\label{sec:work-oriented}
\definition{All threads are assigned $\frac{\textit{Total\ Work}}{\textit{Number\ of\ Processors}}$ work items. Each thread then sequentially processes assigned work items in a loop.}
\characterization{Static, Exact, Flat (+ GPU hierarchy).}
\evaluation{Works well when the upfront cost of performing a prefix sum or binary search is minor compared to the cost of processing the work items; very effective at balancing highly irregular work (such as in scale-free graphs).}

So far we have considered two load-balancing techniques that map \emph{work tiles}, the input, onto compute entities (thread- or group-mapped), however, neither are guaranteed to achieve perfect workload balancing at the device level. Instead, we consider the \emph{work items} as the granularity targetted for load-balancing, where additional computation is done to assign equal amounts of work items to each thread in a GPU, achieving exact workload balancing. \jowens{I don't think you can stress too much in this dissertation that the natural parallelizations are over tiles but that the most load-balanced parallelizations are over work items, and it takes effort to go from the first to the second. Really, you can say this a dozen times and it won't be overemphasized. I think you want to put it early in this section in a giant box and name it something prominent and keep referring to it.} This approach is known as work-oriented scheduling and can be implemented via two predominant methods, which largely differ in the definition of a ``work unit''.

Since the goal is to assign a constant number of work items per thread, the definition of a work item is an important design decision. The first method, known as \emph{non-zero splitting} (used for sparse-linear algebra kernels such as SpMV), considers the total number of nonzero elements within a sparse-matrix or total multiply-accumulate instructions ($y_i \pluseq A_{ij} \times x_j$)
as the total work~\cite{Baxter:2013:MPA,Dalton:2015:OSM,Yang:2018:DPF}. The second method, \emph{merge-path}, considers a work item as either a nonzero element or an output, effectively associating an equal cost to a nonzero and to outputting a value to the GPU's global memory~\cite{Merrill:2016:MPS,Green:2012:GMP}. Fundamentally, the work-oriented assignment is a mapping of the $\frac{Total\ Work}{Number\ of\ Threads}$ to each thread, where each thread is assigned a contiguous range of work items to process in a sequential loop. A binary search per thread (a key low-level algorithm discussed in Section~\ref{sec:challenge:search}) then finds the work tile for each range of work items thread must process (see Algorithm~\ref{alg:work-oriented}).

Merge-path was originally proposed for SpMV kernels. In a merge-path SpMV, the sparse-matrix is stored in a compressed sparse row format (CSR), where the row offsets of the CSR matrix are already known (the row offset array is effectively a prefix-sum array). The schedule aims to perform a 2-D split of the grid created using the row-offsets and the nonzero indices, and the search along the diagonal of the grid allows each thread to find its starting and ending rows and nonzeros. The merge-path schedule ensures that there is constant number of fix-up steps (i.e., the number of threads) required to accumulate the partial rows that were split between two threads. The result of a merge-path schedule is that work is mapped to the smallest possible compute unit in the hierarchy, but merge-path takes advantage of the compute hierarchy by first splitting the work across blocks, and then to threads within a block to reduce the search space each thread has to search through (each thread now only searches through its block's share of the work). If the data is already stored in a CSR representation (as proposed by the original paper by Merrill and Garland~\cite{Merrill:2016:MPS}), the cost of this method is the 2-dimensional binary search along the row offsets and the nonzero indices of the CSR matrix, and the required fix-up step for the partial tiles of work. Beyond load-balancing a CSR sparse matrix, this method can also be generalized to other sparse representations and domains such as graph analytics, with an added cost of a parallel prefix sum to create the required offsets array~\cite{Wang:2017:GGG,Brahmakshatriya:2021:CGA}. Since there is an additional overhead attached to computing this prefix-sum, this method tends to be less performant than other low-overhead methods on an well-structured sparse data~\cite{Yang:2018:DPF}.

\begin{algorithm}
    \caption[SpMV using work-oriented scheduling]{An example \schedule{work-oriented} scheduled kernel for sparse-matrix dense-vector multiplication (\lstinline{spmv}), where each thread is assigned an even share of work~\cite{Merrill:2016:MPS}.}\label{alg:work-oriented}
    \begin{algorithmic}[1]
    \State \textbf{input:} \textbf{$A$}, CSR Matrix. \textbf{$x$}, Dense Vector.
    \State \textbf{output:} $y$, Dense Vector.
    \Procedure{Work-Oriented}{$A, x, y$}
    \State \algcomment{Calculate work division and bounds.}
    \State $t_{id} \gets$ \textbf{ThreadIndex}()
    \State $total\_work \gets A.rows + A.nonzeros$
    \State $items\_per\_thread \gets \frac{total\_work}{num\_threads}$
    \State $ diag \gets$ \textbf{min}($items\_per\_thread \times t_{id}, total\_work$)
    \State $ diag\_end \gets$ \textbf{min}($diag + items\_per\_thread, total\_work$)
    \State $ (row_{start}, nz_{start}) \gets$ \textbf{2DSearch}($diag, A$)
    \State $ (row_{end}, nz_{end}) \gets$ \textbf{2DSearch}($diag\_end, A$)

    \State \algcomment{Perform work on full tiles.}
    \State $ running\_total \gets 0$
    \For{$m \gets row_{start}$, $m{+}{+}$, \textbf{while} $m < row_{end}$}
        \For{$k \gets nz_{start}$, $k{+}{+}$, \textbf{while} $k < A.offsets[m]$}
            \State $running\_total \pluseq x[A.indices[k]] \times A.values[k]$
        \EndFor
        \State $y[row] \gets running\_total$
        \State $ running\_total \gets 0$
    \EndFor

    \State \algcomment{Perform work on partial tiles.}
    \For{$k \gets k$, $k{+}{+}$, \textbf{while} $k < nz_{end}$}
        \State $running\_total \pluseq x[A.indices[k]] \times A.values[k]$
    \EndFor
    \State $row\_carry\_out[t_{id}] \gets row_{end}$
    \State $value\_carry\_out[t_{id}] \gets running\_total$
    \EndProcedure
    \State \algcomment{Fix-up step to accumulate partial tiles.}
    \Procedure{Fix-up}{$A, y, row\_carry\_out, value\_carry\_out$}
    \For{$t_{id} \gets 0$, $t_{id}{+}{+}$, \textbf{while} $t_{id} < num\_threads$}
        \State $y[row\_carry\_out[t_{id}]] \pluseq value\_carry\_out[t_{id}]$
    \EndFor
    \EndProcedure
    \end{algorithmic}
\end{algorithm}

\subsection{Binning and Reordering}
\label{sec:binning}
\definition{Separate work tiles into a fixed number of bins based on heuristics, such that each bin holds work tiles with approximately the same amount of work items.}
\characterization{Dynamic, Approximate, Hierarchical, Non-Cooperative, Centralized or Distributed.}
\evaluation{Effective at balancing workloads with contiguous work tiles with approximately the same number of work items, and with imbalance among these contiguous ranges of work tiles. This method maps well to CUDA's compute hierarchy, where bins can be assigned to a level in the hierarchy depending on the granularity of work within them.}

The static load-balancing techniques discussed above aim for an even work distribution for a given problem; these techniques are inefficient when either the overhead to load-balance is greater than processing the entire computation, or when there is a huge mismatch between the worker size and the work unit. Instead, when a given problem has a known bound on the amount of work, we can implement a load-balancing schedule that is reasonably efficient for a given bounded range, and particularly so if we choose the correct amounts of compute resources that match the given input range (e.g., assign a work tile with exactly 32~work items to be processed by a warp composed of 32~threads.) This is where binning-based load-balancing algorithms, which attempt to dynamically balance a given workload by categorizing each work tile into a fixed number of bins with the goal that each bin will hold work tiles that have approximately the same amount of work items, perform reasonably well. Binning-based techniques can largely be subdivided into three different phases: (1) choosing bin sizes (compile time); (2) dynamically distributing work tiles into bins based on a programmer-specified criteria (runtime); then (3) computing the contents of each bin (runtime).

One approach to binning that specifically targets the GPU compute hierarchy chooses three bins to partition the work tiles: (1) a bin with work tiles that have a number of work items equal to or greater than the size of the GPU block (number of threads per block); (2) a bin with work tiles that have a number of work items less than the size of the GPU block but larger than a warp (32 threads); and (3) a bin with work tiles that have a number of work items less than the size of the warp. For brevity, we will use the terms block-sized, warp-sized, and thread-sized bins. After the initial assignment of work tiles to their respective bins, the processing stage that perform the desired computation on the contents of each bins can be implemented in three different ways:
\begin{enumerate}
    \item Launch three kernels, where each kernel is specialized to process one of the three categories of bins. For the block-sized bin, one kernel uses all threads with the block to cooperatively process the tiles. Similarly, in the warp-sized bin, the kernel uses all threads within a warp to cooperatively process the bin, And finally, the last kernel processes the thread-sized bin, where each work tile has only a few work items, my assigning one thread to each work tile~\cite{Merrill:2012:SGG,Wang:2017:GGG}.
    \item Launch one kernel, where each thread is initially assigned a work tile, determines the amount of work, and assumes control of the block or the warp if the size of work is enough to saturate the compute unit. All threads within the compute unit then cooperatively work together to process each work item within the ``winning'' thread's tile. This process continues until the work tiles are small enough for each thread to individually process them (Algorithm~\ref{alg:twc})~\cite{Davidson:2014:WPG,Wang:2017:GGG}.
    \item Brahmakshatriya et al.\ proposed a variation of the one-kernel approach, where instead of threads competing to assume control of a block, each block first processes a multiple of its size worth of work items of a tile (if any), and then a warp processes multiple of its size worth of work items, and finally the threads complete the work tile by processing the remaining work items~\cite{Brahmakshatriya:2021:CGA}.
  \end{enumerate}

The benefit of the three-kernel specialization is that each kernel is designed to process one bin-type with no added communication or synchronization for threads within the kernel to assume control of a given compute unit or hierarchy. However, the total work now is now split among three kernel launches, making it challenging to fully utilize the device per kernel for a given work distribution. Furthermore, for problems that require a large number of iterations to successfully converge, the kernel launch overhead, which is commonly an insignificant cost and largely ignored, is now 3$\times$ larger.

\begin{algorithm}
    \caption[SpMV using binning-based scheduling]{An example \schedule{binning} scheduled kernel for sparse-matrix dense-vector multiplication (\lstinline{spmv}), where work tiles are placed into bins based on the number of work items within them, and processed with the required compute resource~\cite{Merrill:2012:SGG,Davidson:2014:WPG}.}\label{alg:twc}
    \begin{algorithmic}[1]
    \State \textbf{input:} \textbf{$A$}, CSR Matrix.
    \Procedure{Binning}{$A$}
        \State $row \gets$ \textbf{ThreadIndex}()
        \State $num\_nonzeros \gets A.offsets[row + 1] - A.offsets[row]$
        \If{$num\_nonzeros \geq block\_size$}
            \State $cta\_bin\_ids[cta\_bin\_size] \gets row$
            \State $cta\_bin\_size{+}{+}$
        \ElsIf{$num\_nonzeros \geq warp\_size$}
            \State $warp\_bin\_ids[warp\_bin\_size] \gets row$
            \State $warp\_bin\_size{+}{+}$
        \Else
            \State $thread\_bin\_ids[thread\_bin\_size] \gets row$
            \State $thread\_bin\_size{+}{+}$
        \EndIf
    \EndProcedure

    \State \textbf{input:} \textbf{$A$}, CSR Matrix. \textbf{$x$}, Dense Vector.
    \State \textbf{output:} $y$, Dense Vector.
    \Procedure{CTA\_BIN}{$A, x, y$}
        \State $row \gets cta\_bin\_ids[idx]$
        \State \algcomment{Use an entire CTA to process the row.}
    \EndProcedure
    \Procedure{WARP\_BIN}{$A, x, y$}
        \State $row \gets warp\_bin\_ids[idx]$
        \State \algcomment{Use an entire warp to process the row.}
    \EndProcedure
    \Procedure{THREAD\_BIN}{$A, x, y$}
        \State $row \gets thread\_bin\_ids[idx]$
        \State \algcomment{Sequentially process a row in the thread.}
        \For{$k \gets A.offsets[row]$, $k{+}{+}$, \textbf{while} $k < A.offsets[row+1]$}
            \State $y[A.indices[k]] \pluseq x[A.indices[k]] \times A.values[k]$
        \EndFor
    \EndProcedure
    \end{algorithmic}
\end{algorithm}

Another proposed approach, Logarithmic Radix Binning (LRB), first introduced by Green et al.\ within a triangle counting implementation~\cite{Green:2018:LRB}, allows the load balancer to schedule work items with similar amounts of work within the same spatial and temporal region~\cite{Fox:2019:ISI}. LRB assigns tasks to bins based on the \emph{logarithm} of the amount of work required for a given task. Thus, the range of possible amounts of work across all items in a bin varies by no more than a factor of two. When a work tile is encountered with work items greater than equal to $2^b$ and less than $2^{b+1}$, a counter at location $b$ is incremented in an array of size $B \in{\{32, 64\}}$. Using these bins instead of the ones associated with CUDA's compute hierarchy (threads, warps, blocks) in the previous approach, the total workload is binned at a finer granularity such that work tiles containing a similar number of work items are grouped together for processing. To assign work tiles to threads, a modulo operation is performed on the reordered task index and a constant number of work items, $P$\@. Each thread then processes $(i, P + i, 2 * P + i, \ldots)$ work items~\cite{Fox:2019:ISI,Green:2018:LRB}.

Another widely used technique for reordering or grouping like-sized work together is a simple (but often costly) sort to arrange tiles from most to fewest work items~\cite{Bolz:2003:SMS}. The sort operation reduces the work variance between adjacent tiles and therefore improves load balance. In iterative problems such as sparse-matrix dense-matrix multiplication in a deep learning workload, the overhead of the sort is amortized over the number of runs of an algorithm~\cite{Gale:2020:SGK}.

\subsection{Task-Oriented Scheduling}
\label{sec:task-oriented-scheduling}
\definition{Independent workers fetch work from a work queue or queues, process the work, and optionally add new work to a queue.}
\characterization{Dynamic Approximate, Flat, Cooperative, Centralized or Distributed.}
\evaluation{Works well when it is too expensive to compute a static assignment of work items to processors at the start of the kernel, if the kernel dynamically generates new work items, or if the cost to process each work item is unknown.

Performs best if one thread can fetch chunks of work on behalf of an entire block, to reduce synchronization contention. Powerful when paired with hierarchical load balancing within a block, where a block fetches a chunk of work on behalf of all threads, and then further subdivides this work among its threads.}

The previously-described load balancing methods divide up work at compile time or runtime and associate each chunk of work with a virtual processor, which the hardware scheduler then dynamically maps to a physical processor based on resource utilization. In contrast, task parallelism shifts the job of assigning work to processors to the programmer. Task parallelism methods are more flexible at runtime, and benefit workloads where it would be expensive to compute the optimal work distribution and assign it to individual processors in advance, or where the total amount of work to be generated in a kernel is unknown. Programmers can customize the task generation and consumption behavior to their specific use case, at the cost of some software overhead compared with the GPU's hardware block scheduler. A primary benefit of a task-parallel schedule is that it allows processors that finish their currently assigned task to immediately start work on another task without waiting for synchronization at the end of a kernel, or to generate new irregular tasks that other processors may consume without waiting for a new kernel iteration.

A classic example of task parallelism is a queue-based implementation of Breadth-First Search. After a block removes a vertex from the FIFO queue, it sets the vertex's depth then adds neighbors of that vertex to the back of the queue. In a parallel GPU system, these new work items may be popped and processed by other processors in the system. This process continues until the queue is empty and the processors no longer have new work to add to the queue.

The primary downsides of queue-based schedules are that they suffer from overheads necessary to ensure that each processing element adds or retrieves work at the correct location, and that the queue must be large enough to accomodate the worst-case number of work items in the queue. This is because memory allocations are expensive and happen outside the kernel boundary. Additionally, there is a chance that some processors may have long-running work that leaves other processors sitting idle when the queue is empty. Finally, queue-based schedules replace the hardware work distribution done by the GPU's block scheduler with user-controlled software work distribution, which adds additional overheads.

\subsubsection{Centralized Queue}
\label{sec:centralized-queue}
\evaluation{Useful when the algorithm dynamically creates work at runtime, or when it's undesirable to precompute an assignment of work items to processors.}

\begin{algorithm}
    \caption[BFS using queue-based scheduling]{An example queue-based kernel for breadth-first search, where each worker is assigned an entire vertex.
    }\label{alg:task-oriented}
    \begin{algorithmic}[1]
    \State \textbf{input:} \textbf{$G$}, CSR Graph. \textbf{$Q$}, Work Queue. \textbf{$s$}, Source Vertex.
    \State \textbf{output:} \textbf{$D$}, Depth Array. \textbf{$P$}, Predecessor Array.
    \Procedure{Task-Oriented}{$G, Q, s$}
    \State $t_{id} \gets$ \textbf{ThreadIndex}()

    \State \algcomment{Initialization phase.}
    \State $D[t_{id}] \gets \infty$
    \State $P[t_{id}] \gets None$
    \If{$t_{id} = 0$}
        \State $Q[0] \gets s$
        \State $Q.push\_idx \gets 1$
        \State $Q.pop\_idx \gets 0$
    \EndIf

    \State \algcomment{Loop until the queue is empty.}
    \While{\textbf{not $Q.empty()$}}
        \State $my\_pop\_idx \gets atomicAdd(Q.pop\_idx, 1)$
        \State $my\_pop\_idx \gets $\textbf{mod}$(my\_pop\_idx, Q.size())$
        \State $v \gets Q[my\_pop\_idx]$
        \For{$i \gets 0$, $i{+}{+}$, \textbf{while} $i < G.offsets[v + 1] - G.offsets[v]$}
            \State $n \gets G.indices[G.offsets[v] + i]$
            \State \algcomment{BFS condition: update the neighbor's depth.}
            \If{$D[v] + 1 < D[n]$}
                \State $D[n] \gets D[v] + 1$
                \State $P[n] \gets v$
                \State $my\_push\_idx \gets atomicAdd(Q.push\_idx, 1)$
                \State $my\_push\_idx \gets $\textbf{mod}$(my\_push\_idx, Q.size())$
                \State $Q[my\_push\_idx] \gets n$
            \EndIf
        \EndFor

    \EndWhile
    \EndProcedure
    \end{algorithmic}
\end{algorithm}

The static task list approach presented by Cederman et al.\ is the most basic task-oriented schedule: the kernel consists of two arrays: an ``in-array'' and an ``out-array''~\cite{Cederman:2008:ODL}. The in-array contains tasks to be executed by the processors in the current iteration. As the kernel proceeds, processors iterate over their statically assigned tasks (where a block with index \emph{i} handles all tasks at integer multiples of \emph{i}), perform the task, and potentially add additional tasks to the ``out array''. This process continues until the in-array is empty, at which point the out-array becomes the new in-array. The kernel iterates between the two arrays until there are no remaining tasks in either array. Like all queue methods, this approach requires that the arrays be sized to hold the worst-case number of concurrently-queued tasks, and that the processors use some form of synchronization when adding tasks to the out-array. Note that Cederman et al.\ do not enforce synchronization on reads since blocks execute tasks at predefined positions. This removes the overheads inherent to synchronization, but also loses the greedy consumption property described in the Task-Oriented Scheduling introduction. However, more complex versions of this load-balancing schedule could add synchronization to the in-array to enable greedy task pops from the head of the in-array. Aila et al.\ present a similar method that performs only reads from a single task list with atomic synchronization, and generates no additional work~\cite{Aila:2009:UTE}.

At the next level of complexity, centralized queues allow any processors on the GPU to dynamically consume tasks from the head of the queue when they need more work, or add tasks to the tail of the same queue when they generate new work, which will eventually be consumed by processors without an intermediate synchronization step, such as the one needed for the static task list schedule to swap the in- and out-arrays. Many queue-based load-balancing variants exist, such as variants that use mutual exclusion to control access to the queue, variants that use cheaper atomic increments for synchronization, variants that get an entire block's worth of work with a single thread or warp, variants that issue new work to the GPU's kernel queue as additional kernel launches, variants that use multiple queues to store tasks from different stages of a rendering pipeline, and variants that add tasks from the CPU and consume from the GPU~\cite{Aila:2009:UTE,Chen:2010:DLB,Cederman:2008:ODL,Zhang:2015:DPS,Steinberger:2014:WTS,Chen:2022:AAT}.

The distributed queue schedule presented by Zhang et al.\ assigns per-processor queues with task indices stored in shared memory~\cite{Zhang:2014:CAO}. This allows each processor's threads to add and consume tasks from the block's private queue, and avoids the atomic contention overheads seen in the monolithic queue schedules if the thousands of threads on a GPU were to compete for a single globally-shared atomic. However, as GPU atomics have improved in performance over time, atomic contention in modern centralized queues is no longer a significant bottleneck~\cite{Chen:2022:AAT}. The schedule also receives latency benefits from storing task indices in shared memory. The primary downsides of this method is that the GPU cannot load-balance between blocks, and each block's individual queue must be sized for the worst-case queue size. This may cause the aggregate amount of memory used to be larger than the centralized task queue variant. As part of this work, Zhang et al.\ present the CUIRREE library, which provides general-purpose APIs and implementations for queue-based load balancing.

\subsubsection{Task Stealing}
\label{sec:task-stealing}
\evaluation{Reduces contention compared to the global queue method since processors usually utilize their own private queue. Solves load imbalance between block queues with stealing, and improves locality.}

Task stealing queue variants provide per-processor task queues~\cite{Cederman:2011:DLB,Tzeng:2010:TMF, Zhang:2014:CAO}. Processors typically both consume work from and add work to the tail of their own queue. If a processor has no tasks available to process in its queue, it may ``steal'' a task from the head of another processor's queue. Since stealing is the less common case, such schedules minimize the effect of atomic contention that occurs in the monolithic queue case. Setting the queue so that task pops come from the head, while task additions and steals come from the tail, also ensures that the owning processor almost never needs to synchronize when acquiring a new task. They also improve in data locality, since neighboring tasks in time likely access nearby data.

However, task stealing schedules still have the problem that each individual queue must be sized so that it could hold the worst-case of work that could be simultaneously in the processor's queue, which is not ideal on GPUs with limited memory (both on-chip and off-chip). If the queues are too small, the processor must stall until another processor steals work from a full queue.

\subsubsection{Task Donation}
\label{sec:task-donation}
\evaluation{When considering Task Stealing, Task Donation is even better. It allows processors to offload overflow work and shrinks queue sizes.}

Task Donation extends the work stealing schedule to also allow processors to add overflow work to another processor's queue that still has space available~\cite{Tzeng:2010:TMF}. This improves memory utilization on the system since now only the aggregate capacity of all queues needs to be large enough to hold the maximum amount of work. This also improves latency if queues are small enough to reside in on-chip cache.

In many ways, task donation is the ``ideal'' version of a queue-based load-balancing schedule. It preserves locality between work items, reduces synchronization contention since processors likely work out of their own queues most of the time, uses the same amount of memory as a global queue, and preserves the ability to load balance between blocks using stealing and donation.

\subsubsection{Hierarchical Task Scheduling}
\label{sec:hierarchical-task-scheduling}
\evaluation{Reduces contention at the global queue since a single thread can fetch a chunk of work for the block, and can load balance within the block. Potentially suffers from some load imbalance between blocks.}

The centralized queue, task stealing, and task donation variants previously described use only a single level of load balancing, where threads or blocks retrieve work from the queue, but do not do anything \emph{within} the block to balance the work. Consider, as an example, a breadth-first-search kernel where the first thread in each block acquires a \emph{chunk} of work (multiple vertices) on behalf of all threads in the block, perhaps in an attempt to reduce the overheads of atomic contention that were to occur if each thread in the GPU retrieved its work directly from the queue. This type of kernel would benefit from an addional level of load balancing within the block, such as work-oriented scheduling or binning. Chen et al.\ use a hybrid task/work scheduling approach in Atos~\cite{Chen:2022:AAT}.

Another method presented by Steinberger et al.\ uses task lists in both global memory and shared memory hierarchies~\cite{Steinberger:2014:WTS}. The shared-memory queues are faster, but have limited memory and cause load imbalance between blocks. The global queues have significantly more space and can load balance between blocks, but suffer from high-latency accesses and require more expensive contention. Using two levels of queue hierarchy, where newly-generated work items can go in either level of the hierarchy based on some heuristic or the available storage space, finds a useful balance point. Additionally, Steinberger et al.\ use a binning strategy, with separate queues for block-sized, warp-sized, or thread-sized items to reduce load imbalance within a processor.

%% file: ch_survey/chapters/primitives.tex
\section{Common Low-Level Algorithmic Primitives}
\label{sec:primitives}

In this section we identify the key building blocks that are used to construct the load-balancing implementations discussed in Section~\ref{sec:deeper-dive}. These building blocks are used to address the irregularity/sparsity of their inputs and would likely not be necessary if the inputs were dense or regular. For instance, consider load-balancing an operation on every element of a matrix across two workers. If that matrix is dense, it is likely simple to split it in half knowing only the matrix dimensions such that each worker is assigned half of the matrix elements. With a sparse matrix, determining the split point is more complex; a robust solution must first count all non-zero elements in the matrix, then identify the middle element to be the split point, and finally assign half of the elements to each worker.

We first identify the challenges faced by the load-balancing implementations, then the building blocks needed to address that challenge. Most load-balancing strategies face challenges in two general areas: (1) Discovering work items and work tiles and (2) scheduling/assigning work onto worker units.

\subsection{Challenge: Counting Non-zeros or Work Items}
\textbf{Goal: } Several load balancers must first scan a sparse matrix or graph to determine the number of non-zeros or work-items within each work tile of the input data. For example, SpGEMM implementations using Gustavson's row-wise  implementation~\cite{Gustavson:1978:TFA} first count the number of non-zeros within each row of the left-hand matrix to provide an estimate on the amount of work needed~\cite{Gremse:2015:GSM}. Similarly, kernels such as SpMV, SpMM and higher-order tensors using work-oriented, thread-mapped, or group-mapped approaches first divide the sparse matrix/tensor into regions of equal-sized non-zero blocks~\cite{Yan:2014:YYA}.
\textbf{Target kernels} include nearly all sparse kernels---SpMV, SpGEMM, SpMM, and graph algorithms.
\subsubsection{Implementation Primitives}
When counting regions of work items within a workload, the following algorithms are commonly used:
\begin{itemize}
  \item \textbf{Parallel prefix sum}\label{sec:ps}, i.e., parallel scan, is a widely used algorithm popularized by Guy Blelloch in the parallel setting~\cite{Blelloch:1990:PSA}. Later, Harris et al.\ proposed a parallel GPU implementation~\cite{Harris:2007:PPS}. The algorithm takes as input (1) a binary associative operator---which is $+$ for prefix sum---and (2) an array.
  It then produces a new array where the element at any position is a sum of all previous elements. Prefix-sum when applied to the work items per work tile for all tiles produces an array that not only has the \emph{total number of work} to be processed, but also the indices of this array (i.e. the location of each element) represents the work tile which any given work item belongs to.
  Parallel scan is especially useful when creating a balanced workload within the GPU's hierarchy~\cite{Harris:2007:PPS}. The load balancer performs a parallel prefix sum at a device-, block- or warp-wide level on an input array of expected work items per task. Using the resulting array, the work is effectively partitioned onto GPU threads. This implementation is particularly common in non-zero splitting load balancers~\cite{Baxter:2013:MPA, Dalton:2015:OSM, Yang:2018:DPF,Steinberger:2017:GHL}. It is also common in matrix algorithms splitting work by row length~\cite{Gremse:2015:GSM}. Moreover, schedulers needing the total amount of work, such as work-oriented approaches (Section~\ref{sec:work-oriented}), simply access the last element of the prefix sum array.

  A search on the prefix sum array can be used to determine which task is being processed for any given work item. Examples of this process include a Single-Source Shortest Path algorithms~\cite{Davidson:2014:WPG}, Breadth-First Search algorithms~\cite{Merrill:2012:SGG}, various merging algorithms~\cite{Green:2018:LRB}, and SpMV~\cite{Green:2012:GMP, Dalton:2015:OSM, Merrill:2016:MPS,CUB:2016}. Likewise, the IrGL~\cite{Pai:2016:ACF} compiler and the CUDA-quicksort algorithm~\cite{Manca:2016:CAI} explicitly use prefix sums to allocate space for work items.

  \item \textbf{Segmented scan} and \textbf{segmented reduce}, closely related to prefix sum, reduces data in a work tile or \emph{segments}. Given an array of arrays, the segmented \emph{scan} performs a prefix sum on each array within the larger array. Instead of storing the entire scan array, a segmented \emph{reduce} stores the final, single sum per input sub-array. Binning approaches requiring row-length or vertex output degree information typically employ segmented reduce~\cite{Gale:2020:SGK,Ashari:2014:FSM,Nisa:2019:LSM,Hong:2017:MEG}. Both Yan et al.\ and Liu et al.\ use segmented scan when partitioning sparse 2D, 3D and 4D tensors across threads~\cite{Yan:2014:YYA,Liu:2017:UOA}.
\end{itemize}
\subsection{Challenge: Searching for Work Tiles and Items}
\label{sec:challenge:search}

\textbf{Goal: } For several load-balancing schedules, each thread has a pre-assigned number of work items to process, however, the individual threads may not know exactly where in the total work pool they should begin processing their assigned work, such that each thread has unique work items to process. Furthermore, threads may not know which work tile an assigned work item belongs to. For example, for kernels such as SpMV, if threads are assigned to process non-zero values, they \emph{need} to determine (1) a range of unique non-zeros to multiply and accumulate, and (2) the row and column for the corresponding non-zero. The primitives below help each thread search, in parallel, a sorted list of number of work items to determine their starting position and the corresponding work tiles for a given work item.

\subsubsection{Implementation Primitives}
The following algorithms are commonly used (searches are called with bulk inputs in parallel):

\begin{figure}
  \centering
  \includegraphics[width=\textwidth]{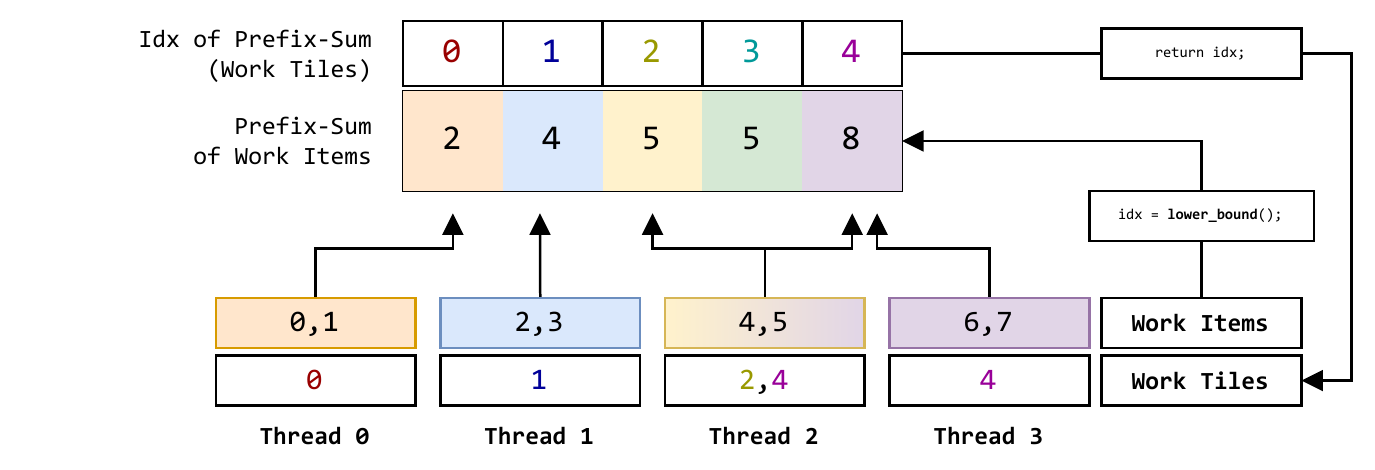}
  \caption[Lower-bound search used for searching load balanced work]{Shows a simple problem where 8~work items are perfectly balanced across 4~threads, and a running sum of work items per work tile (prefix-sum array) is provided as an input. Each thread processes work items range of $[\text{thread\_idx} \times \frac{8}{4}] \dots [(\text{thread\_idx} + 1) \times \frac{8}{4}]$ and uses a lower bound search into the prefix-sum array to determine the index of the work tile for each work item.} \label{fig:lower_bound_search}
\end{figure}

\begin{itemize}
  \item \textbf{Binary Search}, operating on a sorted input, repeatedly divides the input in half until the desired item is found. In approaches using parallel prefix sums to divide work items, each thread performs a lower bound search (using binary search) on the prefix sum array with the work item index as the input. The lower bound search specifically returns the index of the first position where the work item can be inserted. This index corresponds to the work tile the input work item belongs to. Figure~\ref{fig:lower_bound_search} shows an illustration of this process with an example. In graph algorithms, threads may use this method to search on the edge offset array (created using a prefix sum), to determine the source vertex ID of an edge~\cite{Davidson:2014:WPG,Khorasani:2015:SSG,Wang:2017:GGG,Busato:2015:BAE}.
  \item \textbf{Vectorized Sorted Search} or \textbf{Load-balanced Search}, proposed by Sean Baxter in ModernGPU~\cite{Baxter:2013:MPA},
  first sorts the queries to avoid the thread divergence caused by the traditional binary search algorithm. Given a sorted list of queries, $A$, and the sorted database $B$, it recasts the search problem into a merge problem, where the program now linearly searches for $A$ in $B$. This improves work-efficiency over binary search, with complexity $\mathcal{O}(A+B)$ versus $\mathcal{O}(A\log{}B)$. Overall, each thread now has more locality in the search since the keys will be sorted.
  Within a load balancing context, vectorized sorted search is applied to a prefix sum array of work items offsets to determine the index of the work tile that generated the work item.
  Within graph algorithms, and similar to strategies using a binary search, threads cooperatively apply vectorized sorted search to determine source vertex IDs for a given edge~\cite{Wu:2015:PCO, Davidson:2014:WPG,Wang:2017:GGG}.
\end{itemize}
\subsection{Challenge: Efficient Binning of Work Items}
\textbf{Goal: } As mentioned in Section~\ref{sec:binning}, several load balancers use various heuristics to place work items into bins, such that each bin has a similar amount of work. Finding a heuristic with low overhead and near-perfect distribution of work is a critical step in developing an efficient implementation.
\subsubsection{Implementation Primitives}
Some strategies used in the implementation of binning strategies include:
\begin{itemize}
  \item \textbf{Sorting Based on Work Variance}: Parallel sorting algorithms are often used as a means to balance the workload in a given irregular application. To minimize the preprocessing overhead of sorting, load balancers often employ this strategy when the cost can be amortized over multiple runs of the same application for a given input~\cite{Gale:2020:SGK}. For example, in Sparse-Matrix Dense-Matrix Multiplication (SpMM) for sparse deep learning workloads, Gale et al.\ first reorder the sparse matrix by sorting its rows by length before assigning row bundles to each warp~\cite{Gale:2020:SGK}. Several other implementations for SpGEMM, SpMM, triangle counting, and graph pattern mining likewise rely on an initial sorting or reordering by the amount of work, number of non-zeros, or vertex degree before assigning to threads/warps/blocks~\cite{Bolz:2003:SMS,Mehrabi:2021:LSM, Arifuzzaman:2013:PPA, Esfahani:2022:LLO, Jamshidi:2020:PPA, Hu:2021:ATC, Dalton:2015:OSM,Davidson:2014:WPG}.
\end{itemize}
\subsection{Challenge: Assigning Compute Resources}
\textbf{Goal: } In dynamic approaches to load balancing, workers compete for compute resources and/or work items. Such implementations need a way to determine the outcome of the competition.
\subsubsection{Implementation Primitives}
Common primitives in determining resource assignment include:
\begin{itemize}
  \item \textbf{Voting}: This strategy is often used in conjuction with binning approaches, where work is assigned to be processed by individual threads, warps, or CTAs, depending on the work tile size~\cite{Merrill:2012:SGG, Davidson:2014:WPG}. Since there are limited resources\jowens{I don't get this previous phrase}, each worker unit \jowens{what is a ``worker unit''?} competes for the resources of the entire group (warps or CTAs), by writing, atomically, to a shared memory location. \jowens{I understand this to be ``there's a bunch of entities that are part of a group, and all entities negotiate to see who can be the head entity and be in charge'' but I don't think this is explained nearly well enough to be confident of that conclusion.} The winner of the atomic write takes over the resources and proceeds with compute. Originally designed by Merrill et al.~\cite{Merrill:2012:SGG}, several works employ this strategy within their own load balancing implementations~\cite{Dathathri:2018:GCO,Jatala:2020:ALB,Jatala:2020:SGA,Wang:2017:GGG}.

  \item \textbf{Dynamic Parallelism}: Starting with architectures of CUDA capability 3.5 and higher, kernels can be launched using dynamic parallelism, which allows CUDA kernels to recursively launch child kernels. Each child kernel then has complete access over the compute resources. This strategy can be used when processing work tiles containing a large amount of work items. The load balancer dynamically launches one or more kernels sized according to the amount of work within the work tile~\cite{Ashari:2014:FSM, Busato:2015:BAE, Zhang:2015:DPS, Busato:2019:CGT}. \jowens{I'm pretty sure there's work out there that says dynamic parallelism has high overheads and thus isn't very useful, and you need to say that. Programming model good, implementation not good enough.}
\end{itemize}



%% file: ch_survey/chapters/summary.tex
\section{Summary of Load-Balancing Techniques}
\label{sec:summary}

To the best of our knowledge, Table~\ref{tab:annotated-bib}~and~\ref{tab:annotated-bib_2} characterizes all of the predominant approaches practiced for load balancing irregular-parallel problems on the GPU to the load-balancing taxonomy described in the previous section.

\begin{table*}[]
    \tiny
    \begin{tabularx}{\textwidth}{smmbs} 
        \toprule
        \multicolumn{1}{c}{Work(s)} & \multicolumn{1}{c}{Application Space} 
        & \multicolumn{1}{c}{Method(s)} & \multicolumn{1}{c}{Implementation Detail} & \multicolumn{1}{c}{Characterization} \\
        \midrule
        \citet{Green:2012:GMP} \citet{Dalton:2015:OSM} \citet{Merrill:2016:MPS} \citet{CUB:2016} & Merging Algorithm, Sparse-Matrix Dense-Vector Multiplication 
        & Merge-Path Based decomposition & Work (number of outputs and nonzeros) divided equally to number of threads. 2-D binary-search on row-offsets and nonzeros to figure out the active rows and nonzero indices. & Static, Exact, Flat or Hierarchical (to reduce the search space per thread) \\
        \addlinespace[7pt]
        \citet{Baxter:2013:MPA} \citet{Dalton:2015:OSM} \citet{Yang:2018:DPF} \citet{Steinberger:2017:GHL} & SpMV, General Graph Traversal 
        & Nonzero Splitting & Work (number of nonzeros) split equally to number of threads. 1-D binary search on row-offsets to find the active row for any given nonzero/thread. & Static, Exact, Flat \\
        \addlinespace[7pt]
        \citet{Merrill:2012:SGG} \citet{Davidson:2014:WPG} & Breadth-First Search, Single-Source Shortest Path
        & Sequential Gather & Vertices asigned to threads, neighbor-list processed sequentially & Static, Approximate, Flat \\
        \addlinespace[7pt]
        \citet{Merrill:2012:SGG} \citet{Davidson:2014:WPG} & Breadth-First Search, Single-Source Shortest Path
        & Warp Gather & Vertices assigned to warps, paralle prefix sum and binary-search used to asign a neighbor to each thread within the warp & Static, Approximate, Hierarchical \\
        \addlinespace[7pt]
        \citet{Merrill:2012:SGG} \citet{Davidson:2014:WPG} & Breadth-First Search, Single-Source Shortest Path
        & Binning (Scan+Warp+CTA Gather) & Place vertices with similar amount of work in the same bin, process large bins using a block, smaller bins using a warp and smallest using a thread. & Dynamic, Approximate, Hierarchical, Non-Cooperative, Centralized (warp- and block-level) \\
        \addlinespace[7pt]
        \citet{Davidson:2014:WPG} & Single-Source Shortest Path
        & Edge partitioning & Using parallel sorted search on the scanned edge offset queue find an intersection of block's edge-list within the work list. & Static, Exact, Hierarchical \\
        \addlinespace[7pt]
        \citet{Brahmakshatriya:2021:CGA} & General Graph Traversal 
        & Binning & Work items multiple of block-size processed using a block, trickle the remaining work to be (multiple of warp-size) processed by a warp, and finally, if more work is remaining, trickle to a thread. & Dynamic, Exact, Hierarchical, Cooperative, Centralized \\
        \addlinespace[7pt]
        \citet{Green:2018:LRB} \citet{Fox:2019:ISI} & Triangle Counting, Load-Balancing Algorithm 
        & Logrithmic Radix Binning & Assign work to bins based on the logarithmic work estimate. Approximate method of reordering without sort. & Dynamic, Approximate, Hierarchical, Non-Cooperative, Distributed \\
        \addlinespace[7pt]
        \citet{Baxter:2013:MPA} & GPU Parallel Primitives 
        & Vectorized Sorted Search (Load-Balanced Search) & Upper-bound search of work-item indices with exclusive scan of the counts (subtract one). & Static, Exact, Flat \\
        \addlinespace[7pt]
        \citet{Gale:2020:SGK} & Sparse-Matrix Dense-Matrix Multiplication, Sampled Dense-Dense Matrix Product 
        & Row Binning and Bundling & Sort row-indices based on a heuristics over row-length, grouping like-sized rows and tiles together & Static, Approximate, Hierarchical (Block and Warp-wide) \\
        \addlinespace[7pt]
        \citet{Lee:2020:OOG} & Sparse-Matrix Dense-Matrix Multiplication
        & Block Reorganizer & Group blocks into overloaded, normal, underloaded blocks based on the size of their work. Split large blocks, merge underloaded blocks. & Static, Approximate, Hierarchical (Block-wide) \\
        \addlinespace[7pt]
        \citet{Zhang:2015:DPS} \citet{Busato:2015:BAE} \citet{Busato:2019:CGT} & Graph Algorithms
        & Dynamic Kernels & Launch a dynamic parallel kernel configured to a given workload. & Dynamic, Approximate, Hierarchical, Non-Cooperative, Centralized \\
        \addlinespace[7pt]
        \citet{Cederman:2008:ODL} \citet{Aila:2009:UTE} & Ray Traversal 
        & Task List & Work items popped from a read-only queue. If the blocks generate new work, this gets added to a write-only queue. Queues swap when the read-only queue is empty. & Dynamic, Approximate, Flat, Cooperative, Centralized \\
        \addlinespace[7pt]
        \citet{Cederman:2008:ODL} \citet{Chen:2010:DLB} \citet{Zhang:2014:CAO} \citet{Chen:2022:AAT} & Dynamic Load-Balancing (Octree Partitioning), Molecular Dynamics (Atoms-Decomposition), Ray-Tracing, N-queens solver, Potential distribution, Asynchronous Graph Algorithms. 
        & Task Queue & 
        \cite{Cederman:2008:ODL}~Blocking and lock-free (atomic) queue variants. 
        \cite{Chen:2010:DLB}~CPU pushes, GPU blocks pop. \cite{Zhang:2014:CAO}~Device-wide and per-block queue variants (no stealing or donation.) 
        \cite{Chen:2022:AAT}~A single thread acquires tasks on behalf of an entire block to reduce synchronization overheads. 
        & Dynamic, Approximate, Flat, Cooperative, Centralized \\
        \addlinespace[7pt]
        \citet{Tzeng:2010:TMF} \citet{Cederman:2011:DLB} \citet{Zhang:2014:CAO} & General Load-Balancing (Rendering Primitives and Synthetic Workloads) 
        & Task Stealing & Threads consume from per-block queues, can steal from other blocks' queues & Dynamic, Approximate, Hierarchical, Cooperative, Distributed \\
        \addlinespace[7pt]
        \citet{Tzeng:2010:TMF} & General Load-Balancing (Rendering Primitives and Synthetic Workloads) 
        & Task Donation & Per-block queues, can steal from or donate to other queues. & Dynamic, Approximate, Hierarchical, Cooperative, Distributed \\
        \bottomrule
    \end{tabularx}
    \caption[Annotated bibliography of load-balancing techniques]{Annotated bibliography of load-balancing techniques. \emph{Continued in Table~\ref{tab:annotated-bib_2}.}}
    \label{tab:annotated-bib}
\end{table*}

\begin{table*}[]
    \tiny
    \begin{tabularx}{\textwidth}{smmbs} 
        \toprule
        \multicolumn{1}{c}{Work(s)} & \multicolumn{1}{c}{Application Space} 
        & \multicolumn{1}{c}{Method(s)} & \multicolumn{1}{c}{Implementation Detail} & \multicolumn{1}{c}{Characterization} \\
        \midrule
        \citet{Chen:2022:AAT} & Asynchronous Graph Algorithms (BFS and PageRank) 
        & Hybrid task and work, load-balancing schedule & Task queue shared by all blocks, parallel scan within blocks. & Hybrid, Approximate (device), Exact (block), Hierarchical \\
        \addlinespace[7pt]
        \citet{Steinberger:2014:WTS} & General Load-Balancing
        & Hierarchical task queues, task binning & Uses queues at both the device and block level. Queue items are binned into separate block-sized, warp-sized, or thread-sized queues. & Dynamic, Approximate, Hierarchical, Cooperative, Centralized (device-level), Distributed (block-, warp- and thread-level) \\
        \addlinespace[7pt]
        \citet{Nisa:2019:LSM} & Sparse Matricized Tensor Times Khatri-Rao Product 
        & Binning (inter-block load imbalance), Pre-partitioning CSF tensor (inter-warp load imbalance) & 
        Two strategies:
        1.~Preprocessing step where each fiber (equivalent to row/column), is split such that each sub-fiber now contains a predefined maximum number of nonzeros (occupancy-based partitioning). Thread-warps are assigned to each sub-fiber.
        2.~Configuration step where slices are binned (see~\citet{Ashari:2014:FSM}). Thread-blocks are assigned a certain number of nnzs within a slice. ($\frac{\textit{nnz}}{\text{threads per block}})~=~\text{thread blocks}$ per slice.
        & Static, Exact, Hierarchical \\
        \addlinespace[7pt]
        \citet{Ashari:2014:FSM} & Sparse-Matrix Dense-Vector Multiplication
        & Row binning and dynamic parallelism & Target: CSR format, Row binning: place rows into bins with roughly the same number of nonzeros. For rows smaller than a threshold, bin-specific SpMV kernels are launched, where constant number of threads are assigned to each row. For bins with rows larger than a threshold, a separate grid is launched (through CDP) per row. Grid size is based on a parameter that determines compute per thread. Multiple thread-warps are assigned per row, based on nnzs in that row. & Dynamic, Approximate, Hierarchical, Cooperative, Centralized \\
        \addlinespace[7pt]
        \citet{Liu:2017:UOA} & SpMTTKRP, Sparse-Tensor Tensor Multiplication 
        & Group-Mapped (nonzero splitting) & Introduces new F-COO format (variant of COO) to more easily access nnzs. Launches 2-D grids with 1-D thread blocks. On the sparse tensor, each block (x-axis) is assigned a set of nnzs, where each thread has the same number of nnzs. For the dense matrices each block (y-axis) is assigned a partition of the column. Partition size (nnzs and columns) is found by tuning. & Static, Exact, Hierarchical \\
        \addlinespace[7pt]
        \citet{Yan:2014:YYA} & Sparse-Matrix Dense-Vector Multiplication 
        & Pre-partitioning COO matrix and a segmented scan & Introduces new blocked COO format for better reuse, where sparse matrix partitioned into vertically-aligned tiles. Thread blocks assigned equal number of nonzero tiles. Each thread within a thread-block is assigned equal number of consecutive nonzero tiles. Threads perform sequential segmented scan/sum. Blocks perform parallel scan. & Static, Exact, Hierarchical \\
        \addlinespace[7pt]
        \citet{Winter:2019:ASM} & Sparse General Matrix Multiplication  
        & (Thread-block) nonzero splitting of A matrix, (Thread-level) dynamically assign work items to each thread & Partition the A matrix such that each thread block gest the same number of nonzeros. Within a thread block: Every thread is assigned the same number of work items. ESC algorithm is run within each thread block, where each thread works on computing a certain number of multiplies. Threads communicate through scratchpad memory.  Device-wide prefix scan is run to merge all output chunks. & Dynamic, Exact, Hierarchical, Cooperative, Distributed \\
        \addlinespace[7pt]
        \citet{Gremse:2015:GSM} \citet{Liu:2018:RIO} & Sparse General Matrix Multiplication  
        & Nonzero splitting within a row & (RMerge) A sub-warp with N threads is assigned N$\times$2 rows of the left-hand operand, where each row has N nonzeros. If the operand has more than A nonzeros in a row, the matrix is iteratively split into multiple matrices where each sub-matrix has up to N nnzs per row. & Dynamic, Approximate, Flat \\
        \addlinespace[7pt]
        \citet{Parger:2020:SAG} & Sparse General Matrix Multiplication  
        & (Thread-block) binning based on lightweight row-analysis of A and B, (Thread-level) hybrid load balancing, work-oriented & Thread-block (global load balancing) analyzes row patterns of inputs to select one of 5 kernels. Assigns blocks based on binning of rows. Thread-level (local load balancing) Threads are split into groups where each group works on a nonzero of A (several rows of B). Group sizes chosen heuristically based on potential length of B row. & Hybrid, Approximate, Hierarchical \\
        \addlinespace[7pt]
        \citet{Niu:2022:TAT} & Sparse General Matrix Multiplication  
        & Group-Mapped & Divide the matrices into evenly-shaped tiles, and assigns a warp to each sparse output tile. & Static, Approximate, Flat \\
        \bottomrule
    \end{tabularx}
    \caption[Annotated bibliography of load-balancing techniques]{(Continued) Annotated Bibliography---Summary and characterization of research work on load balancing parallel-irregular problems on the GPU\@.}
    \label{tab:annotated-bib_2}
\end{table*}

%% file: ch_survey/chapters/optimizations.tex
\section{Optimizations Orthogonal to Load Balancing}
\label{sec:optimizations}

High-performance load-balancing algorithms often need to take the GPU's unique architectural features into account. Critical characteristics include choice of kernel configuration, the use of synchronization between parallel workers, and management of on-chip memory. In this section, we discuss GPU optimizations that are orthogonal to load balancing, and are often needed to achieve high performance.


\subsection{Kernel Strategy for Load-Balancing Operations}

Although selecting an optimal kernel configuration strategy is independent from a high-level load-balancing algorithm, it nevertheless can have a large impact on its resulting performance. One significant configuration parameter that influences kernel performance is \emph{Occupancy}, defined as the number of blocks concurrently active on each Streaming Multiprocessor (SM) Occupancy depends on several factors including the amount of shared memory used by each thread block, the number of threads per block, the number of registers used per thread, and the hardware capabilities of the targeted GPU\@. In general, the more of a SM's resources an individual block uses, the fewer blocks may run concurrently on a given SM due to resource constraints. If a programmer does not tune the kernel parameters such as block size, shared memory usage, or register consumption correctly, they may find that they are underutilizing their hardware. Ultimately, even with the kernel tuned to provide maximum occupancy, an optimal grid and block size may be difficult to determine without experimentation. Several works perform a simple sweep over block and thread sizes, without any explanation of why one configuration is better than another~\cite{Green:2012:GMP,Aila:2009:UTE}, while other works account for nonobvious architectural factors such as the warp scheduler overheads, register constraints, and synchronization overheads~\cite{Cederman:2008:ODL,Cederman:2011:DLB}, or algorithmic benefits from oversubscribing the number of blocks~\cite{Merrill:2012:SGG,Gale:2020:SGK}. 

After the programmer tunes their kernel's resource usage to achieve maximum occupancy and performance, there are two primary ways to configure the kernel. The first, most common approach is to launch a large number of blocks, often on the order of tens of thousands. The GPU's block scheduler assigns a block to an available SM, where it executes its assigned portion of the parallel work and then terminates, allowing the block scheduler to assign a new block to the SM\@. Programmers often pick the \emph{Many-Blocks} kernel configuration because it is both easy to use and scalable. With hundreds, thousands, or even tens of thousands of blocks available to the SM scheduler, a given kernel is agnostic to the number of SMs present on a GPU as long as the number of blocks reaches the minimum threshold to fully occupy the GPU\@. The same kernel can scale down to mobile GPUs with only a handful of SMs, up to server GPUs with over a hundred SMs (such as the NVIDIA A100), and can continue to scale as future GPU releases include ever more SMs. In this kernel configuration, the programmer is responsible for creating an appropriate division of work among the blocks so that all blocks have approximately equal amounts of work. Additionally, the programmer should pick a number of blocks equal to a multiple of the number of SMs on the GPU if the architecture is known in advance\@. This ensures that all SMs are fully utilized throughout the lifetime of the kernel, and avoids problems of \emph{Wave Quantization}, where the last wave of blocks scheduled onto the GPU occupies only a fraction of the GPU~\cite{NVIDIA-Corporation:2021:DLP,NVIDIA-Corporation:2019:TFO}. By extension, the programmer should also ensure that the size of a tile of work mapped to a block is is a multiple of the number of threads, to avoid \emph{Tile Quantization}.

The other common approach is to use a \emph{Persistent Kernel}, in which the programmer launches just enough blocks to fully occupy all SMs of the GPU and which stay resident on the SMs throughout the lifetime of the kernel, iterating over work items in a loop as needed~\cite{Aila:2009:UTE,Tzeng:2010:TMF,Chen:2010:DLB,Steinberger:2014:WTS}. Using a persistent kernel grants programmers significant benefits such as reduced kernel launch overheads, improved data reuse, and better load balancing. However, this method requires careful management of GPU resources such as shared memory, register usage, and thread count to ensure that the kernel runs at maximum occupancy. To reduce kernel launch overheads and device-wide synchronization overheads at kernel termination, a programmer may use an \emph{uberkernel}, where within the kernel, blocks run in a continuous loop and can select one of multiple paths within the kernel depending on the desired computation. For example, in a multi-stage processing pipeline such as the Reyes renderer proposed by Tzeng et al.~\cite{Tzeng:2010:TMF}, processors may alternate between different section of the processing pipeline without device-wide barriers between stages. A persistent kernel also improves data reuse in two primary ways. First, by never exiting the kernel, a program set up as a pipelined uberkernel may keep processed work in shared memory between stages of the pipeline, rather than performing expensive reads and writes between the GPU and off-chip memory between kernel launches. Secondly, initialization or clean-up tasks required when a block first launches or exits only need to run once, rather than each time a short-lived block acquires residence on a SM\@. The persistent kernel method also gives load-balancing benefits in some situations. If a single warp within a block has significantly more work than the other warps, it will prevent new blocks from being scheduled to that SM\@. If instead the programmer uses a persistent kernel approach, the other warps within the block can move on to consume other work items as long as the algorithm does not require a block-wide synchronization~\cite{Aila:2009:UTE}.

\subsection{Synchronization Avoidance}
Many load-balancing kernels use synchronization to ensure that parallel workers create, consume, and process work correctly. For instance, consider a queue used to sequence work, with many workers that want to push to or pop from the queue~\cite{Tzeng:2010:TMF,Cederman:2008:ODL,Cederman:2011:DLB,Chen:2010:DLB,Aila:2009:UTE,Zhang:2014:CAO}. This algorithm requires each worker to perform an exclusive modification to a shared queue so that all workers on the device know where to add or remove work. Some queue implementations use mutex locks on the queue, which are expensive to acquire and release. Other implementations use atomic operations to ensure that only one worker can access the queue at a time. Such synchronization approaches become increasingly costly as the number of workers accessing the queue increases. Modern GPUs support hundreds of concurrent blocks, equivalent to tens of thousands of concurrent threads. One practical method of minimizing the impact of synchronization is to interact with global memory queues at warp or block granularity, rather than at thread granularity. This reduces the number of atomics necessary for work creation and consumption, since a single thread in a block can fetch a chunk of work for the entire block~\cite{Chen:2010:DLB,Steinberger:2014:WTS}. The primary downside of this method is that there may be some load imbalance within a block if the work obtained by the block-wide chunk is not sufficient to supply all threads with even amounts of work. Aila and Laine solve this problem by periodically replacing finished work items with new work items on some threads while long-running work continues on the busy threads~\cite{Aila:2009:UTE}. In other cases, it is possible to perform a second round of load balancing \emph{within} the block~\cite{Chen:2022:AAT}. Alternatively, Cederman and Tsigas reduce the synchronization overhead of their queue by using lazy updates~\cite{Cederman:2008:ODL}. This queue implementation only updates the queue's head and tail pointers using expensive CAS operations every $n$ accesses, and instead has each requesting block check several consecutive positions in the queue offset from the head or tail to find the true location between CAS updates. One important point to note is that as the performance of atomic operations on modern GPUs continues to improve, globally-synchronized, queue-based methods may be promising areas for continued research.

In some algorithms atomic operations are not needed, and the kernel authors tune their kernel implementations accordingly. For example, rather than using a work queue, Merge Path~\cite{Green:2012:GMP,Merrill:2016:MPS} precomputes the maximum amount of output work items each block may produce and partitions the output storage accordingly. This often requires a larger necessary output size (such as for the output frontier in BFS) and additional computation at the beginning of the load-balancing algorithm to determine such split points, but this method eliminates the synchronization required to add work items to the output work queue. Additionally, Merrill et al.'s BFS implementation~\cite{Merrill:2012:SGG} uses unsynchronized bitmask reads and writes when updating a vertex's ``visited flag''. This causes the kernel to add redundant work to the output frontier if the result of one thread's write is not visible to another thread's read, but the authors claim that the speedup from eliminating atomics is greater than the additional work their kernel must do to remove duplicates. Although this example is algorithm-specific, a programmer can often take one of multiple approaches to implementing a kernel, and it is worth considering whether an atomic-free implementation may show overall performance improvements despite creating redundant work. Even kernel tuning has overlaps with synchronization avoidance. For example, many kernels rely on block-wide thread synchronization methods to ensure that all threads in a block are working on the same work items. However, if a small number of threads within the block are not finished with their assigned work, all other threads must sit idle. Tzeng et al.~\cite{Tzeng:2010:TMF} solve this problem by using blocks sized to be exactly 1 warp (32 threads), which eliminates the need for explicit synchronizations since warps automatically advance in lockstep.

\subsection{Shared Memory}
\label{section:subsection:shared_mem}
GPU kernels need to maximize memory bandwidth to achieve best performance. As a result, the kernel programmer must take advantage of spatial and temporal data locality, and must be aware of the characteristics of each level of the memory hierarchy. Off-chip DRAM has the slowest bandwidth of any GPU memory pool and experiences significant slowdowns from random accesses. To achieve maximum DRAM bandwidth, a programmer must read or write a large sequential chunk of data. On-chip shared memory, however, does not suffer from these issues to the same degree 
On modern NVIDIA GPUs, a programmer may configure a per-block pool of shared memory as either an explicitly-managed scratchpad accessible by all threads, or as a cache where the GPU manages the contents of the storage. For example, consider the merge path load-balancing methods, which require binary searches along the diagonals of the input arrays to determine work decomposition~\cite{Davidson:2014:WPG,Green:2012:GMP,Merrill:2016:MPS,Dalton:2015:OSM}. These would be prohibitively slow in global memory, but by using each block to read sequential chunks of the array into explicitly-managed shared memory, the threads can then perform the random memory accesses required by the binary searches in shared memory with significantly reduced overhead and latency. Meanwhile, the accesses to global memory use a smaller number of more efficient coalesced reads.

Ideally, the programmer should use vectorized reads to achieve maximum throughput. However, sparse data formats make vectorized reads difficult, since a data structure such as a sparse CSR matrix will likely not align with the vector width of the memory system due to irregular row lengths. To solve this challenge, Gale et al.\ introduce \emph{Reverse-Offset Memory Alignment}, which allows the kernel to use vector loads without padding the rows of the sparse matrix with zeros~\cite{Gale:2020:SGK}. Each block decrements its row offset to align it with with the vector width, increases the number of nonzeros it needs to process by the amount it decremented previously, and maintains a bitmask to avoid duplicate work processing for the first values in the row.

Although the queue-based load-balancing methods achieve high-performance load balancing through the use of both work queues and synchronization variables stored in DRAM, blocks must suffer long-latency memory accesses when attempting to synchronize, add work, or consume work. An alternative to a single DRAM queue is to use distributed per-block queues as presented in CUIREE~\cite{Zhang:2014:CAO}. These queues could be stored in either global memory or shared memory (depending on their size), which presents an interesting trade-off where some kernels may benefit from device-wide load balancing, and others may benefit from faster access to the queue in the block's shared memory at the expense of using load balancing only within a block. If the queue is too large to fit in shared memory, distributed queue systems can still take advantage of shared memory by keeping per-block synchronization and indexing variables in the faster shared memory. Whippletree~\cite{Steinberger:2014:WTS} extends this concept to use small, fast, per-block queues stored in each block's shared memory. Under normal runtime conditions, blocks consume work from and add new work to their private queues. If the block runs out of space in the shared memory queue, or if there is no remaining work in the shared memory queue, Whippletree provides a global memory queue used for overflow data.

Explicitly managing shared memory to store frequently-used data is critical for many kernels, but others can instead benefit from configuring on-chip storage as a GPU-managed cache. For example, Gale et al.~\cite{Gale:2020:SGK} evaluate a trade-off between using on-chip memory as explicitly-managed shared memory to store a transpose of the input matrix, versus leaving the memory as a cache while re-computing the matrix transpose in registers as needed. The authors determine that increased L1 cache capacity is more important for their Sampled Dense-Dense Matrix Multiplication (SDDMM) kernel than reducing compute complexity. We hypothesize that this is because other parts of the kernel, separate from the matrix transpose, benefit from the cache's ability to expose spatial and temporal reuse more than the kernel benefits from an explicitly managed transpose scratchpad.
When developing GPU kernels, the programmer should consider whether explicitly managing a portion of the data is beneficial, or if it hurts performance by eliminating caching opportunities in other areas of the program's data accesses.

%% file: ch_survey/chapters/conclusion.tex
\section{Conclusion: A Look Ahead}

The focus of this work is to design a method for characterization of fine-grained GPU load-balancing techniques that are key to high-performance for irregular sparse kernels. Our survey uniquely builds the taxonomy of fine-grained load-balancing, a deeper dive on key techniques practiced in literature today, and an understanding of implementation details and building-block algorithms required to implement these load-balancing algorithms. Our goal is to fasicilitate development of future work in this space that further accelerates irregular problems on the GPUs, and encourages support for load-balancing primitives as a first-class software and hardware citizen.

The aforementioned survey gives us not only a broad, but also a very deep understanding of the load-balancing landscape. In the chaper that follows we bring the knowledge of common load-balancing patterns to bear and build a powerful load-balancing abstraction for GPUs. We strongly believe that this is an important first step towards a future that facilitates ease of programming and portability of code for sparse-irregular problems.


%% file: ch_loadbalance/loadbalance.tex
\input{ch_loadbalance/chapters/introduction}
\input{ch_loadbalance/chapters/design}
\input{ch_loadbalance/chapters/abstraction}
\input{ch_loadbalance/chapters/framework}

\input{ch_loadbalance/chapters/implementation}
\input{ch_loadbalance/chapters/evaluation}
\input{ch_loadbalance/chapters/background}
\input{ch_loadbalance/chapters/conclusion}

%% file: ch_loadbalance/chapters/introduction.tex
The survey in Chapter~\ref{ch:survey} shows that the predominant approach to solving the load imbalance problem is through application-specific load-balancing techniques that aim to evenly distribute the work such that each thread gets the same number of work items to achieve maximum performance (for instance, Merrill and Garland's load-balanced SpMV implementation~\cite{Merrill:2016:MPS}). These load-balancing techniques are often tightly coupled with the application itself.
The load-balancing components within these implementations are both complex and are often collectively the most significant contributor to the performance of an application. The work presented in this chapter generalizes today's application-specific load-balancing algorithms into a clean, modular, powerful abstraction that can be applied to many complex irregular workloads.

In the process of building our abstraction, we identified common load-balancing approaches deployed within sparse, irregular applications on GPUs: application-specific frameworks such as GraphIt~\cite{Brahmakshatriya:2021:CGA}, Gunrock~\cite{Wang:2017:GGG}, and GraphBLAST~\cite{Yang:2021:GAH}; techniques from low-level CUDA libraries such as ModernGPU~\cite{Baxter:2013:MPA} and CUB~\cite{CUB:2016}; and other hand-coded implementations of load-balancing algorithms within applications such as SpMV/SpMM~\cite{Merrill:2016:MPS,Davidson:2014:WPG,Gale:2020:SGK}, triangle counting~\cite{Fox:2019:ISI,Green:2018:LRB} and breadth-first search~\cite{Merrill:2012:SGG,Busato:2015:BAE}.
We show that with a simple, intuitive, powerful abstraction, these load-balancing schedules can be extended to support irregular workloads that are more general than the specific problem for which they were designed. We demonstrate this by using sparse-linear-algebra-based load balancing for data-centric graph traversal kernels.

Writing high-performance load-balancing code is complex, in large part because this code must perform many roles. Among other tasks, it must ingest data from a specific data structure, perform user-defined computation on that data, and schedule that computation in a load-balanced way. The key insight in our abstraction is to separate the concerns between workload mapping (the load-balance task) and work execution (the user-defined computation), where we \emph{map} sparse formats (such as Compressed Sparse Row (CSR)) to simple abstraction components called work \textbf{atoms}, \textbf{tiles}, and \textbf{sets}. These fundamental components are expressed as composable \cpp{} ranges and range-based for loops, and are used to build load-balancing schedules. Programmers can then use these APIs to build load-balanced, high-performance applications and primitives. Expressed in this way, we can reconstruct existing application-dependent load-balancing techniques that address irregularity to be more \emph{general}, \emph{portable}, and \emph{programmable}.
The contributions of our work are as follows:

\begin{enumerate}
    \item We present a novel abstraction for irregular-parallel workloads on GPUs. Our abstraction at a high level allows programmers to develop sparse, irregular-parallel algorithms with minimal code while delivering high performance.
    \item We design and implement a set of intuitive APIs, available in our open-source GPU load-balancing framework, built on the proposed abstraction using CUDA-\cpp{} ranges and range-based for loops.
    \item We show the ease of implementing new load-balancing schedules by implementing a novel cooperative groups-based load-balancing schedule (Section~\ref{sec:load-balancing-schedules}), which is a generalization of previous thread-, warp-, and block-level load-balancing schedules~\cite{Yang:2018:DPF}.
    \item We provide best-in-class SpMV performance as a benchmark with a geomean of speedup of 2.7$\times$ for the SuiteSparse Matrix Collection~\cite{Davis:2011:TUO} over cuSparse's state-of-the-art implementation using simple heuristics and 3~GPU load-balancing schedules.
\end{enumerate}

%% file: ch_loadbalance/chapters/design.tex
\section{Design Goals}
\label{sec:design-goals}
The design goals of our load balancing abstraction are as follows:

\paragraph{Achieve high performance.} First and foremost, the goal of our work is to achieve the high performance of existing load balancing algorithms for irregular applications. Our abstraction cannot come at the cost of significant overhead or performance degradation. We measure our success in achieving high performance by comparing the performance of our abstraction against the performance of existing hardwired implementations.

\paragraph{A composable and programmable interface.} Importantly, we do not want to restrict the user to a library interface that takes control of the larger system. Programmers strongly prefer to adopt new software components that fit into their control structures rather than require them to adopt a new control structure. We want to allow the users to (1) maintain control of GPU kernel boundaries (kernel launches), (2) be able to add new load-balancing algorithms, and (3) compose new load-balanced primitives from existing load-balancing APIs. We measure the programmability of our work by comparing the Lines of Code (LOC) of our abstraction against existing implementations and show composability by implementing a new load-balancing algorithm in terms of our existing APIs.

\paragraph{Extensible to new applications.} We aim to decouple and extend application-specific load-balancing techniques to new irregular-parallel domains. Our abstraction seeks to promote the reuse of existing load-balancing techniques for new applications. We use SpMV as a benchmark application implemented using three different load-balancing techniques, some of which were previously used to implement parallel graph analytics kernels~\cite{Davidson:2014:WPG,Brahmakshatriya:2021:CGA,Wang:2017:GGG,Busato:2015:BAE}.

\paragraph{Facilitate the exploration of optimizations.} A key goal of our abstraction is to facilitate the exploration of optimizations for a given application by switching the underlying load-balancing algorithms used to balance the work. We want to encourage our users to experiment with heuristics and new load-balancing techniques to discover what works best for their application needs. We measure the success of this goal by optimizing SpMV's performance response for a large corpus of sparse matrices across several different load-balancing techniques.

\subsection*{Non-Goals}
In addition to the above design goals, we also define our non-goals:

\paragraph{Targeting other parallel architectures.} Although we believe the lessons learned should apply to other parallel architectures, we explicitly target NVIDIA's CUDA architecture and programming model~\cite{NVIDIA:2016:CUDA}. Many components of our abstraction leverage CUDA's compute hierarchy of threads, warps and blocks mapped onto the physical streaming multiprocessors, the oversubscription model of assigning more work than the number of processors to fully saturate the underlying hardware, and CUDA's Cooperative Groups programming model~\cite{Harris:2017:CG}, described in Section~\ref{sec:load-balancing-schedules}, to achieve high performance.


\paragraph{Multi-GPU support.} This work focuses on load-imbalance issues for a single GPU and does not consider multi-GPU single-node or multi-node systems, although these are interesting directions for future work. 

%% file: ch_loadbalance/chapters/abstraction.tex
\section{Load Balancing Abstraction}
\label{sec:abstraction}

\begin{figure}
    \centering
    \includegraphics[width=\textwidth]{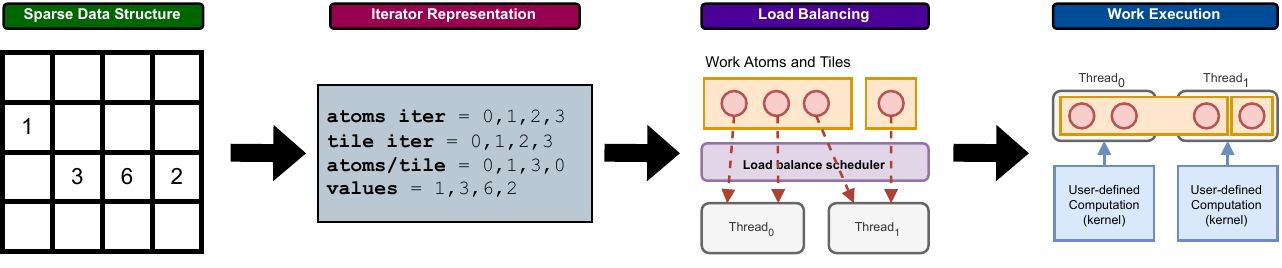}
    \caption[Load balancing abstraction as a simple pipeline]{Load balancing as a simple pipeline of the three key concepts of our abstraction: (1)~sparse data structures represented as iterators, (2)~load-balancing algorithm that partitions the work onto threads, and (3)~user-defined computation consuming the balanced work and executing on each thread.
    \label{fig:load-balance-abstraction}}
\end{figure}

The key insight behind our GPU load balancing abstraction is the \emph{separation of concerns} between the mapping of the work items to processing units and work execution.
We divide our abstraction into three key concepts (illustrated in Figure~\ref{fig:load-balance-abstraction}), each of which describes a different aspect of an implementation: (1)~defining the work;
(2)~defining the workload balance across GPU threads, warps or blocks; and (3)~defining the work execution and computation per thread on the balanced work.
This separation allows us to distinctly divide the work between an application developer and a load-balanced-library developer and facilitates the exploration of optimizations by mixing different load-balancing techniques and sparse-irregular algorithms. Sidebar~\ref{sidebar:cub-sidebar} presents a practical example of the motivation for our load balancing abstraction.


\subsection{Input from Sparse Data Structures}
\label{sec:work-domain}

We begin with our input data expressed in some form of sparse data structure. Examples of such data structures include, but are not limited to, Compressed Sparse Row (CSR) and Coordinate (COO) formats. The goal of the first stage of our abstraction is to map the input data format to a common data framework and vocabulary that is the input to the next stage. This vocabulary has three simple components that together express the input data:

\begin{enumerate}
\item A \textbf{work atom}, a single unit of work that is to be scheduled onto the processors (for example, a non-zero element of a sparse matrix). We assume that all work atoms have an equal cost during execution. 
\item A \textbf{work tile}, a logical entity represented as a set of work atoms (for example, a row of a sparse matrix). Work tiles may have different costs during execution. As we highlighted in the introduction, work is most \emph{logically} parallelized over work tiles but is often most \emph{efficiently} parallelized over work atoms, and mapping between work tiles and work atoms may be expensive and complex.
\item A \textbf{tile set}, a set of work tiles that together comprise the entire working problem (for example, a sparse matrix). In our abstraction, the tiles within a tile set must be independent (and thus can run in parallel across multiple processors).
\end{enumerate}

This mapping between sparse formats and atoms/tiles/tile sets is defined by the user. Though we have not implemented all of them, we believe our mapping abstraction here is flexible enough to express a wide variety of existing sparse data formats in the literature~\cite{Filippone:2017:SMM} in such a way that they are suitable for load balancing in our abstraction's next stage. As well, we have already included several common sparse formats (CSR, CSC, COO) in our load-balancing library implementation so that users can simply select and use them without having to implement them. Given a mapping to atoms/tiles/tile sets, we can next implement a load-balancing algorithm that can parallelize over work atoms or tiles transparently from the computation's perspective.


\begin{sidebar_box}[label=sidebar:cub-sidebar]{A practical example of existing, predominant approach to load balancing sparse-irregular workloads.}
    Consider an SpMV implementation on the GPU provided in the open-source CUDA CUB library~\cite{CUB:2016}. CUB implements and maintains the SpMV algorithm presented in the paper by Merrill and Garland~\cite{Merrill:2016:MPS}. Merge-based SpMV, explained in detail in Section~\ref{sec:merge-path-lb}, is a CSR-based, perfectly load-balanced SpMV, where each thread gets an even share of work, and the amount of work is defined by the total number of matrix rows and the total number of non-zeros, summed. In the reference, this highly efficient, state-of-the-art implementation took 1,100~lines of code (LoC) (or 503~LoC of kernel code) across 3~files (not including a 4th file required for a segmented fixup step of an additional 234 LoC). In contrast, the actual computation of SpMV within this reference implementation is expressed within a \emph{single} for-loop and 4--5 LoC! This disparity between the LoC required to map the work items to processing units in a load-balanced way and the LoC required to express the desired computation is the key motivation behind our work. Additionally, the CUB implementation is specifically dedicated to the SpMV algorithm and would require a significant rewrite to apply it to other algorithms, even within the same computing domain. One such example of this exact rewrite is by Yang et al., where the authors extend merge-path load balancing from SpMV to Sparse-Matrix Dense-Matrix Multiplication (SpMM) algorithm~\cite{Yang:2018:DPF}. The load balancing algorithm in both works is the same but applied to different computations, which motivates the need for reuse.
\end{sidebar_box}

\subsection{Defining Load Balancing}
\label{sec:load-balancing-domain}

By expressing workloads through an abstraction that captures work at differing levels of granularity (i.e., tile set, atoms, and tiles), we can more easily distribute computation evenly across the GPU's available resources. Given a user-defined input tile set and associated sequences of atoms and tiles, along with a user-selected partitioning algorithm, our load balancing stage outputs subsequences of atoms and tiles assigned to processor ids (i.e., where atoms or tiles will be processed). 

The resulting assignment of subsequences to processor ids is critical to effectively balancing workloads across processing elements and is generally problem- and dataset-specific. The user must specify the necessary sequences. Ideally, an \emph{oracle} would take these sequences and select the most optimal subsequences for every processing element. Finding such an oracle is an open problem and as such, we provide the next best thing: the ability for a user to choose and experiment from set of predefined schedules and the ability to implement their own schedules. In general, load-balancing algorithm designers must balance between the cost of scheduling and the benefits from better scheduling. A schedule could be as straightforward as assigning processing elements to tiles with arbitrary numbers of atoms (e.g., rows with an arbitrary number of non-zeros in a sparse matrix) to something more complicated/expensive that takes on a more holistic approach to work (e.g., considering work across multiple rows with a varying number of non-zeros in a sparse matrix).

\subsection{Defining Work Execution}
\label{sec:work-execution}

The final component of our load-balancing abstraction expresses the irregular-parallel computation itself. The previous stage inputs load-imbalanced work and load-balances it; this stage then consumes that load-balanced work by performing computation on it. The scope of what computation can be expressed is extensive, and is only limited by how the load-balanced work, represented as sequences,
can be consumed within a CUDA kernel. Since the framework does not assume control of the kernel, anything you can write in a CUDA kernel will also work in our framework. For instance, programmers can express a mathematical operation performed on each atom or each tile of the work, or build cooperative algorithms that not only consume the work assigned to each thread but also combine the results with neighboring threads to implement more complex algorithms such as parallel reduce or scan. Practical examples that we have implemented in our framework (see Section~\ref{sec:work-execution-representation} and~\ref{sec:application-space-and-optimization-choices}) using this abstraction include, but are not limited to, sparse-linear algebra kernels, such as Sparse-Matrix and Sparse-Tensor contractions, and data-centric parallel graph algorithms, such as Single-Source Shortest Path (SSSP) and Breadth-First Search (BFS) built on a neighborhood traversal kernel.

We expect typical users of our library will \emph{only} write their own code for \emph{this} stage of the abstraction and use standard data structures and load-balancing schedules that are already part of our library. However, those users can also implement custom data formats and load-balancing schedules.


%% file: ch_loadbalance/chapters/framework.tex
\section{High-Level Framework Implementation}
\label{sec:framework}

Our GPU load-balancing framework implements the abstraction described in Section~\ref{sec:abstraction} using \cpp{}17 and CUDA\@. In our system, programmers use CUDA/\cpp{} to develop irregular-parallel algorithms and implement new load-balancing schedules. Per our design goals of composable APIs, extensibility and reuse,
this and the following section introduce the implementation details of our API, and how it is used to develop new applications that promote the reuse of high-performance load-balancing techniques available within the framework. We also explore a new load-balancing method (Section~\ref{sec:load-balancing-schedules}) built on CUDA's Cooperative Groups model. Furthermore, we identify how our work can be used to facilitate the exploration of optimizations for a given application such as SpMV\@.

\subsection{Implementing Sparse Data Structures}
\label{sec:sparse-data-structure-representation}

Our framework translates sparse data structures (e.g., COO, CSR, CSC) into work atoms, work tiles, and tile sets (Section~\ref{sec:work-domain}) using simple \cpp{} iterators. \cpp{} iterators are objects that point to some element in a range of elements and enable iteration through the elements of that range using a set of operators. For example, a \lstinline{counting_iterator} is an iterator that represents a pointer into a range of sequential values~\cite{CPPREFERNCE}. Our framework requires the user to define three important iterators using \cpp: (1)~an iterator over all work atoms; (2)~an iterator over the work tiles; and (3)~an iterator over the number of atoms in each work tile. (Our library already supports several common sparse data structures.)
Using these iterators, the load-balancing schedule can then determine and distribute load-balanced work across the underlying hardware. Listing~\ref{lst:csr-tile-set} shows how our abstraction expresses the commonly used CSR format as a tile set within our framework. 

\begin{listing}
  \caption[CSR format expressed within our framework]{Compressed-Sparse Row (CSR) format expressed within our framework using C++17.
  The CSR format describes a matrix using three arrays: (1)~column indices of nonzero values; (2)~the extent of rows (row offsets); and (3)~the nonzero values. Since the CSR data structure does not contain arrays that point to indices of atoms and tiles (nonzeros and rows), in the listing above we define atom and tile iterators as simple counting iterators from 0 to the total number of nonzeros ($\textit{nnz}$) and from 0 to the total rows in the matrix ($\textit{rows}$), respectively (Lines~2~and~3). The iterator over the atoms-per-work-tile is expressed using a transform iterator, which computes the expression within a provided function for each tile id. For CSR, this is simply the row offset of the current tile subtracted from the offset of the next tile (Lines~5--11).
  }

  \label{lst:csr-tile-set}
  \begin{minted}[
      obeytabs=true,
      tabsize=4,
      linenos=true,
      numbersep=-1pt]{c++}
  // Simple iterators for atoms and tiles.
  counting_iterator<int> atoms_iter(0, nnz);
  counting_iterator<int> tile_iter(0, rows);
  // Iterator over the atoms within tile i.
  auto atoms_per_tile = make_transform_iterator(
    tile_iter,
    [tile_iter, row_offsets]
    __host__ __device__(const int& i) {
      return (row_offsets[tile_iter[i + 1]] -
              row_offsets[tile_iter[i]]);
  });
  \end{minted}
\end{listing}

\subsection{Implementing Load-Balancing Schedules}
\label{sec:load-balancing-representation}

Perhaps the most straightforward schedule is scheduling each work tile onto one GPU thread. This approach is common in the literature and practice~\cite{Merrill:2012:SGG,Davidson:2014:WPG,Steinberger:2017:GHL,Baxter:2013:MPA,Yang:2018:DPF}; although this strategy is ineffective in the presence of significant load imbalance across tiles, we use it here as an example to illustrate how load balancing is defined within our framework.

The inputs are the three iterators from the last stage plus an atom and tile count. The load-balance algorithm developer, then, implements \lstinline{tiles()} and \lstinline{atoms()} procedure calls, which return the \cpp{} range of tiles and atoms to be processed by the current thread, effectively creating a map between assigned processor ids and segments of the workload. Listing~\ref{lst:thread-mapped} shows a complete example of the thread-mapped schedule. Although a simple algorithm, it can deliver high performance for well-balanced workloads with coarse-grained parallelism (a small number of atoms per tile), such as multiplying a sparse vector by a dense vector. Furthermore, our abstraction is not limited to only simple scheduling algorithms, as Section~\ref{sec:load-balancing-schedules} provides examples of more complex load balancing algorithms.


\begin{listing}
  \caption[Thread-mapped load-balancing algorithm]{A thread-mapped load-balancing algorithm expressed as C++ ranges, incorporating the atoms and tiles defined as iterators from Listing~\ref{lst:csr-tile-set}.
  Each tile is mapped to a thread, where the thread id corresponds to the index of the tile in the tile set. All atoms within a tile are sequentially processed by the thread. After a tile is processed, a thread is mapped to the next tile, obtained by striding the index by the grid size of the kernel.
  }
  \label{lst:thread-mapped}
  \begin{minted}[
      obeytabs=true,
      tabsize=4,
      linenos=true,
      numbersep=-1pt]{c++}
  class schedule_t {
    // Construct a thread-mapped schedule.
    __host__ __device__ schedule_t(atoms_it_t atoms_it,
      tiles_it_t tiles_it,
      atoms_it_t atoms_per_tile_it,
      size_t num_atoms, size_t num_tiles) :
      m_atoms_it(atoms_it), m_tiles_it(tiles_it),
      m_atoms_per_tile_it(atoms_per_tile_it),
      m_num_atoms(num_atoms), 
      m_num_tiles(num_tiles) {}
    // Returns a range of tiles to process in "this" thread.
    // Stride by grid dimension.
    __host__ __device__ auto tiles() {
      auto begin = m_tiles_it(blockDim.x * blockIdx.x + threadIdx.x);
      auto end = m_tiles_it(m_num_tiles);
      return range(begin, end).step(gridDim.x * blockDim.x);
    }
    // Returns a range of atoms to process in "this" thread.
    __host__ __device__ auto atoms(
        const std::size_t& tile) {
      auto begin = m_atoms_per_tile_it[tile];
      auto end = m_atoms_per_tile_it[tile + 1];
      return range(begin, end).step(1);
    }
  };
  using schedule_t = thread_mapped_schedule_t;
  \end{minted}
\end{listing}

\subsection{Implementing Work Execution}
\label{sec:work-execution-representation}


Our framework is designed to explicitly let the user \emph{own} the kernel launch boundary. Owning a CUDA kernel boundary means that the user is responsible for maintaining and configuring launch parameters and implementing the CUDA kernel used to define the application. Although this design decision comes at a cost of convenience and simplicity, it offers significant flexibility in what users can express through our abstraction. This design decision is motivated by the following reasons. (1)~Users are not required to add a complex dependency to their existing workflow/libraries, therefore making code maintenance simpler and more scalable as they do not have to rely on our framework to incorporate new CUDA constructs and features. (2)~Users are free to express anything and everything CUDA allows within their kernels while consuming our load-balanced \cpp{} ranges. This allows for versatility in what can be expressed, as the users can now specify multiple load-balanced work domains, range-based for loops, and even fusing multiple computations to build more complex algorithms within a single kernel. (3) Higher-level APIs can be used to build simpler higher-level abstractions that \emph{do} own the kernel boundary and provide simpler APIs at the cost of flexibility.


As an input to this stage, users consume the load-balanced \cpp{} ranges to implement their computation. This can be done in multiple ways, but one of the most common patterns is a nested range-based for loop that loops over all the assigned tiles and atoms ranges. Listing~\ref{lst:spmv} shows a simple example of a CUDA kernel that implements the SpMV algorithm using CSR format and thread-mapped load-balancing algorithm constructed in Listings~\ref{lst:csr-tile-set} and~\ref{lst:thread-mapped}. In this example, the outer \texttt{for} loop within each thread iterates over the assigned rows of the sparse matrix (tiles), and the inner loop sequentially processes the assigned nonzeros (atoms) within each row. In Section~\ref{sec:application-space-and-optimization-choices} we implement and discuss more complex kernels and computations.

\begin{listing}
  \caption[SpMV implemented within our load-balancing abstraction]{Sparse-Vector Matrix Multiplication (SpMV) implemented within our load-balancing abstraction using range-based nested \texttt{for} loops. The sparse matrix is represented using a CSR-based format, where $x$ is the dense input vector and $y$ is the dense output vector ($y = Ax$).  Lines~11--14 use the load-balancing schedule implemented in Listing~\ref{lst:thread-mapped} and the iterators defined in Listing~\ref{lst:csr-tile-set} to construct the load-balanced work to be processed. Lines~16 and~19 show the \texttt{for} loops within each thread, which iterate over the assigned rows of the sparse matrix and sequentially process the assigned atoms within each row. Line~20 shows the actual computation performed on each work atom (nonzero), and Line~21 writes the result to the dense output vector $y$.}
  \label{lst:spmv}
  \begin{minted}[
      obeytabs=true,
      tabsize=4,
      linenos=true,
      numbersep=-1pt]{c++}
  // Implements load-balanced SpMV kernel.
  __global__ void spmv(const size_t rows, 
    const size_t cols, 
    const size_t nnz,
    const int* offsets, 
    const int* indices, 
    const float* values, 
    const float* x, float* y) {
    // Configure load-balancing.
    // Input: iterators defined for CSR format.
    schedule_t config(
        atoms_iter, tile_iter,
        atoms_per_tile_it,
        nnz, rows);
    // Consume rows using a range-based for loop.
    for (auto row : config.tiles()) {
      type_t sum = 0;
      // Consume atoms using a range-based for loop.
      for (auto nz : config.atoms(row))
        sum += values[nz] * x[indices[nz]];
      y[row] = sum;
    }
  }
  // Launches SpMV kernel.
  constexpr size_t blocks = 256;
  size_t grid = (rows + blocks - 1) / blocks;
  spmv<<<grid, blocks>>>(rows, cols, nnz, offsets, indices, values, x, y);
  \end{minted}
\end{listing}


%% file: ch_loadbalance/chapters/implementation.tex
\section{Implementation Details}
\label{sec:implementation-details}


\subsection{Flexible, Composable CUDA-enabled Ranges}
\label{sec:composable-api-ranges}

Composability of load-balanced primitives and applications using our API is
a conscious design choice within our framework supported through the use of CUDA-enabled \cpp{} ranges. Our framework does not \emph{own} the kernel boundary (kernel launch), which forces our APIs to be focused and contained within the kernels. This allows programmers to build and maintain their own kernels while still benefiting from our framework's load-balancing capabilities. This is largely implemented using device-wide \cpp{} functions and classes tagged with CUDA's \lstinline{__device__} keyword.\footnote{A method decorated with the \lstinline{__device__} keyword allows the CUDA compiler to generate a device-callable entry point. This allows the code to be called from within kernels~\cite{NVIDIA:2016:CUDA}.} We implemented and expose several different types of specialized ranges that were particularly useful in implementing load-balanced schedules:

\begin{itemize}
    \item \lstinline{step_range}: A range that iterates from \lstinline{begin} to \lstinline{end} in steps of \lstinline{step}. Useful for defining load balancing schedules that require a custom stepping range or process a constant number of work items per thread (which can be defined using \lstinline{step}).
    \item \lstinline{infinite_range}: A range that iterates from \lstinline{begin} to infinity. Useful for defining load balancing schedules in persistent kernel mode~\cite{Zhang:2022:PKI}, where the kernel persistently runs until all work is consumed or an algorithm has converged.
    \item \lstinline{grid_stride_range}: A specialized case of \lstinline{step_range} that iterates from \lstinline{begin} to \lstinline{end} in steps of \lstinline{step} using the CUDA kernel's grid size. Also supports \lstinline{block_stride} and \lstinline{warp_stride} variants that iterate in steps of the block or warp size, respectively.
\end{itemize}

\subsection{Implementing Non-Trivial Load-Balancing}
\label{sec:load-balancing-schedules}

As we describe in Section~\ref{sec:composable-api-ranges}, we can decouple and express existing load-balancing techniques as a set of \cpp{} ranges.
To illustrate the potential of this abstraction, we begin by decoupling and expressing a state-of-the-art load-balancing algorithm known as merge-path~\cite{Green:2012:GMP} previously used for balancing CSR-based SpMV and SpMM~\cite{Merrill:2016:MPS,Yang:2018:DPF}, and implement three additional load balancing algorithms (warp-, block- and group-mapped), all of which are available in our library for programmers to use. Our new group-mapped algorithm is a tile-per-group-based schedule, where a group is defined as a collection of threads of any arbitrary size (not limited to a warp or block size). Our group-mapped schedule is a generalization of the tile-per-thread, -warp or -block schedules~\cite{Merrill:2012:SGG,Brahmakshatriya:2021:CGA} using CUDA's Cooperative Groups programming model~\cite{Harris:2017:CG}.

\subsubsection{Merge-path load balancing}
\label{sec:merge-path-lb}

In the language of a sparse matrix, merge-path assumes that each non-zero in the matrix and each new row in the matrix are an equivalent amount of work, then evenly divides $\text{nnzs} + \text{rows}$ work across the set of worker threads. Each thread then performs a 2-D binary search within the nonzero indices and row offsets of a CSR matrix to find the starting position of the row and nonzero it needs to process. Threads then sequentially process the rows and nonzeros from the starting position until they reach the end of their assigned work~\cite{Merrill:2016:MPS}.

We implement this algorithm as a load-balancing schedule in our abstraction by expressing it in two steps:
(1)~\textbf{Setup:} The initialization step of the \cpp{} schedule class computes the number of work units per thread, conducts a binary search as described above, and stores the starting position of each tile and atom in a thread-local variable.
(2)~\textbf{Ranges:} The second step of the algorithm builds the ranges for each thread to process as ``complete'' tiles and ``partial'' tiles~\cite{Merrill:2016:MPS}.
If a thread's atom range lies entirely within one tile, it is ``complete'', and is processed in a simple nested loop. If a thread's range crosses a tile boundary, the thread processes its work in a separate nested loop.

Because we decouple the load-balancing method (Section~\ref{sec:load-balancing-representation}, and above) from the work execution (Section~\ref{sec:work-execution-representation}), we can use this merge-path implementation to implement not only SpMV but also any other algorithm whose work can be divided into tiles and atoms, e.g., a graph neighborhood-traversal algorithm used to implement breadth-first search~\cite{Wang:2017:GGG}. Just as importantly, the merge-path schedule is now no longer limited to a CSR-based sparse format. Supporting other formats only requires building the necessary slightly more complex iterators that are able to count atoms per tile (the computation that the CSR implementation achieves with the row offsets array in Listing~\ref{lst:csr-tile-set}). 

\subsubsection{Warp- and block-level load balancing}
The goal of a warp- or block-level load-balancing schedule is to assign an equal share of tiles to each warp or block, which are then sequentially processed. The work atoms within each tile will be processed in parallel by the available threads within a warp or a block. Each thread strides by the size of the warp or block to process a new work atom until the end of work is reached.

The imbalance across different processing units is left for the hardware scheduler to handle. This scheduler depends on the oversubscription model of CUDA, where the programmer can launch a larger number of warps or blocks than the GPU can physically schedule at any given time. As the processing units finish processing their work, new ones are scheduled from the oversubscribed pool~\cite{Merrill:2012:SGG,Brahmakshatriya:2021:CGA}.

\subsubsection{Group-level load balancing}
Group-level load balancing generalizes warp- and block-level schedules. Instead of requiring that group sizes are the size of a warp or block, as above, this method leverages CUDA's Cooperative Groups (CG) programming model~\cite{Harris:2017:CG} to allow programmer-specified dynamically sized groups of arbitrary size. Within these groups, the CG model permits detailed control of the group's synchronization behaviors as well as simple parallel group-level collectives such as reduce or scan. We leverage this powerful tool to implement a generalized group-level load balancing schedule, effectively giving us the warp- and block-level schedules above for free when the group size equals that of a warp or a block.

Our schedule assigns work tiles to a group, and each group looks at its equal share of tiles and computes the number of atoms for each tile and stores it in a scratchpad memory (CUDA's \emph{shared memory}). The group then performs a parallel prefix-sum, a widely used parallel algorithm that inputs an array and produces a new array where the element at any position is a sum of all previous elements~\cite{Blelloch:1990:PSA}. We use this prefix-sum array for two purposes: (1)~the last element of a prefix-sum array indicates the aggregated number of work atoms that a group has to process, and~(2) the position of each sum in the prefix-sum array corresponds to the work tile to which those atoms belong. The setup phase of the schedule builds the prefix-sum arrays per group in the scratchpad memory, and the ranged-loop of the schedule returns the atom to process in each thread. The corresponding tile, if needed, is obtained by a simple \lstinline{get_tile(atom_id)} operation, which executes a binary search within the prefix-sum array to find the tile corresponding to the atom being processed.

Relying on the CG model for this load-balancing schedule has a unique advantage of configuring the group size (effectively software constructs that directly map onto the hardware) per the shape of the problem and the underlying hardware architecture. For example, targeting GPUs where the warp size is not 32~threads (AMD's GPU architecture supports a warp size of 64~\cite{AMD:2022:HIP}) is now possible with a simple compile-time constant, or configuring the group size to perfectly align with the structure of the problem.

\subsection{Application Space}
\label{sec:application-space-and-optimization-choices}

Our work definition (Section~\ref{sec:work-domain}), composable APIs (Section~\ref{sec:composable-api-ranges}), and multiple sophisticated, high-performance load-balancing schedules (Section~\ref{sec:load-balancing-schedules}) together provide for a versatile and extensible framework with plenty of room for application-specific optimizations. In Listing~\ref{lst:spmv} we already demonstrated how to implement the SpMV algorithm using our framework. A simple and natural extension is to implement Sparse-Matrix Matrix Multiplication (SpMM)\@. Listing~\ref{lst:spmm} shows the minor change necessary, which adds another loop over the columns of the $\textbf{B}$ matrix around the existing code from Listing~\ref{lst:spmv} to implement SpMM\@. This implementation could also be extended to support Gustavson's General Sparse Matrix-Matrix Multiplication (SpGEMM), using two kernels and an allocation stage; the first kernel would compute the size of the output rows used to allocate the memory for the output sparse matrix and the second kernel would perform the multiply-accumulation.

\begin{listing}
    \caption[Load-balanced SpMM computation]{A simple loop wrapped around SpMV introduced in Listing~\ref{lst:spmv} allows us to represent the slightly more complex SpMM load-balanced computation. 
    }
    \label{lst:spmm}
  \begin{minted}[
      obeytabs=true,
      tabsize=4,
      linenos=true,
      escapeinside=||,
      numbersep=-1pt]{c++}
  // ... Inside the CUDA kernel.
  // Loop over all the assigned rows.
  for (auto row : config.tiles()) {
    // Loop over all the columns of Matrix B.
    for (auto col : range(size_t(0), B.cols)
      .stride(size_t(1))) { /// < New Loop
      float sum = 0;
      // Loop over all the assigned nonzeros.
      for (auto nz : config.atoms(row))
          sum += values[nz] * B(nz, col);
      // Output the sum to Matrix-C.
      C(row, col) = sum;
    }
  }
  \end{minted}
\end{listing}

Beyond sparse linear algebra, we can use our framework to address applications in other domains. Listing~\ref{lst:sssp} implements the graph primitive Single-Source Shortest Path (SSSP) using our group-level load-balancing schedule. SSSP's performance on GPUs is largely gated by good load balancing~\cite{Wang:2017:GGG,Brahmakshatriya:2021:CGA}, but if the programmer chooses a load-balancing schedule from our library, the details of load balancing are completely hidden. Moreover, the same schedules that were used in one application domain (e.g., sparse linear algebra) are easily reusable in this different application domain.

\begin{listing}
  \caption[SSSP graph primitive implementation]{The parallel single-source shortest path (SSSP) graph primitive expressed using our load-balanced schedule.}
  \label{lst:sssp}
  \begin{minted}[
      obeytabs=true,
      tabsize=4,
      linenos=true,
      numbersep=-1pt]{c++}
  // ... Inside the CUDA kernel.
  // Loop over all the assigned edges to process.
  for (auto edge : config.atoms()) {
    auto source = config.get_tile(edge);
    // G is the graph data structure
    auto neighbor = G.get_neighbor(source, edge);
    auto weight = G.get_edge_weight(edge);
    float source_dist = dist[source];
    float neighbor_dist = source_dist + weight;
    // Check if the destination node has been
    // claimed as someone's child.
    float recover_distance =
      atomicMin(&(dist[neighbor]), neighbor_dist);
    // Add the neighbor to the frontier.
    if (neighbor_dist < recover_distance)
      out_frontier[neighbor] = true;
  }

  // ... Outside the CUDA kernel.
  // Loop until the frontier is empty.
  \end{minted}
\end{listing}



%% file: ch_loadbalance/chapters/evaluation.tex
\section{Evaluation}
\label{sec:evaluation}

We aim to show that our framework, built on our load balancing abstraction, enables both high performance and better programmability for sparse-irregular problems. Our evaluation below uses our SpMV implementation as a benchmark against state-of-the-art implementations provided within NVIDIA's (open-source) CUB library and production (closed-source) cuSparse library. We considered (and implemented) several additional applications for evaluation, including SSSP, BFS, and SpMM\@. We found they led to similar high-level conclusions. Thus our evaluation here focuses on SpMV\@. Our test corpus consists of approximately the \emph{entire} SuiteSparse Matrix Collection~\cite{Davis:2011:TUO} with a broad scope of sparse matrices from many different high-performance computing domains. We ran all experiments on a Ubuntu 20.04 LTS-based workstation with an NVIDIA Tesla V100 GPU and CUDA 11.7.

\subsection{Performance Overhead}
\label{sec:performance-overhead}

\begin{figure}
    \centering
    \includegraphics[width=\columnwidth]{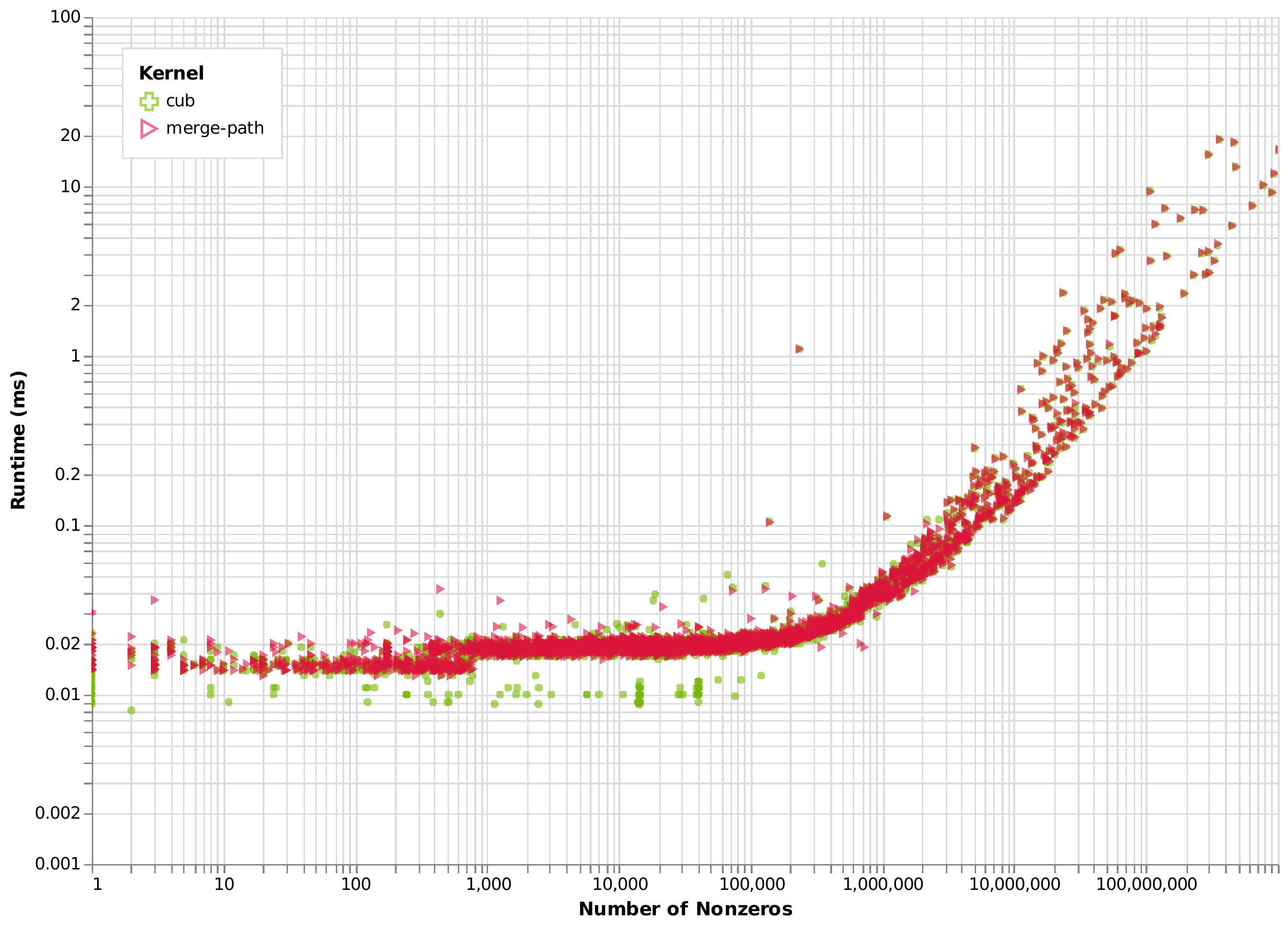}
    \caption[Load balancing abstraction performance overhead analysis vs.\ CUB]{Load balancing abstraction performance overhead analysis using our Merge-path SpMV vs.\ CUB across all SuiteSparse datasets. Our runtimes almost perfectly match CUB's for all datasets. The small number of datasets where CUB is faster is due to a simple heuristic that CUB uses for sparse matrices with the number of columns equal to 1 (i.e., a vector).}
    \label{fig:performance-overhead}
\end{figure}

Our first and foremost goal is to ensure that the elements within our abstraction do not add any additional performance overhead to the existing load balancing techniques and algorithms developed using them. To verify this, we compare the runtime performance of our SpMV implementation using the merge-path schedule to the implementation provided by NVIDIA's CUB library~\cite{CUB:2016} (also used for Merrill and Garland's merge-path SpMV paper~\cite{Merrill:2016:MPS}) on the SuiteSparse collection. As previously mentioned, and in contrast to our design, CUB contains a hardwired implementation of the merge-path scheduling algorithms and does not decouple workload balancing from the actual SpMV computation. CUB's approach is not reusable for any other irregular parallel problem without significant changes to the implementation.

Figure~\ref{fig:performance-overhead} plots the number of nonzeros (i.e., the total work) vs.\ runtime for our work vs.\ CUB's implementation. Our implementation has minimal performance overhead when using our abstraction: a geomean slowdown of 2.5\% vs.\ CUB, with 92\% of datasets achieving at least 90\% of CUB's performance. Figure~\ref{fig:performance-overhead} shows our implementation almost perfectly matches CUB for all datasets, except for some datasets with fewer than 100,000 nonzeros. Upon further investigation, we identify that CUB uses a simple heuristic to launch a thread-mapped SpMV kernel where the number of columns of a given input matrix equals 1 (i.e., a sparse vector). Unlike our more general implementation, CUB's simple (but specialized) thread-mapped SpMV kernel has no load-balancing overhead for a perfectly balanced workload such as SpVV computation.

\subsection{Improved Performance Response}
\label{sec:improved-performance-response}

\begin{figure}
    \centering
    \includegraphics[width=\textwidth]{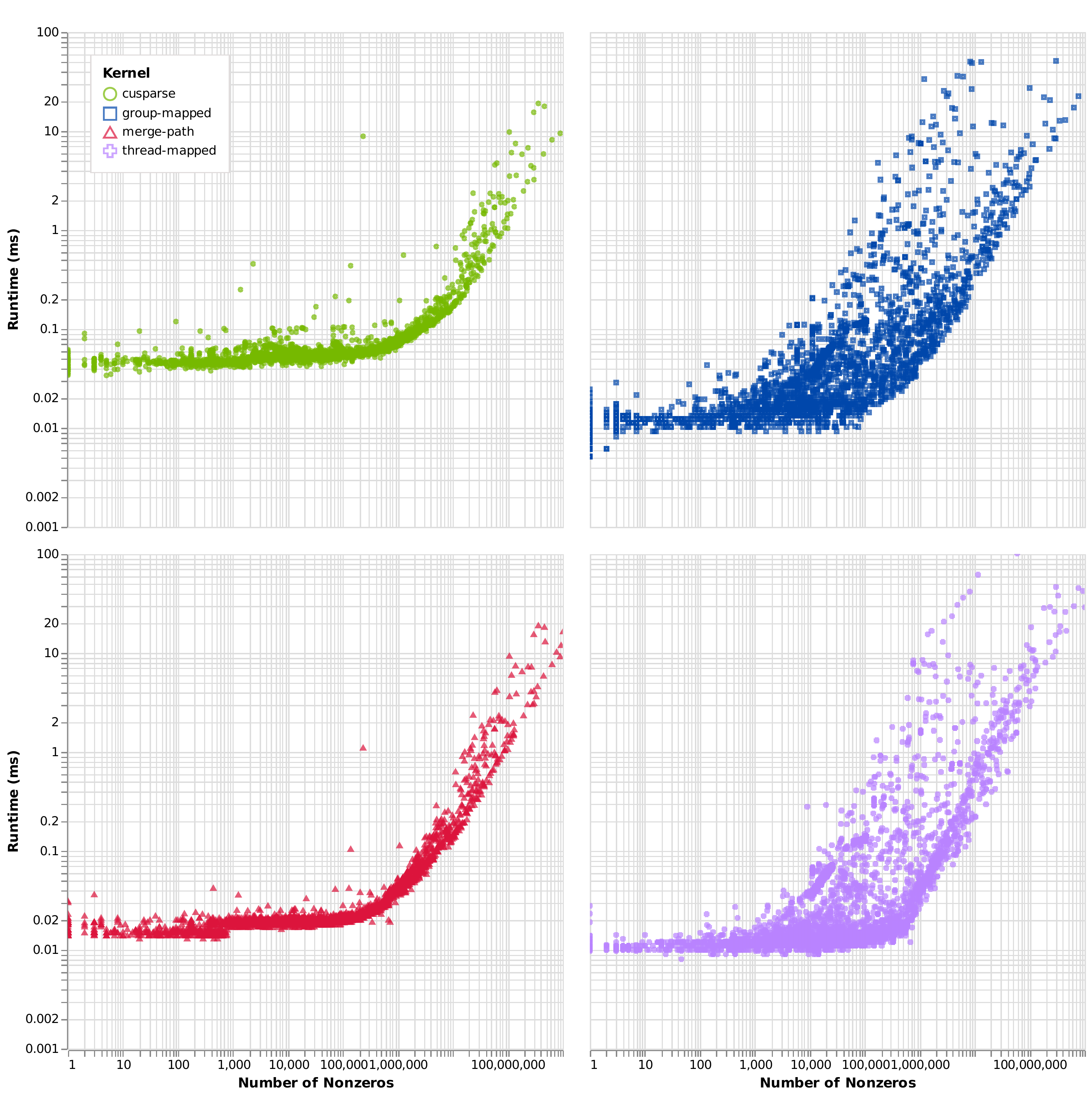}
    \caption[Complete SpMV performance landscape with improvements]{Complete performance landscape of SpMV across all SuiteSparse datasets using 3~load balancing schedules vs.\ NVIDIA's cuSparse library. Our 3~different SpMV implementations are made possible with very little code change.}
    \label{fig:performance-response}
\end{figure}

We also compare our work to NVIDIA's vendor library for sparse computations, cuSparse. Figure~\ref{fig:performance-response} shows the performance response of our SpMV implementation using each of our scheduling algorithms individually vs.\ cuSparse's state-of-the-art implementation. Switching between any of our implementations requires very little code change; in the case of merge-path and thread-mapped, we need only update a single \cpp{} enum (identifier) to select the desired load balancing schedule.

\begin{figure}
    \centering
    \includegraphics[width=\columnwidth]{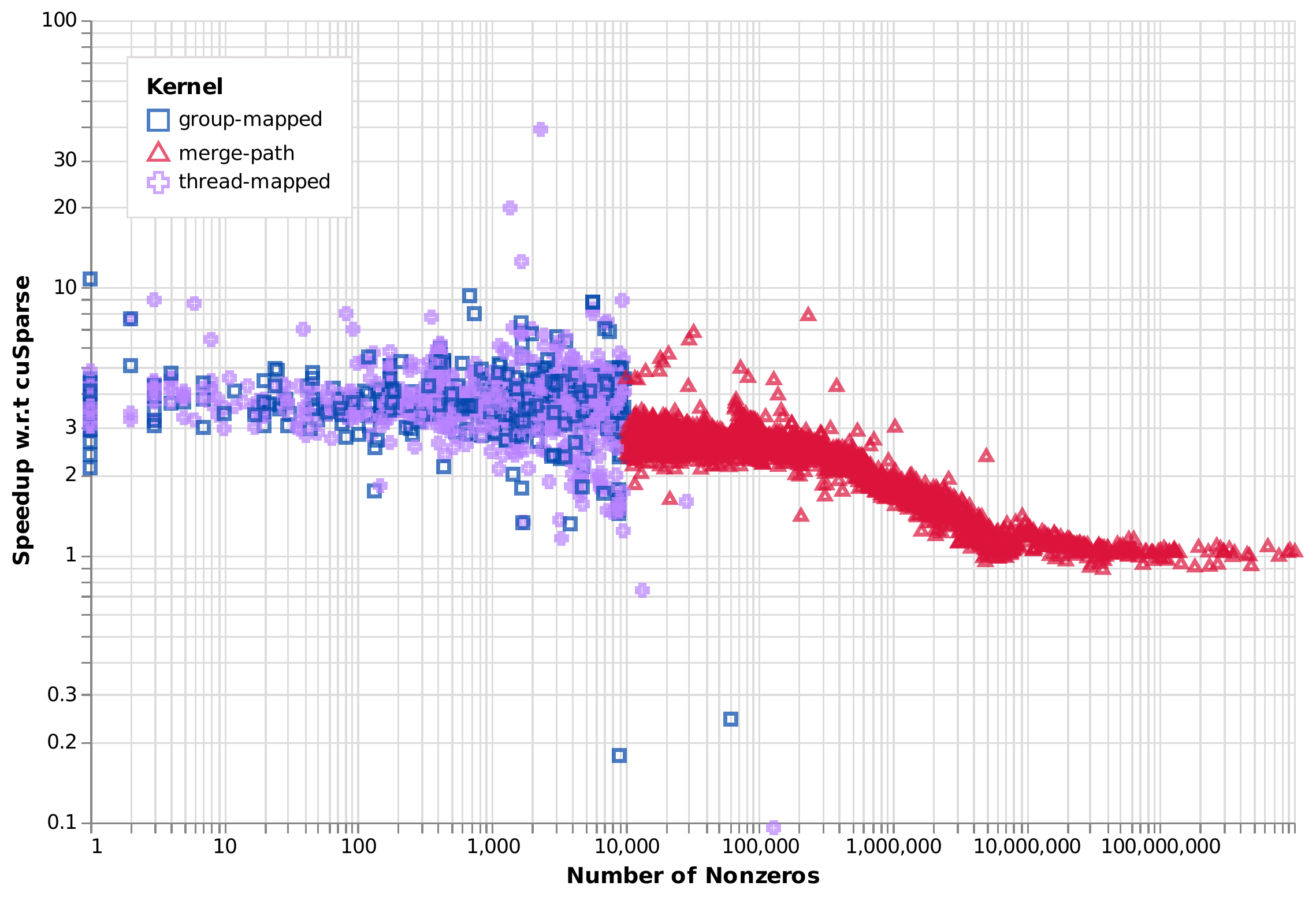}
    \caption[Speedup of our framework's SpMV vs.\ cuSparse's SpMV]{Speedup of our framework's SpMV vs.\ cuSparse's SpMV across SuiteSparse using a heuristic (Section~\ref{sec:improved-performance-response}) to choose the appropriate load-balancing schedule.}
    \label{fig:spmv-speedup}
  \end{figure}

We then combine our scheduling algorithms into one implementation for SpMV, demonstrating noticeable performance improvements over cuSparse (Figure~\ref{fig:spmv-speedup}). This is primarily possible due to our ability to quickly experiment with different heuristic schemes with a variety of available load-balancing schedules. Here, we use merge-path unless either the number of rows or columns are less than the threshold $\alpha$ and the nonzeros of a given matrix are less than threshold $\beta$ (we choose $\alpha = 500$ and $\beta = 10000$ for SuiteSparse). In this case, we use thread-mapped or group-mapped load balancing instead of merge-path. Our system shows a peak performance speedup of 39$\times$ and a geomean performance speedup of 2.7$\times$ vs.\ cuSparse.

Our framework not only allows programmers to express computations efficiently and simply (i.e. without worrying about the load-balancing algorithms), but also quickly optimize a given application using a range of scheduling algorithms, both with minor code changes.

\subsection{Lines of Code (LOC)}
\label{sec:loc}

We are able to achieve these performance gains with minimal code complexity. Table~\ref{tab:spmv_loc} shows lines of code (LOC) for our framework when compared to the state-of-the-art open-source implementation of merge-path and thread-mapped within NVIDIA's CUB library. We deliver the same performance results as highlighted in the previous section with 14$\times$ and 1$\times$ fewer lines of code for merge-path and thread-mapped scheduling algorithms, respectively. Using our merge-path implementation only requires $\sim$15 additional LoC to the trivial thread-mapped schedule.

Furthermore, we extend the same SpMV computation to our novel \emph{group-mapped} load balancing schedule (that can also be specialized to perform \emph{block-} and \emph{warp-mapped} load balancing) within the same 30 LoC\@.

\begin{table}
  \centering
  \begin{small}
    \begin{tabular}{@{}lcc@{}}
      \toprule
      Load Balancing Algorithm & NVIDIA/CUB & Our Work \\ \midrule
      Merge-Path & 503 & 36 \\
      Thread-Mapped & 22 & 21 \\
      Group-Mapped & N/A & 30 \\
      Warp-Mapped & N/A & 30 (free) \\
      Block-Mapped & N/A & 30 (free) \\
      \bottomrule
    \end{tabular}
    \end{small}
  \setlength{\belowcaptionskip}{0pt}
  \caption[Lines of code (LoC) comparison versus CUB]{Lines of code (LoC) comparison for NVIDIA's CUB library versus our work for SpMV application implemented using merge-path, thread-mapped and group-mapped (warp- and block-mapped use the \emph{exact} same code for group-mapped) load balancing algorithms. We report only non-commented lines of code, formatted using \lstinline{clang-format} tool with the Chromium style guide that contributes to the kernel implementation~\cite{Google:Chromium}.}
  \label{tab:spmv_loc}
\end{table}

%% file: ch_loadbalance/chapters/background.tex
\section{Related Work}
\label{sec:related-work}

Load balancing is the key to achieving high performance on GPUs for sparse, irregular parallel problems. Several high-performance computing applications deploy sophisticated load balancing algorithms on the GPUs. For instance,  high-performance sparse-matrix vector multiplication (SpMV) leverages merge-path~\cite{Merrill:2016:MPS} (discussed in detail in this paper) and nonzero splitting algorithm, which partitions the number of non-zeros in a sparse-matrix evenly across the number of threads~\cite{Baxter:2013:MPA,Dalton:2015:OSM,Steinberger:2017:GHL}. Sparse-matrix matrix multiplication (SpMM) and sparse matricized tensor times Khatri-Rao product (SpMTTKRP) use binning and bundling algorithms~\cite{Yang:2018:DPF,Gale:2020:SGK,Nisa:2019:LSM}, which attempt to bin like-length work together such that they are processed together.

While some applications actively perform work to load balance a given input, others store the input in more efficient, already-load-balanced/partitioned formats. These include the F-COO format (a variant of coordinate format) used for SpMTTKRP and Sparse-Tensor Tensor Multiplication (SpTTM), where each thread gets the same number of nonzeros to process~\cite{Liu:2017:UOA}.

Many of the above GPU load-balancing algorithms, along with other novel techniques, were first described in the graph analytics domain. Davidson et al.\ and Merrill and Garland were the first to present Warp, Block-level and Thread-Warp-CTA dynamic load balancing techniques for Single-Source Shortest Path (SSSP) and Breadth-First Search (BFS) respectively~\cite{Davidson:2014:WPG,Merrill:2012:SGG}. Logarithmic Radix Binning (LRB) is a particularly effective technique for binning work based on a logarithmic work estimate, used for the Triangle Counting graph algorithm and more~\cite{Green:2018:LRB,Fox:2019:ISI}. Gunrock, GraphIT, and GraphBLAST are graph analytics libraries that implement several different graph algorithms such as BFS, SSSP, PageRank, Graph Coloring, and more, built on these previously mentioned load-balancing techniques~\cite{Wang:2017:GGG,Brahmakshatriya:2021:CGA,Yang:2021:GAH}. Although many of these are effective load balancing techniques with high-performance implementations, they all tightly couple workload scheduling with the application itself. Our framework is designed to separate these two concerns, allowing the application to be independent of the load-balancing algorithm, and therefore be expressed simply. Our approach also allows these previously proposed techniques to be implemented within our framework, and be used for applications beyond those originally targeted.

Relatively few GPU works target generalized load balancing for irregular workloads. Most of these are focused on providing a singular, dynamic load-balancing solution centered on task parallelism, often using a GPU queue-based data structure. Cederman and Tsigas proposed a task-based approach to load balancing an octree partitioning workload using lock-free and lock-based methods~\cite{Cederman:2008:ODL}. Two Tzeng works provide task-management frameworks that implement load balancing of tasks using a single monolithic task queue and distributed queues with task stealing and donation~\cite{Tzeng:2010:TMF,Tzeng:2012:AGT}.
CUIRRE, a framework for load balancing and characterizing irregular applications on GPUs, also uses a task-pool approach~\cite{Zhang:2014:CAO}, and more recently, Atos, a task-parallel GPU dynamic scheduling framework, targets asynchronous algorithms~\cite{Chen:2022:AAT}. All of these works deploy either a centralized or a distributed queue-like data structure on the GPUs, each making design decisions on how the queue is to be partitioned and updated. Except for the most recent Atos work, most earlier works focus on a coarse-grained parallelism approach of effectively distributing tasks to the GPU\@. Our work takes advantage of more modern GPU architectures, which are more effectively utilized by a fine-grained parallelism approach (parallelizing over work atoms instead of work tiles). Unlike our abstraction, these aforementioned works also rely on a singular load balancing solution, whereas our abstraction flexibly adapts to many different load balancing techniques, static and dynamic, and allows for new schedules to be implemented within our framework.




%% file: ch_loadbalance/chapters/conclusion.tex
\section{Conclusion}
\label{sec:conclusion}

In this paper, we present a programming model for GPU load balancing for sparse irregular parallel problems. Our model is built on the idea of separation of concerns between workload mapping and work execution. In the future, we are interested in expanding our model to a multi-GPU environment, and implementing load balancing schedules that span across the GPU boundary covering multiple devices and nodes for massive parallel problems.
Our current work focuses solely on load balancing, but we also identify locality to be another key factor for high performance. We are interested in identifying an orthogonal model that builds an abstraction for caching and locality into our existing load balancing framework.

In part, this chapter has shown the importance of load-balancing to the performance of irregular problems. While regular problems typically map in a more straightforward way to GPUs, and thus suffer from fewer load-balancing issues, the next chapter shows that even dense, arithmetically intense, operations such as matrix-matrix product (GEMM) can suffer from poor resource utilization due to load
imbalance. However, whereas load imbalance for sparse problems typically arises from variable-length data and unpredictable access patterns, load imbalance in GEMM arises from work decompositions that are a poor fit to the physical resources of the architecture. In the following chapter we introduce \emph{Stream-K}, a work-centric parallel decomposition for dense operations that addresses the resource imbalance issues.

%% file: ch_streamk_v2/streamk_v2.tex
\input{ch_streamk_v2/chapters/introduction}

\input{ch_streamk_v2/chapters/background}
\input{ch_streamk_v2/chapters/decomposition}
\input{ch_streamk_v2/chapters/optimization}
\input{ch_streamk_v2/chapters/performance}
\input{ch_streamk_v2/chapters/conclusion}

%% file: ch_streamk_v2/chapters/introduction.tex
General matrix-matrix product (GEMM), convolution, and other similar 
computations constitute the dominant workloads in many deep learning and 
scientific computing applications. High-performance processors such as GPUs, 
for example, are designed to achieve nearly 100\% of their theoretical peak 
math throughput when computing GEMM.  Doing so, however, requires a work 
decomposition that perfectly occupies the underlying physical cores.  As we 
show, attaining such high levels of processor utilization across a broad 
landscape of problems shapes and sizes can be challenging.

Classically, GEMM implementations block their computation using a 
\emph{data-parallel} tiling of the output matrix, assigning the independent
production of output tiles among concurrent threads (or thread groups)
~\cite{Abdelfattah:2016:KAO,Kerr:2017:CUTLASS,Nath:2010:AIM}.  The work 
per output tile is regular, and tile production tends to dispatch across 
idle physical cores in ``waves''.  The overall workload is well-balanced 
and processor utilization is highest when there are many waves, 
i.e., the number of output tiles greatly oversubscribes the number of cores. 

However, such oversubscription has shrunk considerably as processors have grown 
in size.  An increased core count will require fewer waves to produce a given 
tile count.  Bigger cores will compel larger matrix blocking factors, leading 
to fewer waves of larger tiles.  In general, execution schedules with fewer 
waves are much more likely to suffer from \emph{quantization inefficiency}, 
i.e., the processor underutilization that occurs when the number of output 
tiles is not an even multiple of the number of processor cores.  When the last 
wave is partially full, the unused cores must wait for the remaining threads 
to execute millions (if not billions) of multiply-accumulate (MAC) instructions
before they are able to execute any dependent work.

\begin{figure}
    \centering
    \begin{subfigure}[t]{0.49\textwidth}
        \centering
        \includegraphics[width=\columnwidth]{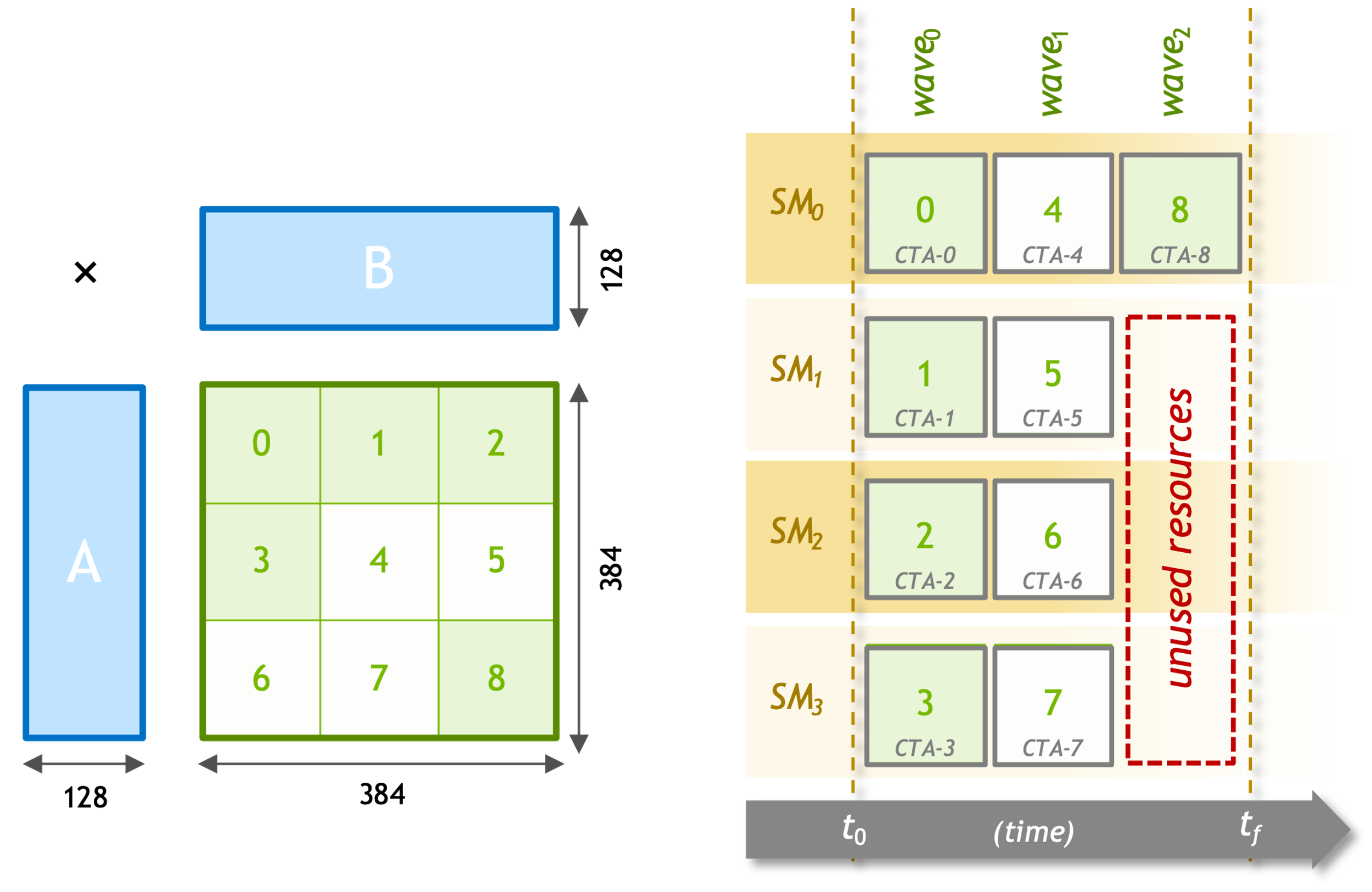}
        \caption[\textit{Data parallel} decomposition (grid size=9~CTAs)]{\textit{Data parallel} decomposition with grid size $g$=9~CTAs,\\
          large $128\times128\times128$ CTA work volumes,\\
          and 75\% processor utilization ceiling} \label{fig:data-parallel-efficient}
    \end{subfigure}
    \hfill%
    \begin{subfigure}[t]{0.49\textwidth}
        \centering
        \includegraphics[width=\columnwidth]{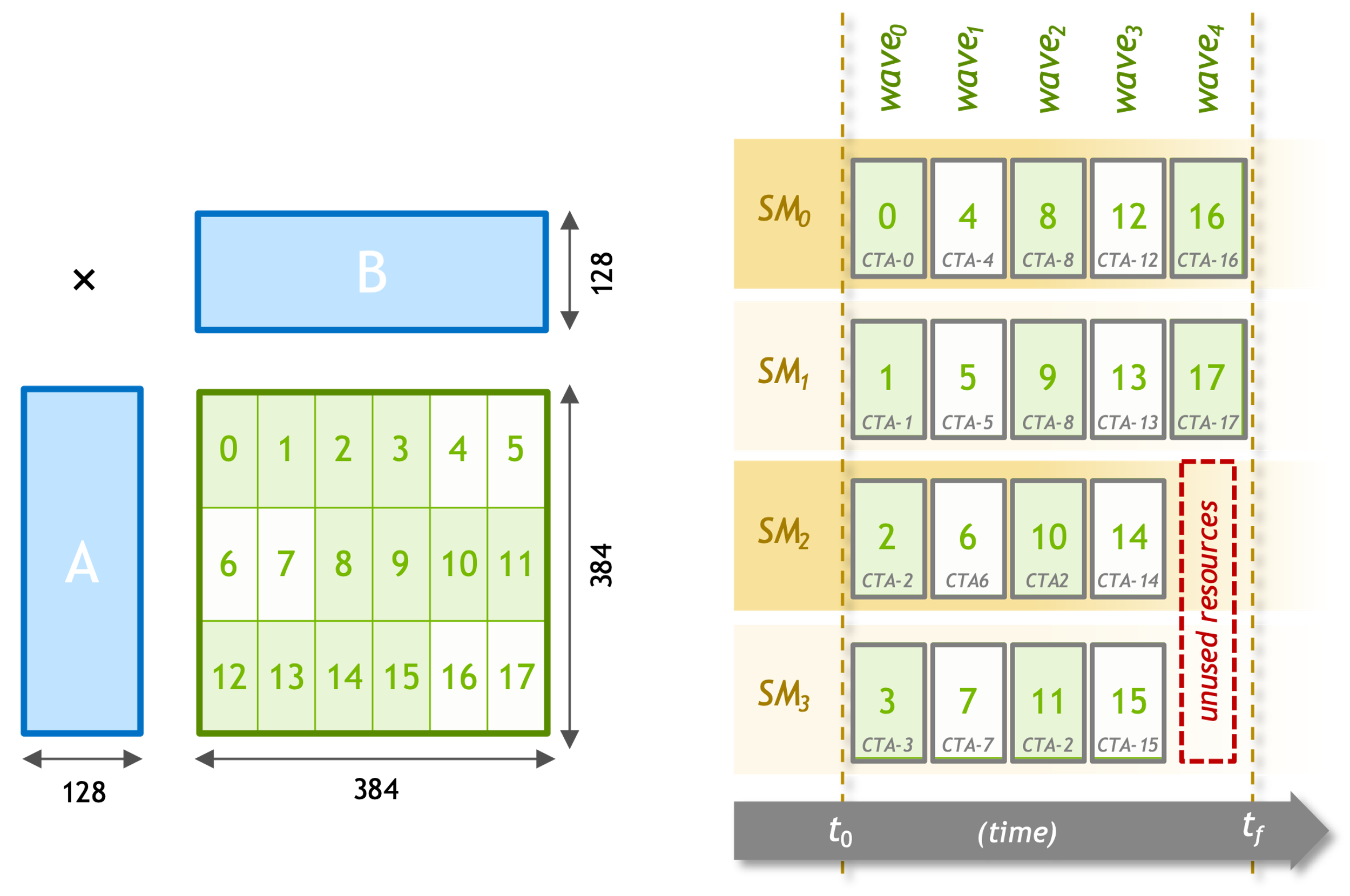}
        \caption[\textit{Data parallel} decomposition (grid size=18~CTAs)]{\emph{Data parallel} decomposition with grid size $g$=18~CTAs,\\
          smaller $128\times64\times128$ CTA work volumes,\\
          and 90\% processor utilization ceiling} \label{fig:data-parallel-inefficient}
    \end{subfigure}
    \caption[\emph{Data-parallel} execution schedules]{\emph{Data-parallel} execution schedules for $384\times384\times128$ GEMM across a hypothetical four-SM GPU.} 
    \label{fig:data-parallel}
\end{figure}

Figure~\ref{fig:data-parallel-efficient} illustrates such a scenario
on a hypothetical GPU with four streaming multiprocessor cores (SMs).
If we block a $384\times384\times128$ GEMM computation into nine
$128\times128$ output tiles, a \emph{data-parallel} decomposition cannot achieve
more than 75\% of the processor's rated throughput.  This theoretical
utilization ceiling can be improved to 90\% by halving the tile size as
shown in Figure~\ref{fig:data-parallel-inefficient}.  However, the
finer-grained blocking factor will be less cache and scratchpad efficient,
and may preclude any practical performance improvement.

\begin{figure}
  \centering
  \begin{subfigure}[t]{0.49\textwidth}
    \includegraphics[width=\columnwidth]{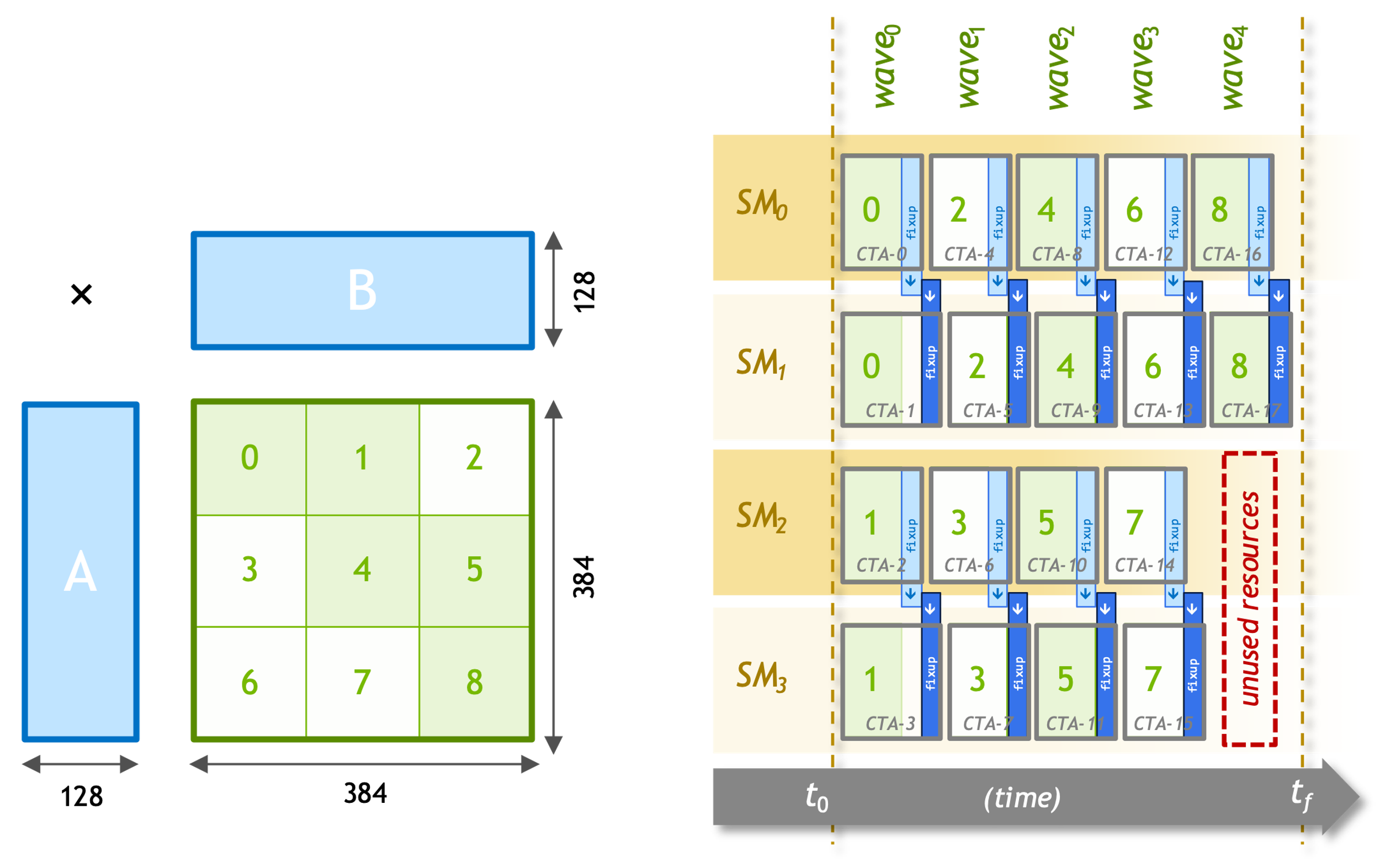}
    \caption[\emph{Fixed-split} decomposition]{\emph{Fixed-split} decomposition with splitting factor $s$=2,\\
      grid size $g$=18~CTAs, smaller $128\times128\times64$ CTA work volumes,\\
      and 90\% quantization efficiency} \label{fig:fixed_split}
  \end{subfigure}
  \hfill%
  \begin{subfigure}[t]{0.49\textwidth}
    \includegraphics[width=\columnwidth]{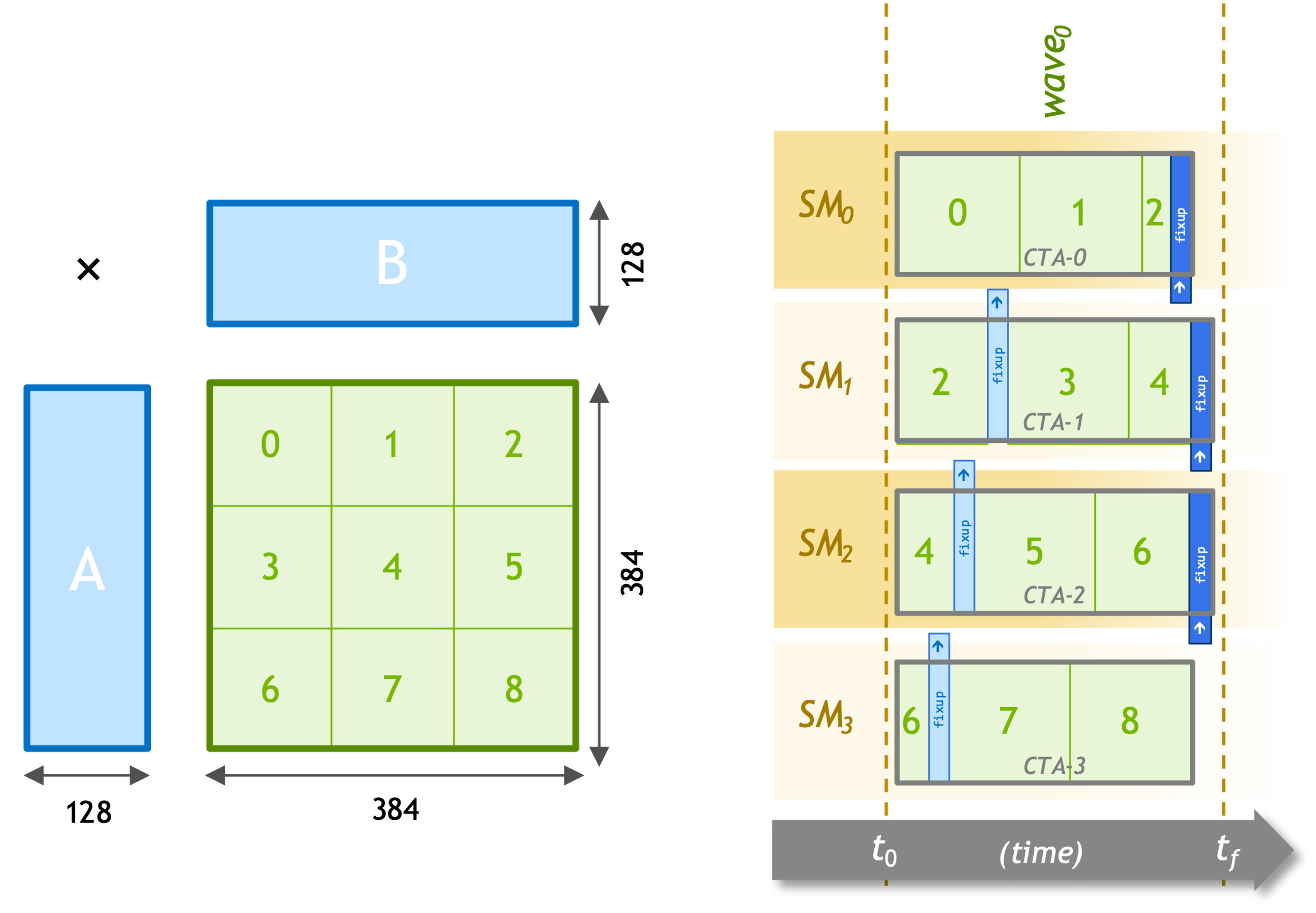}
    \caption[Basic \emph{Stream-K} decomposition]{Basic \emph{Stream-K} decomposition with grid size $g$=4~CTAs,\\ 
      larger $128\times128\times288$ CTA work volumes,\\
      and nearly 100\% quantization efficiency} \label{fig:stream_k}
  \end{subfigure}
  \caption[Tile-splitting execution schedules]{Tile-splitting execution schedules for $384\times384\times128$ GEMM across a hypothetical four-SM GPU.} 
\end{figure}

Quantization inefficiency is a concern for increasingly wide processors such as GPUs, 
where ALUs-per-core and cores-per-processor both currently number in the 
hundreds. Consequently, many common GEMM-like workloads now exhibit a final, 
partially full wave that comprises a significant fraction of the total computation time.

The current remedy employed by GPU-based math and deep learning libraries is to 
deploy an ensemble of tiling configurations.  When the ideal blocking factor does 
not quantize well, the library chooses among tiling alternatives with smaller 
concurrent work volumes, such as those illustrated in 
Figure~\ref{fig:data-parallel-inefficient} and Figure~\ref{fig:fixed_split}.

Tile-based ensembles, however, present performance and logistical challenges 
for math libraries seeking to deliver the best-achievable performance across diverse 
problem sizes and shapes.  Distributable code size can be problematic for large 
ensembles.  For example, NVIDIA's cuBLAS library~\cite{NVIDIA:2020:CUBLAS} is hundreds 
of megabytes, often providing more than twenty pre-compiled kernel specializations 
per architecture for a given API entry point.  Large ensembles also require 
sophisticated selection heuristics.  In our evaluation, we show these heuristics 
can struggle to consistently identify the optimal configuration for arbitrary 
problems.

Unlike these tile-based methods, our \emph{Stream-K} decomposition always 
distributes an even share (within one) of the aggregate multiply-accumulate loop 
iterations required by the GEMM computation across SMs.  Because the instruction 
workload of a single MAC-loop iteration is far smaller than that of an entire 
output tile, any variance in core workload is practically negligible.  \emph{Stream-K} 
uses the ideal blocking factor regardless of problem shape, has communication 
overheads that scale with processor width (rather than output tiles), and 
compiles to a single kernel.

We use an enormous corpus of 32,824~GEMM shapes and sizes to evaluate \emph{Stream-K}, 
which we implemented within NVIDIA's CUTLASS library~\cite{Kerr:2017:CUTLASS}.  In 
comparison with CUTLASS's \emph{data-parallel} implementation of the same blocking 
factor, \emph{Stream-K} provides a substantially higher performance response across 
our landscape of GEMM problems, demonstrating up to 14$\times$ speedup on NVIDIA A100 
GPUs.

To highlight the practical challenges of ensemble-based solutions, we also 
evaluate NVIDIA's cuBLAS library as well as an oracle-driven 
ensemble of \emph{data-parallel} CUTLASS tilings.  Relative to both ensembles, we 
show that our single-kernel \emph{Stream-K} achieves both (1) higher average 
performance, and (2) higher performance consistency. Versus cuBLAS, \emph{Stream-K} 
demonstrates up to 6.7$\times$ speedup and virtually no instances of slowdown 
for compute-bound problems.

%% file: ch_streamk_v2/chapters/background.tex
\section{Background}

General Matrix Multiplication (GEMM) is defined as the product $\textbf{C} = \alpha\textbf{AB} + \beta\textbf{C}$ where $\alpha$ and $\beta$ are scalar values and $\textbf{A}$, $\textbf{B}$, and $\textbf{C}$ are matrices. (For simplicity,
we assume $\alpha = 1$, $\beta = 0$ throughout this paper.) We refer to the \emph{shape} of a given GEMM problem by the volumetric extents of its computation. For example, a $m\times n \times k$ GEMM consumes $m \times k$ and $k \times n$ input matrices \textbf{A} and \textbf{B}, respectively, performs $m\times n\times k$ multiply-accumulate operations, and produces an $m \times n$ output matrix \textbf{C}.

GEMM is a performance-critical subroutine in many large-scale engineering and scientific applications. It plays an important role in matrix factorization methods such as LU, QR, and Cholesky decomposition. High-performance modeling and simulation applications in engineering, climate simulation, cosmology, quantum chemistry, and other scientific domains rely on these factorization methods.

Matrix multiplication is also the fundamental building block of modern deep learning (DL) methods. The training of deep neural networks (DNNs) is often performed on massive datasets across large distributed systems~\cite{Mattson:2020:MTB}. Many DL training and inference operations are cast as matrix multiplications. For example, image recognition and computer vision models rely on convolution, which can be implemented directly as the product of filter and image datasets~\cite{Chetlur:2014:CEP}. 
Transformer architectures, which have come to dominate natural language
processing and other applications, are almost entirely limited by the
performance of large matrix products.
%

Early work on GPU matrix-matrix multiplication from Larsen and McAllister framed the computation as a multi-texture multiplication and blending operation~\cite{Larsen:2001:FMM}. The user programmable shared memory provided by subsequent GPU architectures enabled higher-performing \emph{data parallel} schemes with two levels of blocking (shared memory and registers) with tile sizes informed via extensive microbenchmarking analysis~\cite{Barrachina:2008:EAT,Nath:2010:AIM,Tan:2011:FID,Tillet:2019:TAI} and auto-tuning~\cite{Cui:2010:ATD,Jiang:2005:ATM,Li:2009:ANA}.

The MAGMA GPU math library was perhaps the first to optimize for diverse GEMM problem shapes~\cite{Kurzak:2012:AGK}. Their solution applied a constrained set of tiling parameters to a templated CUDA C++ code stencil, generating several hundred \emph{data-parallel} variants per API primitive (e.g., \lstinline{hgemm_tt()} for half-precision transpose-transpose GEMM). They evaluated these variants to distill a small ensemble of typically three to five kernels that collectively perform well across a diversity of problem shapes. Kernel selection and dispatch for a given problem was governed by size thresholds expressed via simple handwritten rules.

Subsequent GPU math libraries have employed more sophisticated code-generation 
and kernel-selection components.  For example, the ISAAC project uses machine learning 
techniques to predict an optimal tiling and/or splitting parameterization 
for a given GEMM shape, which can then be instantiated either online or 
offline via a PTX-level code generator~\cite{Tillet:2017:IAA}.

NVIDIA's cuBLAS~\cite{NVIDIA:2020:CUBLAS} library has provided an extended 
\verb|cublasGemmEx| interface that allows the caller to select from among 
24 different GEMM ``algorithms''. Carefully trained heuristics choose between this 
large space of alternatives when using the default interface.  These algorithms 
implement a variety of different \emph{data-parallel} and \emph{fixed-split} variants, 
and it is common for cuBLAS to have assembled each variant into its own 
architecture-specific kernel program for code optimization purposes.  The cross product 
of GEMM API functionality, strategic variants, and microarchitecture has resulted 
in distributions that are increasingly enormous, exceeding hundreds of megabytes of executable code.

Given the fast-paced and rapidly changing nature of contemporary deep learning, recent work has focused on programming models for simplifying the expression and construction high performance kernels that alter or supplement the GEMM computation. The CUTLASS C++ library provides data-movement and multiply-accumulation classes for composing custom GEMM-like computations at all levels of the GPU thread hierarchy~\cite{Kerr:2017:CUTLASS}. Triton~\cite{Tillet:2019:TAI} is a domain-specific language for tensor programming centered on the expression, transformation, and optimization of block/tile concepts. Other domain-specific programming languages such as Halide~\cite{Ragan-Kelley:2013:HAL} and TVM~\cite{Chen:2018:TEO} separate the expression of pointwise operators from that of loop scheduling. Fireiron~\cite{Hagedorn:2020:FAS} further adds data movement constructs into the scheduling grammar.

%% file: ch_streamk_v2/chapters/decomposition.tex
\section{Work Decomposition Strategies}

Modern processors typically store \textbf{A}, \textbf{B}, and \textbf{C}
in a large, slow, distant memory and have access to a small, fast,
scratchpad or cache memory. A primary goal for any GEMM implementation
is to leverage these local storage resources so that the resulting
implementation is computation-bound.

\subsection{Sequential Cache-Blocked}
The classic cache-blocked formulation of GEMM divides its computational volume 
into blocks and chooses a traversal order that exposes memory locality. 
Algorithm~\ref{alg:sequential} presents a simplified implementation comprising 
six loops. The innermost three loops iterate within the blocking factors BLK\_M, 
BLK\_N, and BLK\_K, while the outermost three iterate across them. If the cache 
can capture one block from each of the three matrices, the resulting data reuse
among those elements will significantly reduce the number of last-level memory 
accesses~\cite{Lam:1991:TCP}.

\input{ch_streamk_v2/code/sequential.tex}

\subsection{Data-parallel}

As shown in Algorithm~\ref{alg:data_parallel}, the \emph{data-parallel} 
GPU formulation of GEMM is decomposed across a grid of parallel thread blocks, 
or \emph{cooperative thread arrays} (CTAs)\footnote{Blocks of GPU threads are 
coscheduled in CTAs, which virtualize the hardware's streaming multiprocessor 
cores (SMs).}. The grid is sized such that each CTA produces its own (BLK\_M 
$\times$ BLK\_N) output tile. 

For exposition, the \lstinline{MacLoop()} subroutine of Algorithm~\ref{alg:macloop}
encapsulates the multiply-accumulate (MAC) workloads that compute the values of the 
CTA's output tile.  It performs a sequence of \emph{MAC-loop} iterations in the 
accumulation domain, e.g., the \emph{k}-axis for GEMM.  Each \emph{MAC-loop} 
iteration comprises a per-thread volume of (BLK\_M $\times$ BLK\_N $\times$ BLK\_K) 
$/$ CTA\_THREADS MAC operations. As the computation proceeds, fragments of the input 
matrices are staged through the SM's shared memory for local reuse among individual 
threads.

Although this particular presentation of \lstinline{MacLoop()} deploys one thread 
per output tile element, the sophisticated implementations in CUTLASS~\cite{Kerr:2017:CUTLASS} 
and cuBLAS~\cite{Kerr:2017:CUTLASS} will: (1) fully unroll the per-thread MAC-loop 
iteration; (2) implement additional blocking at the warp and/or thread levels; 
and (3) orchestrate a software pipeline of shared memory data movement across 
MAC-loop iterations.

Unfortunately, this classic \emph{data-parallel} decomposition is liable to suffer from 
quantization inefficiency on modern GPUs, as illustrated in Figure~\ref{fig:data-parallel}. 
Although an ensemble of diverse blocking factors may uncover opportunities for greater 
processor utilization, it is unlikely to facilitate perfect quantizations for 
arbitrary problem sizes.  Furthermore, smaller blocking factors have two drawbacks: 
(1)~fewer instructions per MAC-loop iteration for covering the latencies of global and 
shared memory transfers in pipelined implementations; and (2)~a higher proportion of 
memory operations relative to MAC instructions, which may prevent them from being 
computation-bound.

\input{ch_streamk_v2/code/dataparallel.tex}
\input{ch_streamk_v2/code/macloop.tex}

\subsection{Fixed-split}
Alternatively, the granularity of work assigned to each CTA can be reduced via 
parallelization across the accumulation dimension. For a given output tile, the 
associativity of addition allows the iteration domain to be split among multiple
concurrent CTAs, followed by a dependent ``fixup'' step to reduce the partial sums 
computed by each CTA\@. We highlight this \emph{fixed-split} approach in 
Algorithm~\ref{alg:fixed_split}, where each output tile is cooperatively produced
by $s$ CTAs. Notably, it functions identically to the \emph{data-parallel} 
decomposition when the splitting factor $s = 1$.

The \emph{fixed-split} decomposition is also featured in CUTLASS and cuBLAS. The 
splitting factor is implemented as a runtime parameter, allowing a single kernel 
executable to support multiple work volumes while retaining the ideal blocking 
factors for optimal data sharing and latency hiding. However, as illustrated in 
Figure~\ref{fig:fixed_split}, the prospect of achieving a perfect quantization from
a uniform tile-splitting is unlikely.  Furthermore, the extra overheads of 
communication and synchronization scale with both the overall problem size as well 
as the splitting factor.

\input{ch_streamk_v2/code/fixedsplit.tex}

\subsection{Stream-K}
Our \emph{Stream-K} decomposition is a tile-splitting parallelization in which 
the splitting seams are completely dissociated from the tiling structure 
itself. Although we employ familiar blocking and tiling strategies for data 
reuse, we instead quantize the GEMM computation into MAC-loop iterations, 
i.e., small volumes of CTA-wide BLK\_M $\times$ BLK\_N $\times$ BLK\_K work. As 
presented in Algorithm~\ref{alg:streamk}, \emph{Stream-K} evenly partitions the GEMM's 
aggregate workload of MAC-loop iterations across a constant-sized grid of $g$ 
CTAs. Each CTA's range of MAC-loop iterations is mapped contiguously into the 
$m \rightarrow n \rightarrow k$ linearization of the GEMM shape, crossing 
output-tile boundaries as it may.

\input{ch_streamk_v2/code/streamk.tex}

Should a given CTA's starting and/or ending iterations not coincide with
tile boundaries (as is expected to be the common case), it must
consolidate its partial results with those of the other CTA(s) also
covering that tile. In this basic implementation, each output tile in
\textbf{C} is written by the CTA that performed that tile's $k=0$
MAC-loop iteration. Before it can do so, however, it must accumulate any
partial sums shared from other CTAs in temporary global storage.
Notably, \emph{Stream-K'}s communication, synchronization, and global
storage overheads are independent of problem size, scaling instead with
the number of CTAs.

A secondary benefit of \emph{Stream-K} is that
synchronization-waiting~is likely negligible when the number of output
tiles is greater than the number of CTAs. In this regime, each output
tile is covered by at most two CTAs, and the tile-processing skew
ensures that the accumulating CTA will not need its peer contributions
until well after those collaborators have finished producing them.

Continuing our earlier example, Figure~\ref{fig:stream_k} illustrates the basic
\emph{Stream-K} execution schedule of the $384\times384\times128$ GEMM problem on a
hypothetical four-SM GPU\@. To fully occupy the GPU, we launch $g=4$
CTAs. Assuming BLK\_M~$=128$, BLK\_N~$=128$, and BLK\_K~$=4$, each CTA is tasked
with a $128\times128\times288$ work volume comprising 72~MAC-loop iterations. This
results in a 100\% quantization efficiency, as all four SMs will execute
the same number of MAC instructions.

Additionally, the work volume of a single MAC-loop iteration is 32$\times$
smaller than that of an entire output tile. Consequently, a 32-way
\emph{fixed-split} decomposition would also provide a 100\% quantization
efficiency, but at the expense of an 8$\times$ larger ``fixup'' overhead.
Furthermore, \emph{Stream-K} is better able to hide the latency of inter-CTA 
synchronization due to the temporal skew between writers and readers when 
sharing partial sums.

\emph{Stream-K} also generalizes to both \emph{fixed-split}
and \emph{data-parallel} decompositions. When the grid size $g$ is
an even multiple of the number of output tiles, \emph{Stream-K}
functions exactly as the \emph{fixed-split} decomposition. Similarly,
when $g$ equals the number of output tiles, \emph{Stream-K} behaves
identically to the \emph{data-parallel} decomposition. We take advantage
of this generalization to create an optimized hybridization of the \emph{Stream-K}
decomposition in following section (\ref{sec:data-parallel-hybridization}).

%% file: ch_streamk_v2/code/sequential.tex

\begin{algorithm}
    \caption[Sequential cache-blocked GEMM]{Sequential cache-blocked GEMM\@.}\label{alg:sequential}
    \begin{algorithmic}[1]
    \State \algcomment{tile-processing outer loops}
    \For{mm $\gets$ 0 \textbf{to} m \textbf{step} BLK\_M}
        \For{nn $\gets$ 0 \textbf{to} n \textbf{step} BLK\_N}
            \State \algcomment{zero-initialize output tile}
            \For{mmm $\gets$ mm \textbf{to} (mm + BLK\_M)}
                \For{nnn $\gets$ nn \textbf{to} (nn + BLK\_N)}
                    \State C[mmm,nnn] $\gets$ 0
                \EndFor
            \EndFor
            \State \algcomment{perform the MAC iterations for this tile}
            \For{kk $\gets$ 0 \textbf{to} k \textbf{step} BLK\_K}
                \State \algcomment{MAC iteration (fully unrolled)}
                \For{mmm $\gets$ mm \textbf{to} (mm + BLK\_M)}
                    \For{nnn $\gets$ nn \textbf{to} (nn + BLK\_N)}
                        \For{kkk $\gets$ kk \textbf{to} (kk + BLK\_K)}
                            \State C[mmm,nnn] $\gets$ C[mmm, nnn] +
                            \State \quad (A[mmm,kkk] $\times$ B[kkk,nnn])
                        \EndFor
                    \EndFor
                \EndFor
            \EndFor
        \EndFor
    \EndFor
    \end{algorithmic}
\end{algorithm}

%% file: ch_streamk_v2/code/dataparallel.tex


\begin{algorithm}
  \caption[\textit{Data-parallel} GPU GEMM]{\textit{Data-parallel} GPU GEMM\@.}\label{alg:data_parallel}
  \begin{algorithmic}[1]
  \State \shared accum[BLK\_M,BLK\_N]
  \State iters\_per\_tile $\gets \lceil$k/BLK\_K$\rceil$
  \State \algcomment{instantiate one CTA per output tile}
  \Fork{CTA$_{[x]}$ in [ $\lceil$m/BLK\_M$\rceil$ $\times$ $\lceil$n/BLK\_N$\rceil$ ]}
    \State \algcomment{perform the MAC iterations for this tile}
    \State accum $\gets$ MacLoop(x, 0, iters\_per\_tile)
    \State \algcomment{store accumulators to output tile}
    \State StoreTile(C, x, accum)
  \EndFork
  \end{algorithmic}
\end{algorithm}

%% file: ch_streamk_v2/code/macloop.tex

\begin{algorithm}
    \caption[CTA-wide \lstinline{MacLoop()} subroutine]{CTA-wide \lstinline{MacLoop()} subroutine for performing a sequence of MAC-loop iterations.}\label{alg:macloop}
    \begin{algorithmic}[1]
    \Procedure{MacLoop}{tile\_idx, iter\_begin, iter\_end}
    \State \shared accum[BLK\_M,BLK\_N]
    \State \shared frag\_a[BLK\_M,BLK\_K]
    \State \shared frag\_b[BLK\_K,BLK\_N]
    \State \algcomment{determine output tile coordinates}
    \State mm $\gets$ BLK\_M $\times$ (tile\_idx / $\lceil$m/BLK\_M$\rceil$)
    \State nn $\gets$ BLK\_N $\times$ (tile\_idx \% $\lceil$m/BLK\_M$\rceil$)
    \State \algcomment{zero-initialize local accumulator storage}
    \State accum $\gets$ {0}
    \State \algcomment{perform the specified range of MAC iters for this tile}
    \For{iter $\gets$ iter\_begin \textbf{to} iter\_end}
        \State kk $\gets$ iter $\times$ BLK\_K
        \State \algcomment{copy global matrix fragments to local storage}
        \State frag\_a $\gets$ LoadFragment(A, mm, kk)
        \State frag\_b $\gets$ LoadFragment(B, kk, nn)
        \Fork{THREAD$_{[mmm,nnn]}$ in [BLK\_M, BLK\_N]}
            \State \algcomment{MAC iteration per thread (fully unrolled)}
            \For{kkk $\gets$ 0 \textbf{to} BLK\_K}
                \State accum[mmm, nnn] $\gets$ accum[mmm,nnn] + 
                \State \quad (frag\_a[mmm,kkk] $\times$ frag\_b[kkk,nnn])
            \EndFor
        \EndFork
    \EndFor
    \State \textbf{return} accum
    \EndProcedure
    \end{algorithmic}
\end{algorithm}

%% file: ch_streamk_v2/code/fixedsplit.tex


\begin{algorithm}
    \caption[\textit{Fixed-split} GPU GEMM with splitting factor $s$]{\textit{Fixed-split} GPU GEMM with splitting factor $s$.}\label{alg:fixed_split}
    \begin{algorithmic}[1]
    \State \shared accum[BLK\_M,BLK\_N]
    \State iters\_per\_tile $\gets \lceil$k/BLK\_K$\rceil$
    \State iters\_per\_split $\gets \lceil$iters\_per\_tile/$s$$\rceil$
    \State \algcomment{instantiate $s$ CTAs per output tile}
    \Fork{CTA$_{[x,y]}$ in [ $\lceil$m/BLK\_M$\rceil$ $\times$ $\lceil$n/BLK\_N$\rceil$, $s$]}
        \State \algcomment{perform the range of MAC iterations for this split}
        \State iter $\gets$ y $\times$ iters\_per\_split
        \State iter\_end $\gets$ min(iters\_per\_tile, iter + iters\_per\_split)
        \State accum $\gets$ MacLoop(x, iter, iter\_end)
        \State \algcomment{consolidate partial-sums across CTAs}
        \If{y $\neq$ 0}
            \State \algcomment{store accumulators to temporary global storage}
            \State StorePartials(partials[x,y], accum)
            \State Signal(flags[x,y])
        \Else
            \State \algcomment{accumulate partial sums from other CTAs contributing to this tile}
            \For{cta $\gets$ 1 \textbf{to} $s$}
                \State Wait(flags[x,cta])
                \State accum $\gets$ accum + LoadPartials(partials[x,cta])
            \EndFor
            \State \algcomment{store accumulators to output tile}
            \State StoreTile(C, tile\_id, accum)
        \EndIf
    \EndFork
    \end{algorithmic}
\end{algorithm}

%% file: ch_streamk_v2/code/streamk.tex
\begin{algorithm}
    \caption[Basic \textit{Stream-K} GPU GEMM with grid size $g$]{Basic \textit{Stream-K} GPU GEMM with grid size $g$.}\label{alg:streamk}
    \begin{algorithmic}[1]
    \State \shared accum[BLK\_M,BLK\_N]
    \State iters\_per\_tile $\gets \lceil$k/BLK\_K$\rceil$
    \State total\_iters $\gets \lceil$m/BLK\_M$\rceil \times \lceil$n/ BLK\_N$\rceil$ $\times$ iters\_per\_tile
    \State iters\_per\_cta $\gets \lceil$total\_iters / g$\rceil$
    \State \algcomment{instantiate g CTAs}
    \Fork{CTA$_{[x]}$ \textbf{in} [g]}
        \State iter $\gets$ x $\times$ iters\_per\_cta
        \State iter\_end $\gets$ iter + iters\_per\_cta
        \State \algcomment{iteration-processing outer loop}
        \While{iter $<$ iter\_end}
            \State tile\_idx $\gets$ iter / iters\_per\_tile
            \State tile\_iter $\gets$ tile\_idx $\times$ iters\_per\_tile
            \State tile\_iter\_end $\gets$ tile\_iter + iters\_per\_tile
            \State \algcomment{perform the range of MAC iterations for this tile}
            \State local\_iter $\gets$ iter - tile\_iter
            \State local\_iter\_end $\gets$ min(iter\_end, tile\_iter\_end) - tile\_iter
            \State accum $\gets$ MacLoop(tile\_id, local\_iter, local\_iter\_end)
            \State \algcomment{consolidate partial-sums across CTAs}
            \State tile\_started $\gets$ iter = tile\_iter
            \State tile\_ended $\gets$ (iter\_end $\geq$ tile\_iter\_end)
            \If{$\neg$tile\_started}
                \State \algcomment{store accum to temporary global storage}
                \State StorePartials(partials[x], accum)
                \State Signal(flags[x])
            \Else
                \State \algcomment{store accumulators to output tile}
                \If{$\neg$tile\_ended}
                \State \algcomment{accumulate partial sums from other CTA}
                \State \algcomment{contributing to this tile}
                \State cta\_end $\gets$ tile\_iter\_end / iters\_per\_tile
                \For{cta $\leftarrow$ (x+1) in cta\_end}
                    \State Wait(flags[cta])
                    \State accum $\gets$ accum + LoadPartials(partials[cta])
                \EndFor
                \EndIf
                \State StoreTile(C, tile\_id, accum)
            \EndIf
            \State iter $\gets$ tile\_iter\_end
        \EndWhile
    \EndFork
    \end{algorithmic}
\end{algorithm}

%% file: ch_streamk_v2/chapters/optimization.tex
\section{Implementation Details}
\label{sec:practical-usage}   

\begin{figure}
    \centering
    \begin{subfigure}[t]{0.15\textwidth}
        \includegraphics[width=\linewidth]{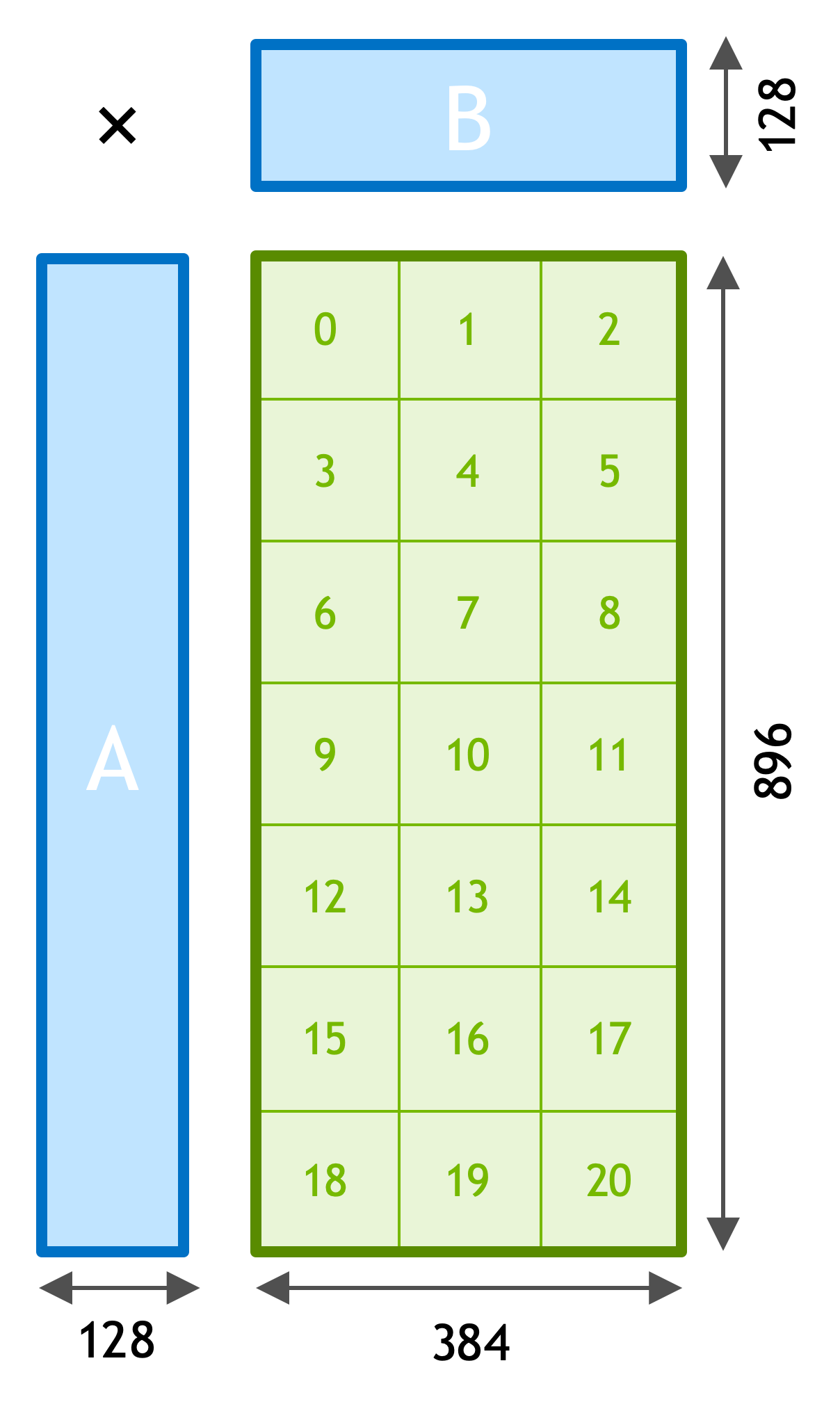}
    \end{subfigure}
    \hfill%
    \begin{subfigure}[t]{0.25\textwidth}
        \includegraphics[width=\linewidth]{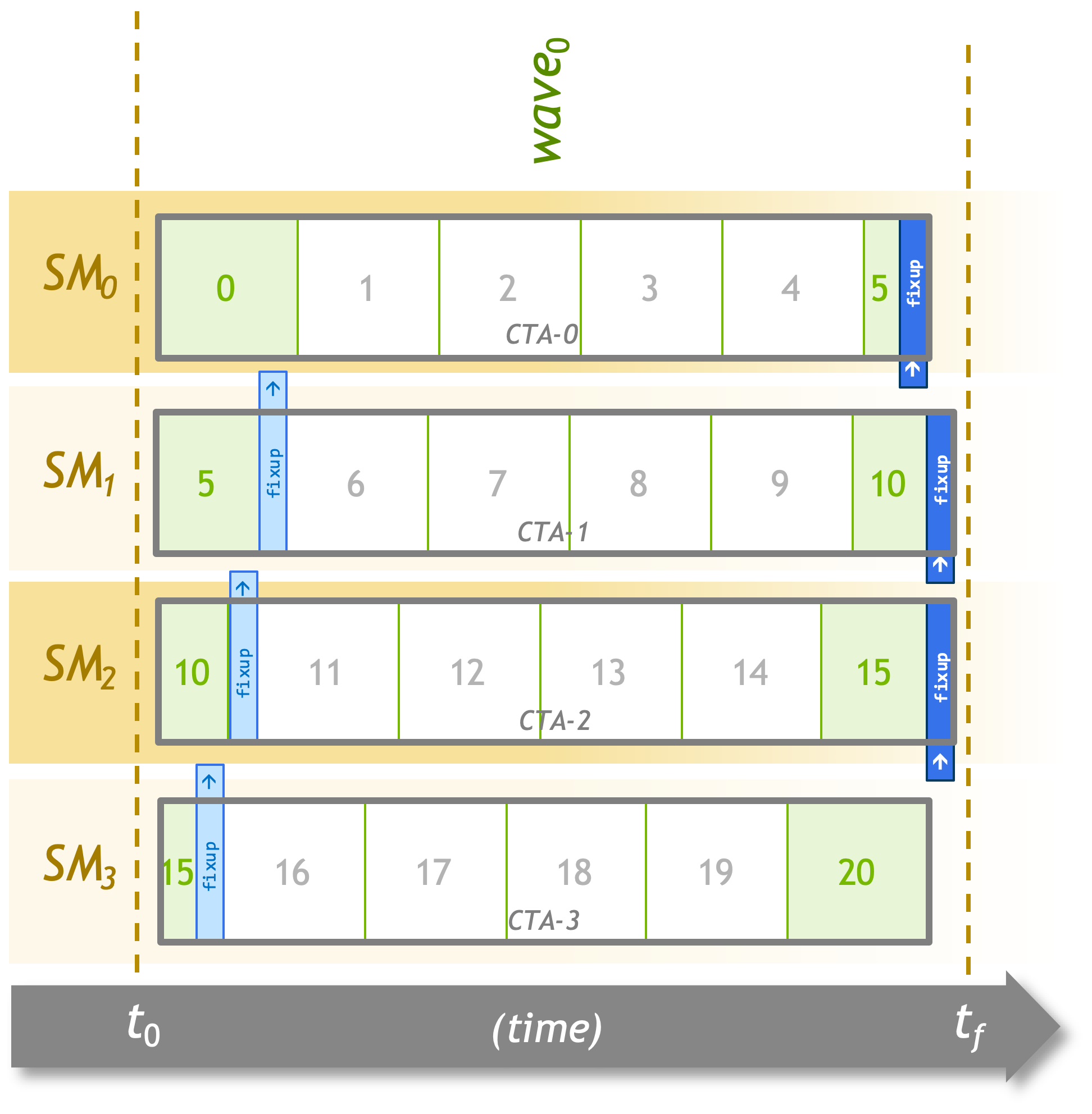}
        \caption{Basic Stream-K} \label{fig:basic_streamk}
    \end{subfigure}
    \hfill%
    \begin{subfigure}[t]{0.25\textwidth}
        \includegraphics[width=\linewidth]{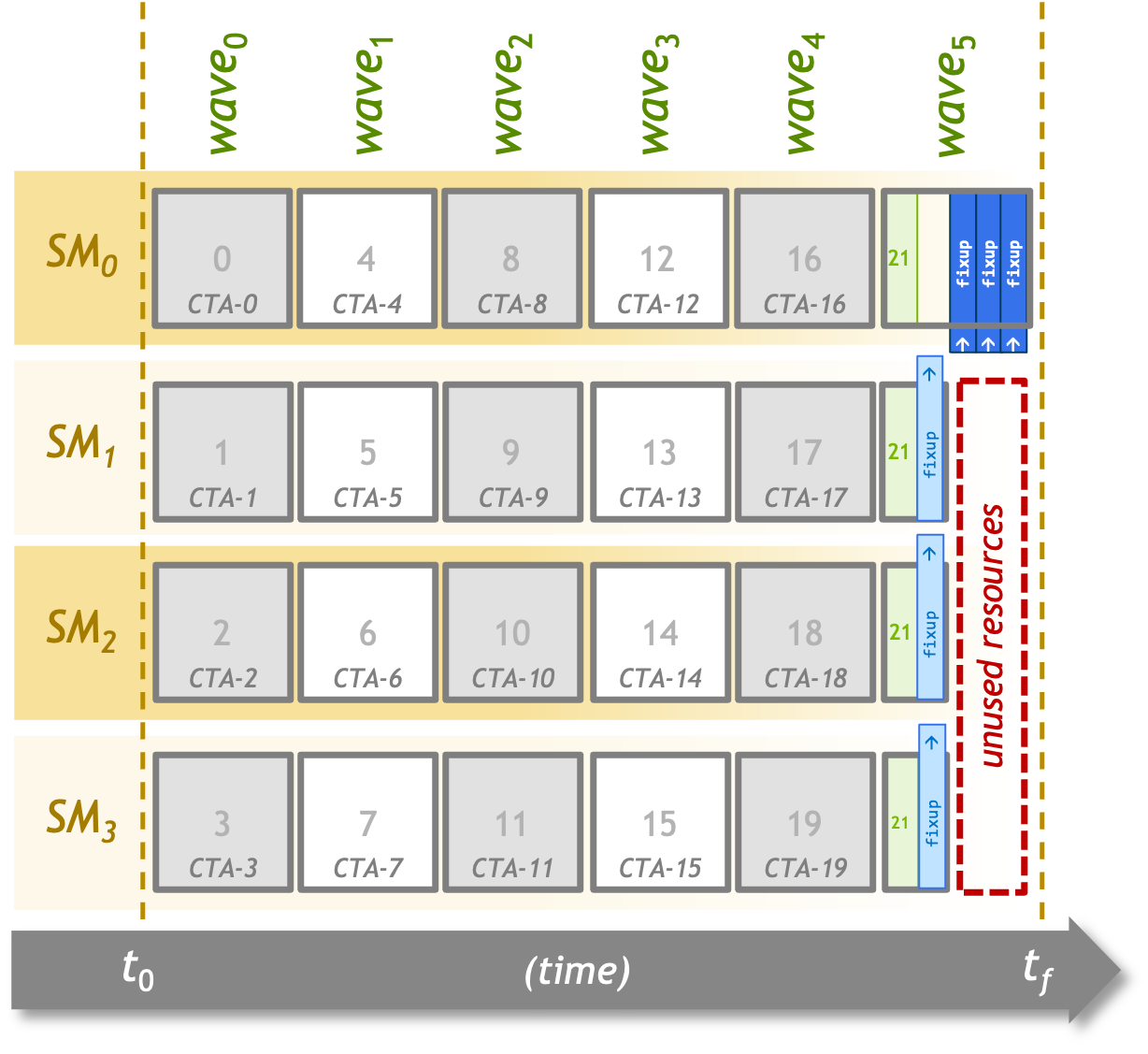}
        \caption{DP + one-tile SK} \label{fig:dp_one_tile_sk}
    \end{subfigure}
    \hfill%
    \begin{subfigure}[t]{0.25\textwidth}
        \includegraphics[width=\linewidth]{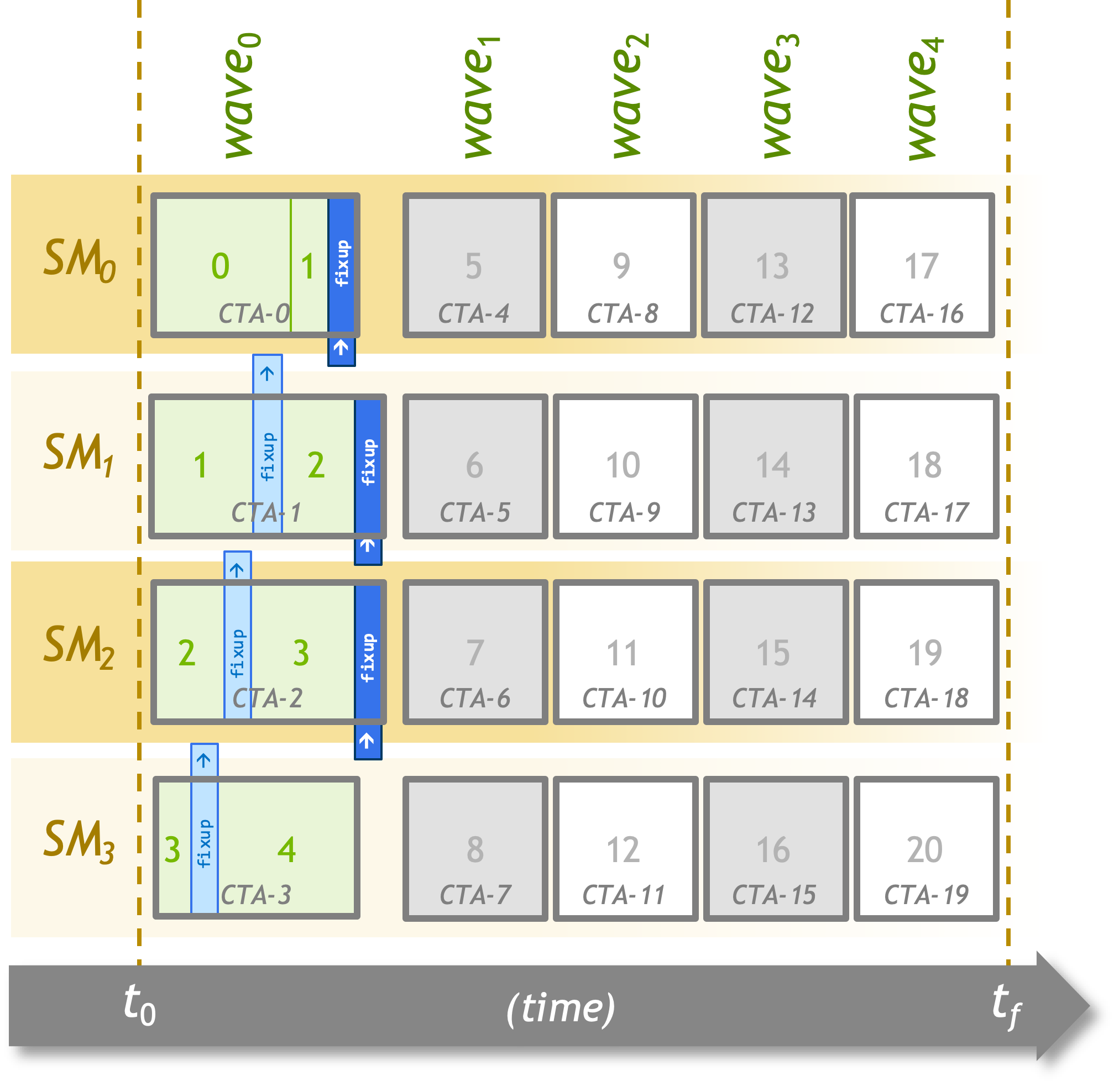}
        \caption{Two-tile SK + DP} \label{fig:two_tile_sk_dp}
    \end{subfigure}
    \caption[Basic \textit{Stream-K} vs.\ hybrid execution schedules]{Basic \textit{Stream-K} vs.\ hybrid execution schedules for $896\times384\times128$ GEMM across a hypothetical four-SM GPU\@.}
\end{figure}

The work decomposition we introduced in the last section can be
instantiated in a number of different ways to suit the needs of
different hardware architectures and software library designs.  Our
implementation targets NVIDIA GPUs and is designed to be integrated into
existing libraries like cuBLAS and CUTLASS.  In this section, we
describe how we configure the kernels we launch and introduce a
hybridization scheme that helps ensure users achieve maximum GEMM
performance across the widest possible range of problem shapes.

We also emphasize that these are truly internal implementation details.
They are completely transparent to the user of a BLAS-like library and
do not alter the library's interface.  The only observable impact is the
improved performance characteristics that we analyze in
Section~\ref{sec:stream-k-evaluation}.

\subsection{Kernel Configuration}
\label{sec:blocking-factor-and-grid-configuration}

The tile size chosen for blocking the GEMM computation is, of course, a
critical parameter controlling the performance of the GEMM kernel.  For
modern NVIDIA GPUs, appropriate tile sizes are determined by the shape
of matrices supported by the GPU's Tensor Cores.  Based on extensive
empirical experience, we selected the smallest CTA-wide tile size
capable of achieving 99\% of the GPU's peak TFLOP/s for very large GEMM
volumes for each supported precision.  For the NVIDIA~A100 GPU used in
our experiments, these sizes are 64$\times$64$\times$16 for FP64
problems and 128$\times$128$\times$32 for FP16$\rightarrow$32 problems.

Achieving maximal GEMM performance from \emph{Stream-K} parallelization
also requires some degree of dynamic problem-specific configuration.
Before launching a kernel we choose a grid size likely to yield the best
performance on the specific problem shape at hand.  This is in contrast
to ensemble-based approaches which accommodate diverse problem shapes
through the static generation of many kernel variants based on workload
decomposition and blocking factor.

Our grid size selection heuristic is based on a simple analytical model
that minimizes the cost of reading, writing, and accumulating partial
sums while equally distributing the MAC-loop iterations per CTA.
Details of this analytical model are provided in the section below. 
Parameters to the
model are trivially chosen with empirical measurements and need only be
done once per target architecture.  The resulting parameters can then be
compiled statically into the library.  Again, this is in contrast to
ensemble-based approaches that rely on potentially complex
heuristics and machine learning models for kernel selection at run time.

\input{ch_streamk_v2/chapters/appendix}

\subsection{Data-parallel Hybridization}
\label{sec:data-parallel-hybridization}

The basic \emph{Stream-K} decomposition can, in certain cases, exhibit
tile-processing skew that leads to potentially adverse effects on cache
performance.
When
the number of output tiles $t$ is not an even multiple of the
grid size $g$, the starting $k$-offset for the first MAC-loop
iteration in each CTA will be different. Depending on the sizes and
shapes of the input matrices and blocking factors, this skew may
preclude these fragments from seeing reuse across CTAs in the GPU's
cache structure. In Figure~\ref{fig:basic_streamk}, for example, the initial $k$-axis
fragment offsets for each of the four CTAs will be $k=0$,
$k=32$, $k=64$, and $k=96$, respectively. Furthermore,
this 32-element skew between CTAs will persist for the duration of the
GEMM computation.

Tile-processing skew is a direct consequence of \emph{Stream-K}'s
workload balancing strategy.  However, we can take measures to limit its
duration by applying \emph{Stream-K}'s iteration balancing to a
smaller, tile-aligned region of the total iteration domain such that the
remaining tiles can be produced in full, temporally aligned waves.

The simplest hybrid scheme is the
``\emph{data-parallel} + one-tile \emph{Stream-K}''
schedule illustrated in Figure~\ref{fig:dp_one_tile_sk}.  It applies iteration balancing
only among the tiles otherwise remaining for a final, partially full
\emph{data-parallel} wave.
The total number of full waves is
$w = \lfloor t/p \rfloor$, where $t$ is the number of output
tiles and $p$ is the number of SM cores in the GPU\@. Consequently,
each \emph{Stream-K} CTA receives an even share of iterations that is
less than one tile's worth.
Unfortunately, this strategy has little ability to hide the
synchronization latency for the exchange of partial sums when three or
more CTAs cover the same tile. In these scenarios, the accumulating CTA
may be forced to wait for the contributions of other CTAs to become
visible, as all but the last will be completing their final iterations
at roughly the same time. Furthermore, the basic version of our scheme 
for aggregating partials is serialized within a single CTA, and thus will 
likely cause SM workload imbalance when the number of contributing CTAs 
per tile is large.

We address these problems with
our ``two-tile \emph{Stream-K} + \emph{data-parallel}'' hybrid schedule,
illustrated in Figure~\ref{fig:two_tile_sk_dp}.
It performs
one fewer full data-parallel wave in exchange for each \emph{Stream-K}
CTA receiving more than one tile's worth of iterations (but fewer than
two). This provides much better latency hiding when $w\geq2$,
and each accumulating CTA will only need to receive partials from
one other contributing CTA\@. Otherwise, it behaves identically to the
``\emph{DP + one tile SK}'' schedule.
This hybrid approach results in both improved memory access patterns and latency hiding.
It also shows the versatility of the generic \emph{Stream-K} looping structure to implement different scheduling policies within the same kernel instance.

%% file: ch_streamk_v2/chapters/appendix.tex
\subsubsection{Analytical Modeling for Stream-K Configuration}
\label{sec:analytical-modeling}
In practice, it is not always
advantageous to invoke the \emph{Stream-K} decomposition with as many CTAs
as can be actively resident on the GPU\@. Because it is a tile-splitting
approach, it incurs fixup costs above and beyond the simple
\emph{data-parallel} decomposition. Consequently, the fundamental
proposition is one of strong scaling: how much additional parallelism
can be expressed before the extra overhead causes a negative return on
investment. Depending on the problem shape, the optimal splitting could
be enough to fill the entire processor (i.e., $g \leftarrow p$), no
splitting at all (i.e., $g \leftarrow t$), or somewhere in between.

To predict this inflection point, we present a simple approach
to model the runtime of
\emph{Stream-K} as a function of grid size $g$. In the absence of
other work on the GPU, the runtime of the entire \emph{Stream-K}
schedule will be the same as that of one of its tile-outputting CTAs,
which we formulate as follows:

\begin{align*}
    time_{CTA}(g) \leftarrow & \mathpzc{a} + \mathpzc{b} (FixupPeers(g) > 1) \\
                            & + \mathpzc{c} (ItersPerCta(g)) + \mathpzc{d} (FixupPeers(g) - 1) \notag
\end{align*}

where:

\begin{align*}
    ItersPerCta(g) \leftarrow & \left\lceil
        \frac{
            \lceil \frac{m}{\text{BLK\_M}} \rceil \times
            \lceil \frac{n}{\text{BLK\_N}} \rceil \times
            \lceil \frac{k}{\text{BLK\_K}} \rceil}
        {g}\right\rceil \notag \\
    FixupPeers(g) \leftarrow &
        \left\lceil
            \frac{\left\lceil\frac{k}{\text{BLK\_K}} \right\rceil}
            {IterationsPerCta(g)}
        \right\rceil \notag
\end{align*}

This CTA runtime model comprises four components. The $\mathpzc{a}$ workload
encompasses the one-time, fixed-size costs incurred by each CTA, e.g.,
the grid launch latency, the initial compulsory cache misses, the cost
of writing the final output tile to \textbf{C}, etc. The second
component $\mathpzc{b}$ incorporates the conditional costs of outputting temporary
partial sums for scenarios where the number of output tiles does not
quantize perfectly across the processor. The third---the per-iteration
workload $\mathpzc{c}$---represents the instruction and stall workload of each
MAC-iteration. The final, per-collaborator workload $\mathpzc{d}$ is the cost of
reading and accumulating the partial sums from another CTA covering the
same tile. The set of workload constants \{$\mathpzc{a}$, $\mathpzc{b}$, $\mathpzc{c}$, $\mathpzc{d}$\} will be
unique to each combination of blocking factors, matrix data type, and
GPU microarchitecture, and can be determined empirically via
microbenchmarks.

\begin{figure}
    \centering
    \begin{subfigure}[t]{0.49\textwidth}
        \includegraphics[width=\linewidth,valign=b]{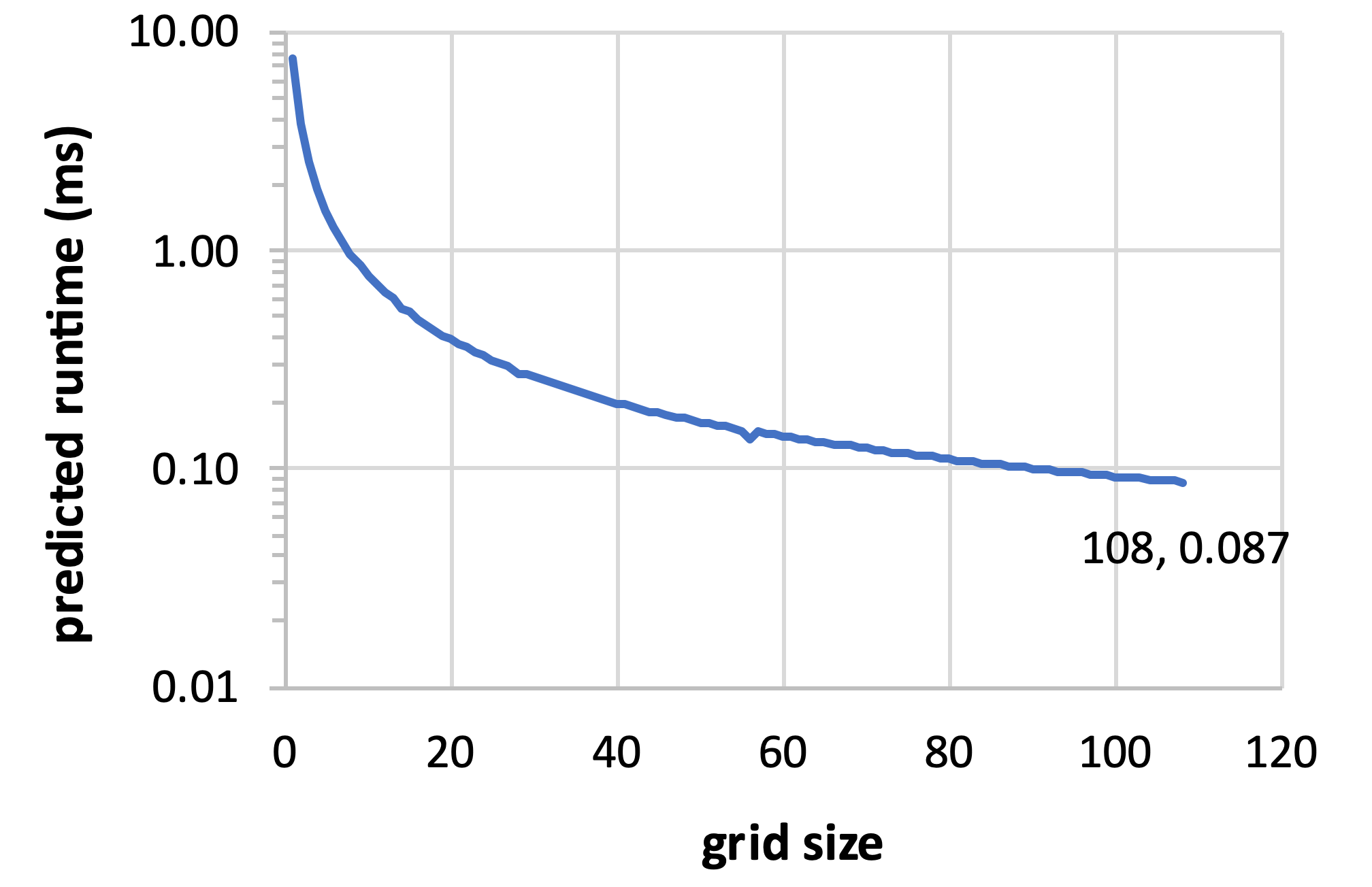}
        \caption[CTA runtime model for $256\times3584\times8192$ GEMM size]{GEMM $256\times3584\times8192$\\
        56~output tiles, 256~iterations per tile\\
        $g_{best} \leftarrow 108$ CTAs, 132/133~iterations per CTA} \label{fig:model_smallm}
    \end{subfigure}
    \hfill%
    \begin{subfigure}[t]{0.49\textwidth}
        \includegraphics[width=\linewidth,valign=b]{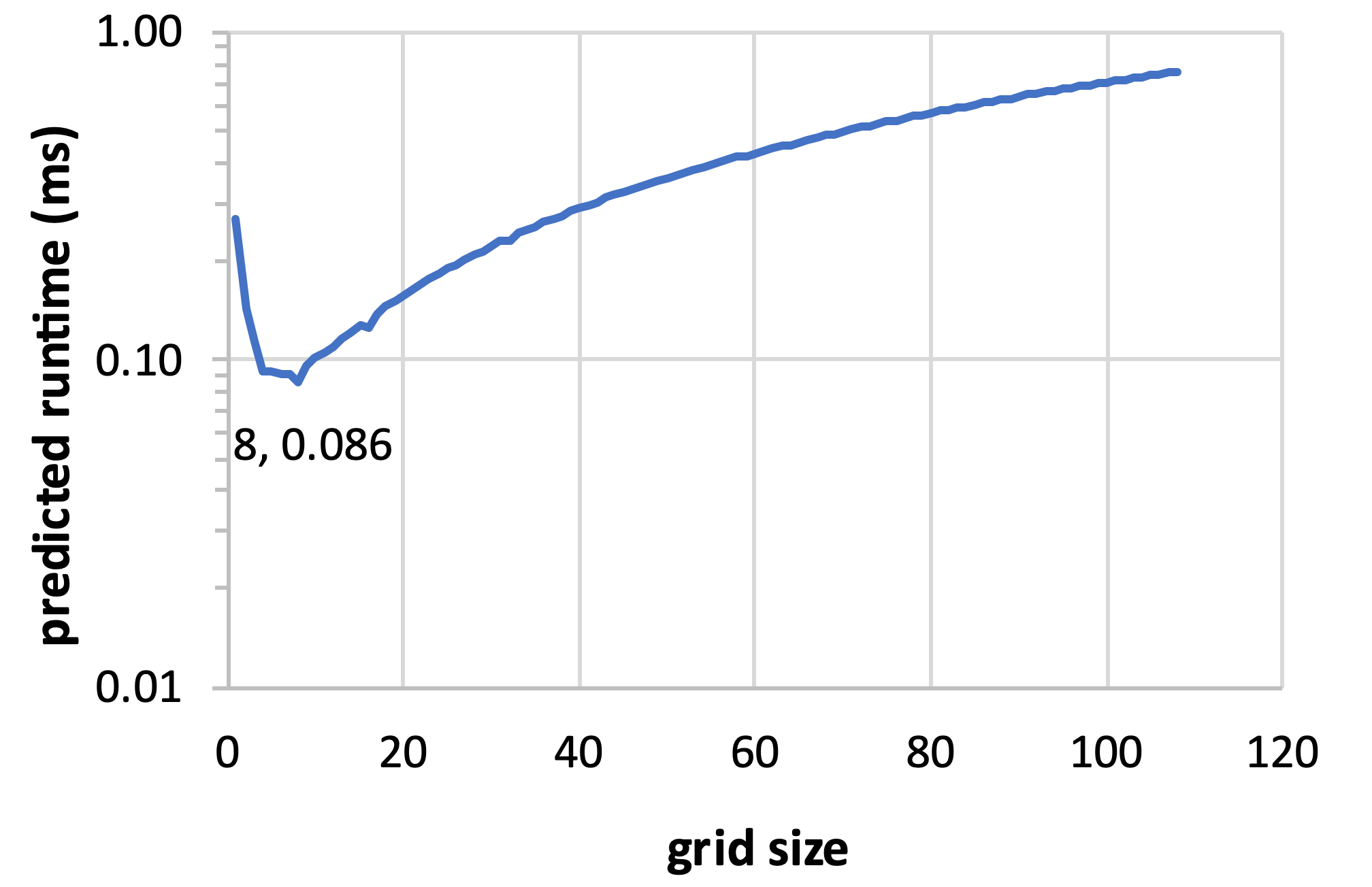}
        \caption[CTA runtime model for $128\times128\times16384$ GEMM size]{GEMM $128\times128\times16384$\\
        1~output tile, 512~iterations per tile\\
        $g_{best} \leftarrow 8$ CTAs, 64~iterations per CTA} \label{fig:model_largek}
    \end{subfigure}
    \vfill%
    \begin{subfigure}[t]{0.5\textwidth}
        \includegraphics[width=\linewidth,valign=b]{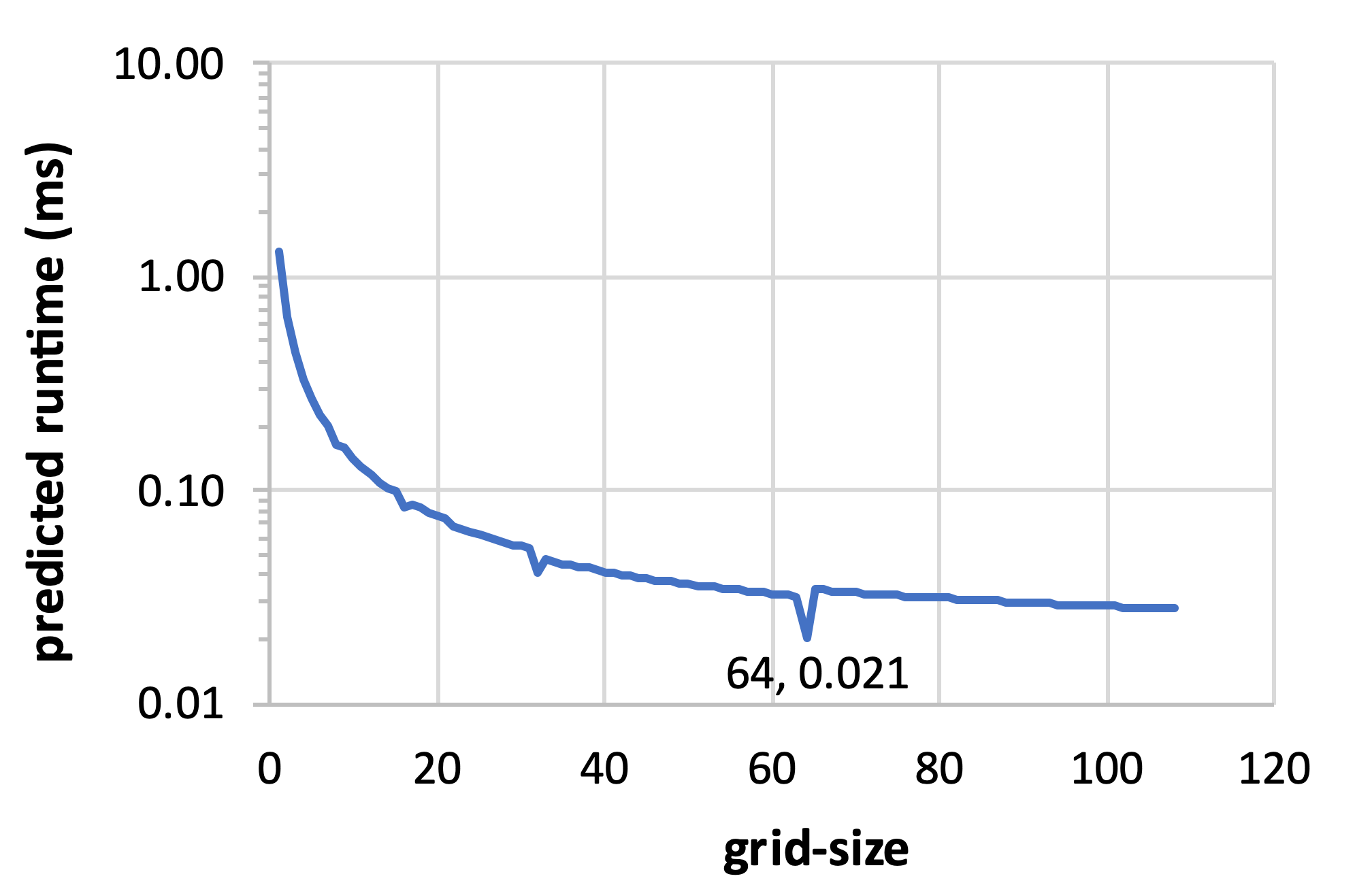}
        \caption[CTA runtime model for $1024\times1024\times1024$ GEMM size]{GEMM $1024\times1024\times1024$\\
        64~output tiles, 32~iterations per tile\\
        $g_{best} \leftarrow 64$ CTAs, 32~iterations per CTA} \label{fig:model_largemnk}
    \end{subfigure}
    \caption[Modeled \textit{Stream-K} performance on NVIDIA A100.]{Modeled \textit{Stream-K} performance on NVIDIA A100 (108~SMs) for BLK\_M=128, BLK\_N=128, BLK\_K=32} \label{fig:model_streamk}
\end{figure}

Figure~\ref{fig:model_streamk} illustrates the behavior of our grid size selection model as
parameterized for fp16-precision GEMM on NVIDIA's A100 GPU using
blocking factors BLK\_M~$=128$, BLK\_N~$=128$, and BLK\_K~$=32$. Specifically, we
highlight three strong-scaling GEMM scenarios where the number of output
tiles is insufficient to produce a single full wave across the
processor's 108~SM cores.

The first GEMM shape accumulates through a large-sized
$k$-dimension to produce a short, wide output matrix. In this
scenario, the reduction in MAC-loop time relative to the increasing
costs of seam fixup is monotonically improving. Consequently, the
optimal grid size coincides with maximal parallelism at $g = 108$
CTAs.

The second shape accumulates through a medium-sized $k$-dimension
to produce a square matrix with 64 output tiles. In this case, the fixup
costs of $\mathpzc{b}$ and $\mathpzc{d}$ outweigh any reduction in MAC-loop iteration count, as
seen by the global minima ``dip'' at $g = 64$ CTAs.

\begin{figure}
    \centering
    \includegraphics[width=\columnwidth]{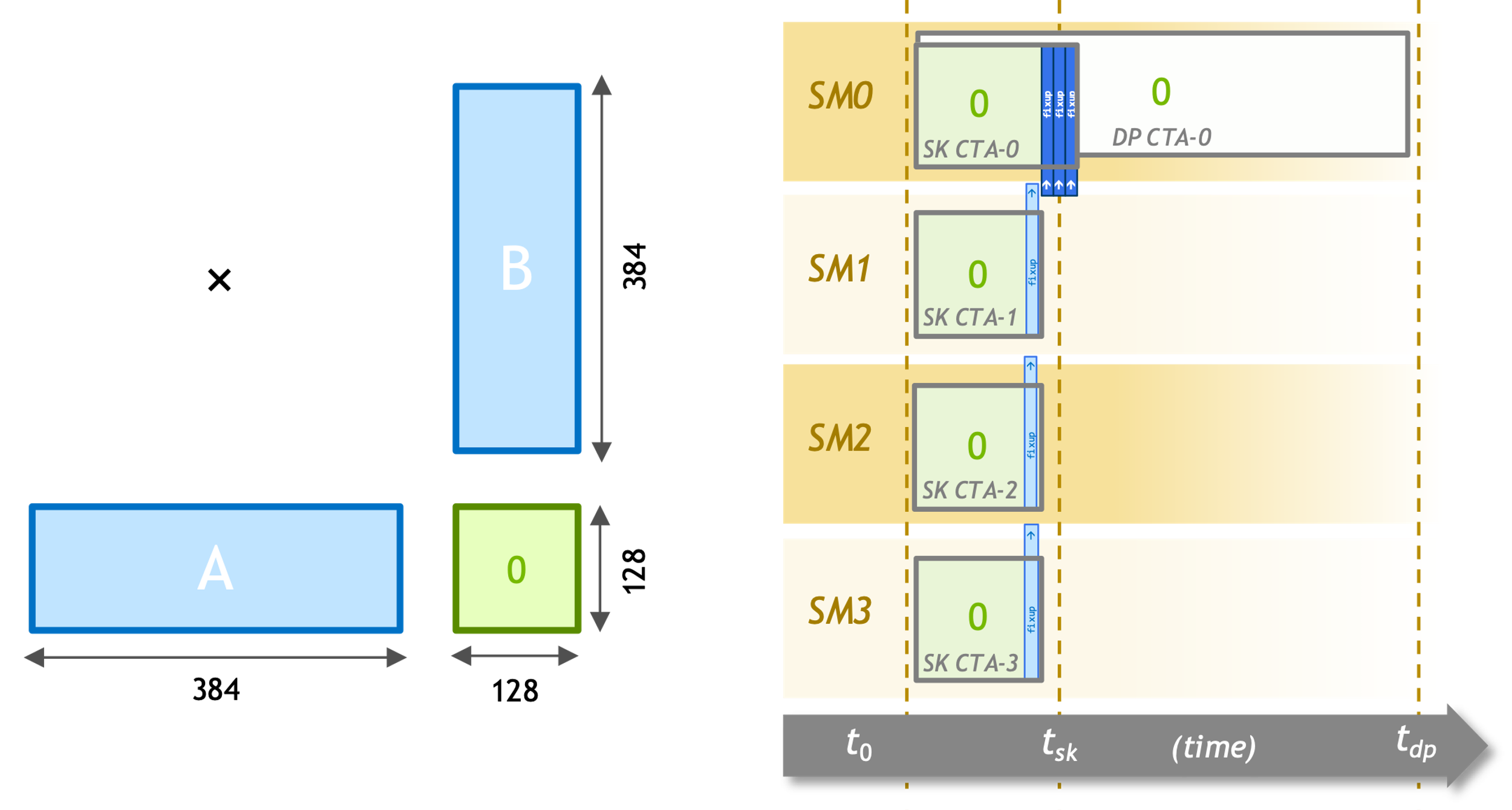}
    \caption[Strong-scaling comparison of \emph{data-parallel} and \emph{Stream-K}]{Strong-scaling comparison of \emph{data-parallel} and \emph{Stream-K} execution schedules for $128\times128\times384$ GEMM across a hypothetical four-SM GPU\@. \emph{Data-parallel} causes the enormous $k$-dimension to be sequentially processed within single CTA, whereas \emph{Stream-K} is able to take advantage of the parallelism available across the $k$-dimension.} \label{fig:comparison_dp_streamk}
\end{figure}

The third shape produces a single output tile after accumulating through
an enormous $k$-dimension, analogous to the execution schedule in
Figure~\ref{fig:comparison_dp_streamk}. Although the opportunity for strong scaling is quite large,
the per-peer cost of serial reduction is entirely incurred by a single
CTA\@. These accumulation costs begin to outweigh any further reductions
in iteration count for grid sizes $g>8$.


%% file: ch_streamk_v2/chapters/performance.tex
\section{Performance Evaluation} 
\label{sec:stream-k-evaluation}

We have implemented our \emph{Stream-K} decomposition using NVIDIA's
CUTLASS library of CUDA C++ template abstractions for authoring 
GEMM-like computations. CUTLASS provides the optimized equivalent 
of the CTA-wide \lstinline{MacLoop()} subroutine in Algorithm~\ref{alg:macloop}, 
which performs blocking, tiling, and software-pipelined 
data movement that is analogous to the closed-source cuBLAS and cuDNN 
implementations. 
Our evaluation encompasses both (1) double-precision FP64 GEMM, and 
(2) mixed-precision FP16$\rightarrow$32 GEMM. For the latter, the input 
matrices \textbf{A} and \textbf{B} comprise half-precision FP16 values, 
yet the internal accumulation and output matrix \textbf{C} values are 
single-precision FP32. 

\paragraph{Hardware environment.} Our test GPU is the NVIDIA A100, which contains 108~SM cores and can issue 13,824 data-parallel instructions per cycle. It provides 40~MB of L2 cache and 1,555~GB/s of global memory bandwidth. 
For measurement stability, we lock the power envelope at 400~W and SM clocks at 1005~MHz ($\sim$71\% of their dynamic peak). This establishes FP64 tensor-core 
peak throughput of 13.9~TFLOP/s, and mixed FP16$\rightarrow$32 
tensor-core peak throughput of 222.3~TFLOP/s.

\begin{figure}
    \centering
    \includegraphics[width=\columnwidth]{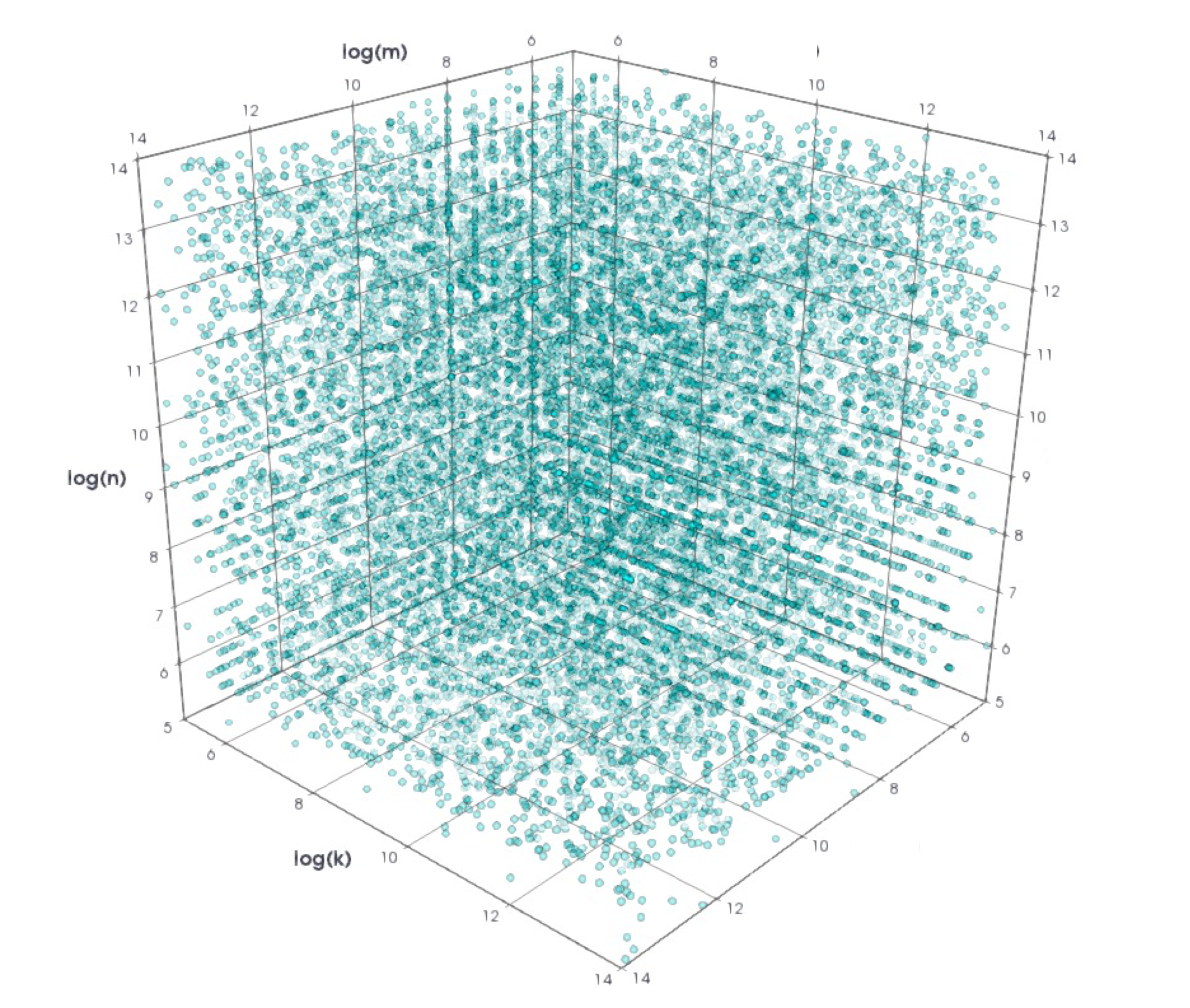}
    \caption[The test domain of 32,824 GEMM problem shapes]{The test domain of 32,824 GEMM problem shapes and sizes used for performance evaluation.
    $\{m\} = \{128 \ldots8192\}$,~
    $\{n\} = \{128 \ldots8192\}$,~
    $\{k\} = \{128 \ldots8192\}$} \label{fig:test_corpus}
\end{figure}

\paragraph{Dataset.} Our test corpus intends to approximate the enormous breadth
and scope of device-wide GEMM problems that GPU math kernel libraries are 
designed to accommodate. As shown in Figure~\ref{fig:test_corpus}, we evaluate
32,768 different problem sizes and shapes, log-sampled at random within a 
domain of $m$, $n$, and $k$ matrix dimensions whose volume spans six orders 
of magnitude.

\paragraph{Methodology.} For both GEMM precisions, 
we build a single \emph{Stream-K} kernel 
that has been specialized per the guidelines in the Section~\ref{sec:practical-usage}.
Furthermore, these kernels implement our ``two-tile \emph{Stream-K} +
\emph{data-parallel}'' hybrid decomposition.
Our evaluation compares each \emph{Stream-K} kernel with: 
\begin{enumerate}
    \item the default \emph{data-parallel} CUTLASS kernel of the same blocking factor; 
    \item the cuBLAS ensemble for that precision (CUDA 11.6); and
    \item an idealized oracle that will always select the highest
    performing \emph{data-parallel} CUTLASS blocking factor to execute 
    for a given GEMM instance.  
\end{enumerate}
For FP64 problems, this oracle 
selects among the ensemble of (32$\times$32$\times$16), (32$\times$64$\times$16), 
(64$\times$64$\times$16), (64$\times$128$\times$16), and (128$\times$128$\times$16) 
blocking factor specializations.  For FP16 $\rightarrow$ 32, it selects 
among the ensemble of (64$\times$64$\times$64), (64$\times$128$\times$32), 
(128$\times$128$\times$32), and (128$\times$256$\times$32) blocking factor 
specializations. These specific specializations
are an open-sourced strict subsets alternative of the corresponding 
cuBLAS GEMM kernel ensembles.

\begin{figure}
    \centering
    \begin{subfigure}[t]{0.49\textwidth}
        \includegraphics[width=\columnwidth]{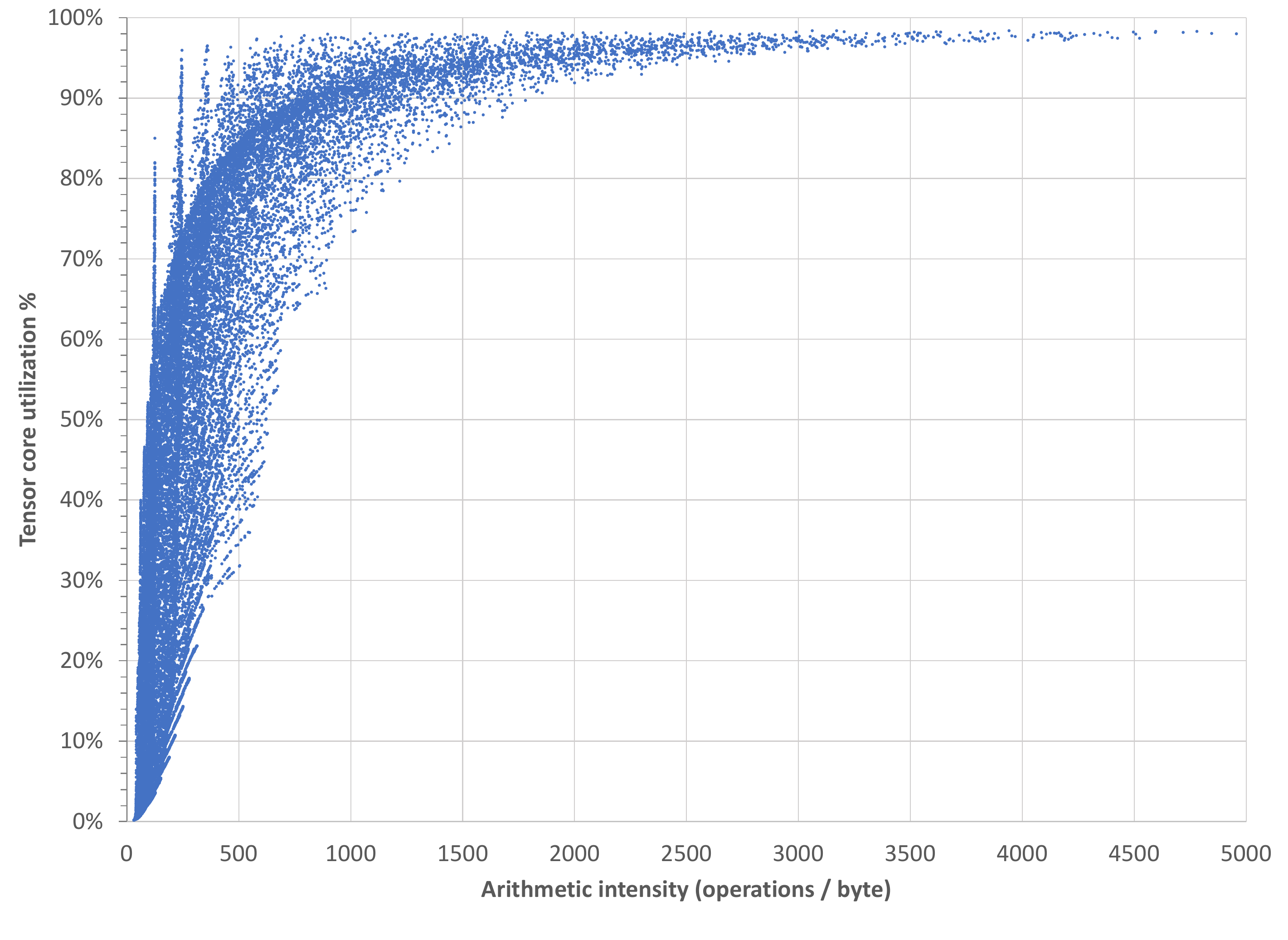}
        \caption{CUTLASS FP16$\rightarrow$32 \emph{data-parallel} ``roofline''\\
        performance ($\text{blocking factors} = $ 128$\times$128$\times$32).} \label{fig:hgemm_roofline_cutlass}
    \end{subfigure}
    \hfill%
    \begin{subfigure}[t]{0.49\textwidth}
        \includegraphics[width=\columnwidth]{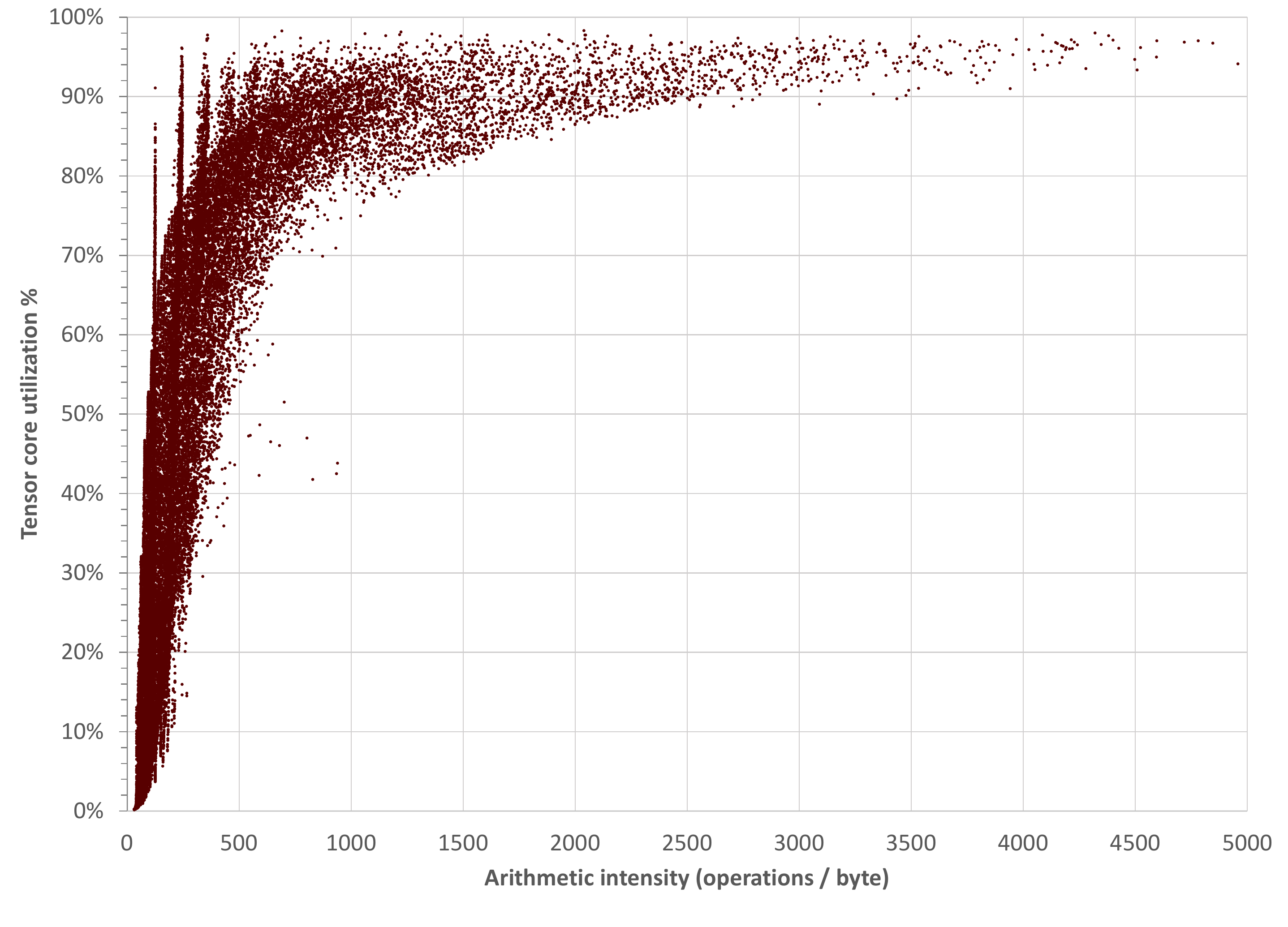}
        \caption{cuBLAS
        (ensemble)} \label{fig:hgemm_roofline_cublas}
    \end{subfigure}
    \vfill%
    \begin{subfigure}[t]{0.49\textwidth}
        \includegraphics[width=\columnwidth]{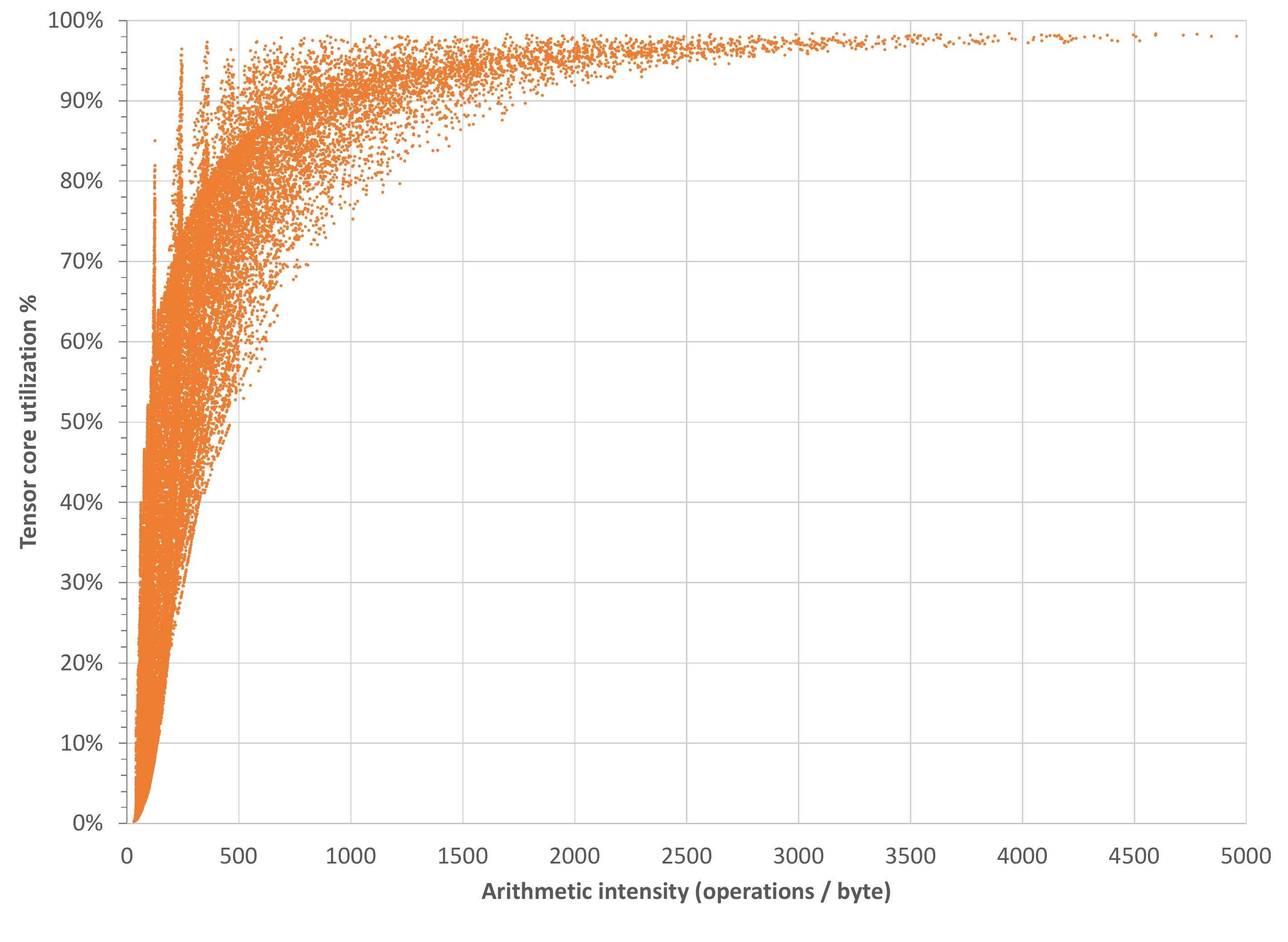}
        \caption{Idealized CUTLASS oracle
        (ensemble)} \label{fig:hgemm_roofline_oracle}
    \end{subfigure}
    \hfill%
    \begin{subfigure}[t]{0.49\textwidth}
        \includegraphics[width=\columnwidth]{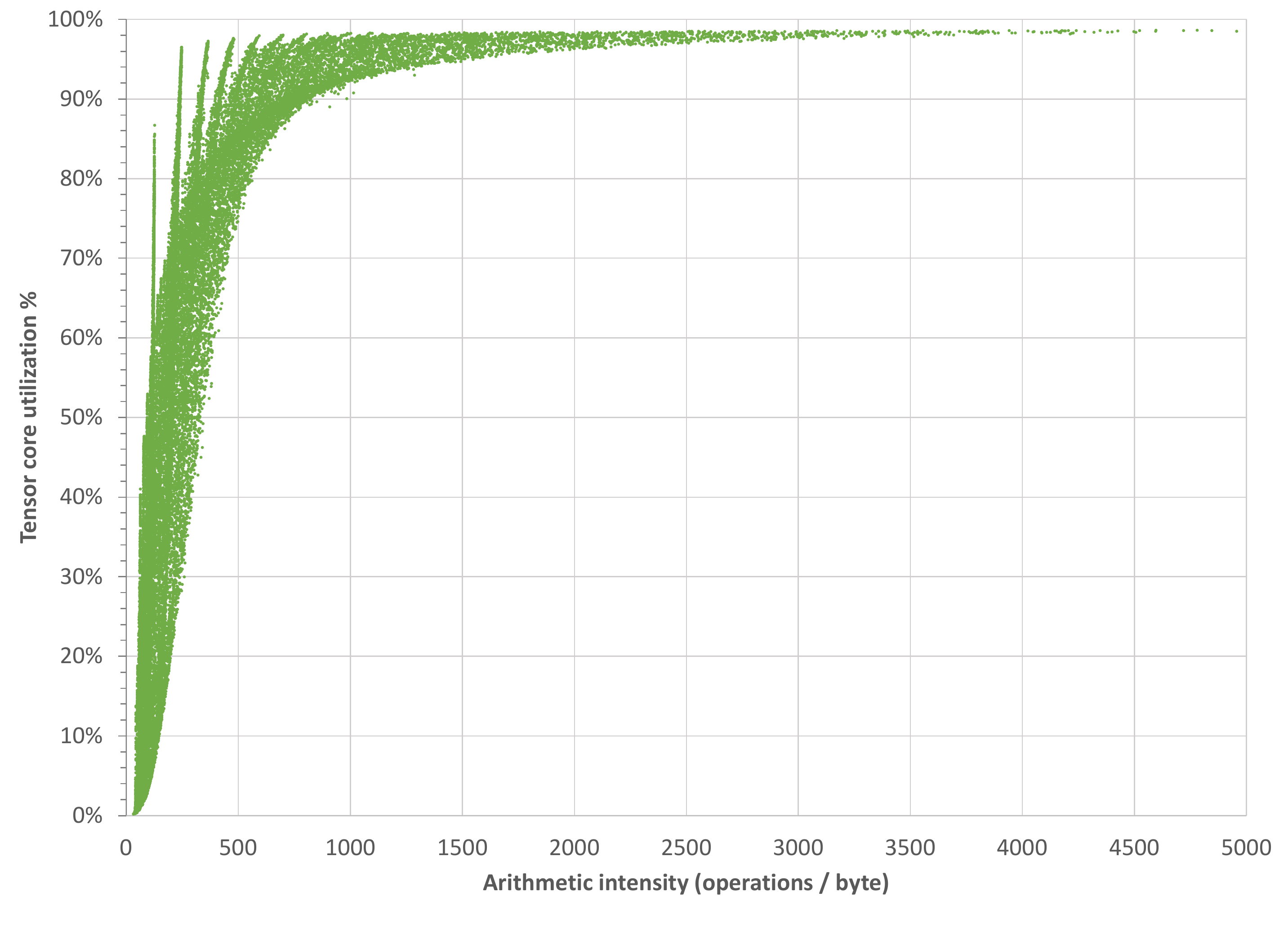}
        \caption{\emph{Stream-K}
        ($\text{blocking factors} = $ 128$\times$128$\times$32)} \label{fig:hgemm_roofline_streamk}
    \end{subfigure}
    \caption[FP16$\rightarrow$FP32 GEMM performance landscape]{FP16$\rightarrow$FP32 GEMM ``roofline'' performance utilization landscapes on NVIDIA A100 across 32K GEMM problem shapes and sizes.} \label{fig:hgemm_comparison}
\end{figure}

\begin{figure}
    \centering
    \begin{subfigure}[t]{0.49\textwidth}
        \includegraphics[width=\columnwidth]{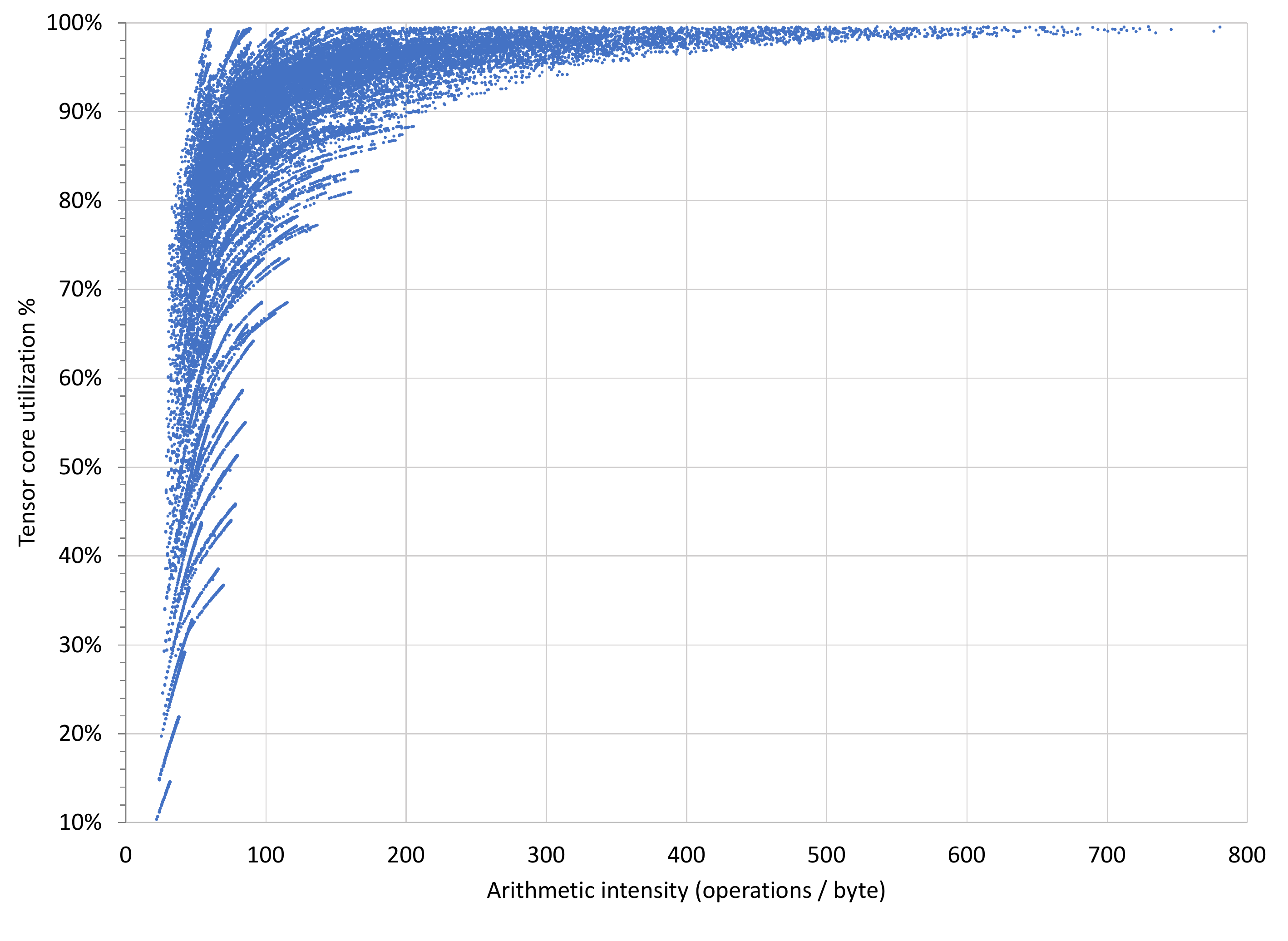}
        \caption{CUTLASS \emph{data-parallel} \\
        ($\text{blocking factors} = $ 64$\times$64$\times$16)} \label{fig:dgemm_roofline_cutlass}
    \end{subfigure}
    \hfill%
    \begin{subfigure}[t]{0.49\textwidth}
        \includegraphics[width=\columnwidth]{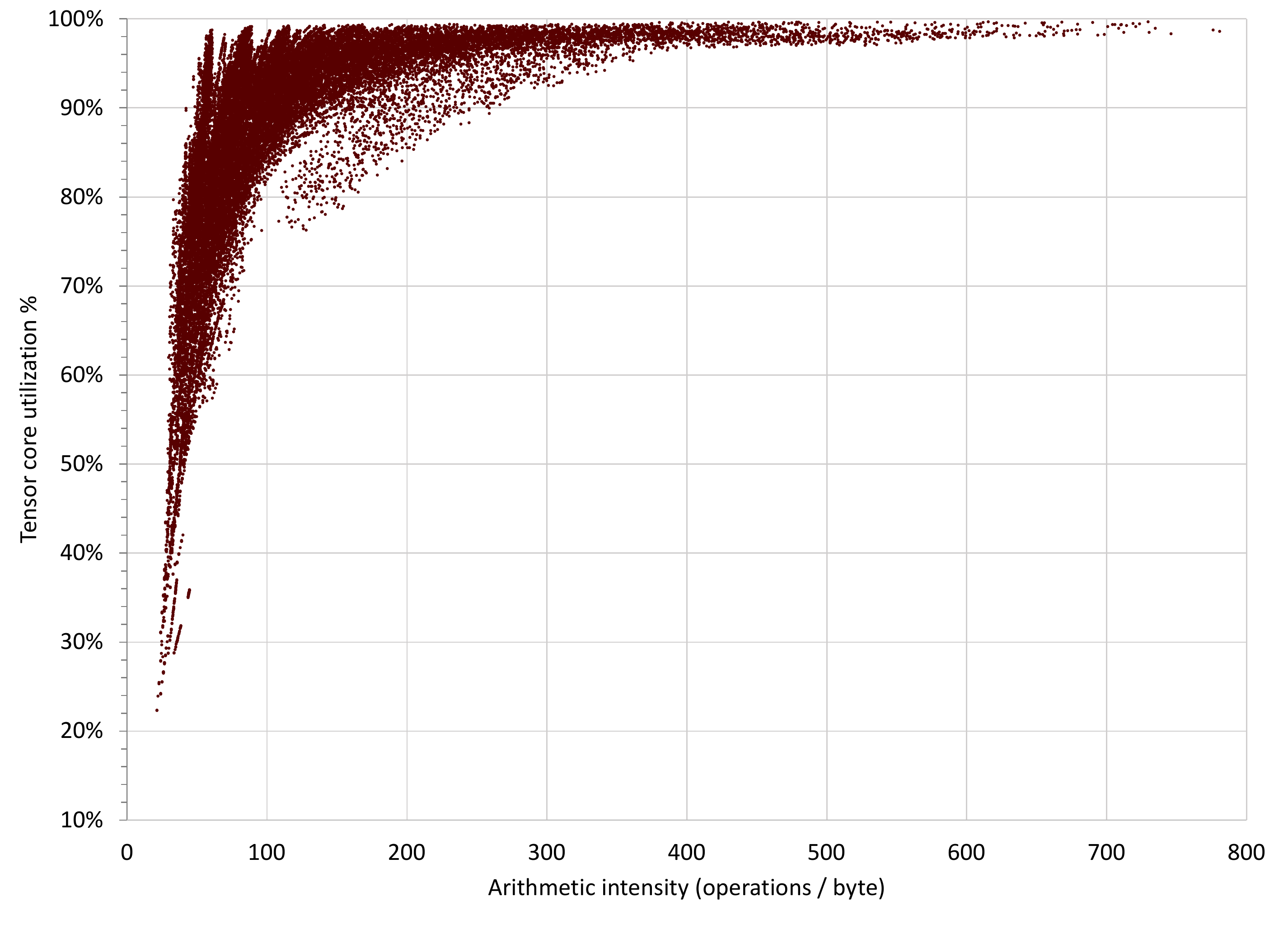}
        \caption{cuBLAS 
        (ensemble)} \label{fig:dgemm_roofline_cublas}
    \end{subfigure}
    \vfill%
    \begin{subfigure}[t]{0.49\textwidth}
        \includegraphics[width=\columnwidth]{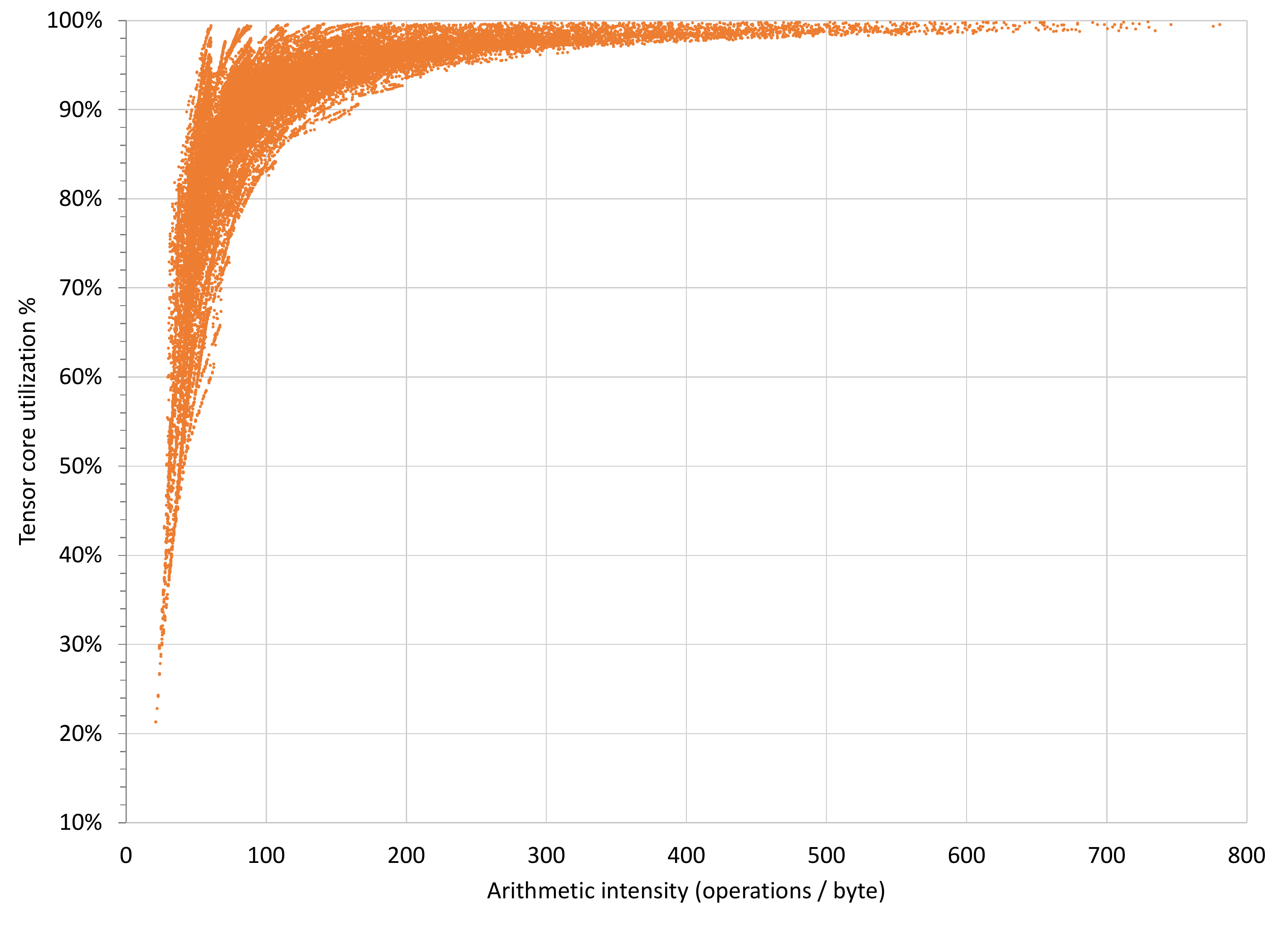}
        \caption{Idealized CUTLASS oracle
        (ensemble)} \label{fig:dgemm_roofline_oracle}
    \end{subfigure}
    \hfill%
    \begin{subfigure}[t]{0.49\textwidth}
        \includegraphics[width=\columnwidth]{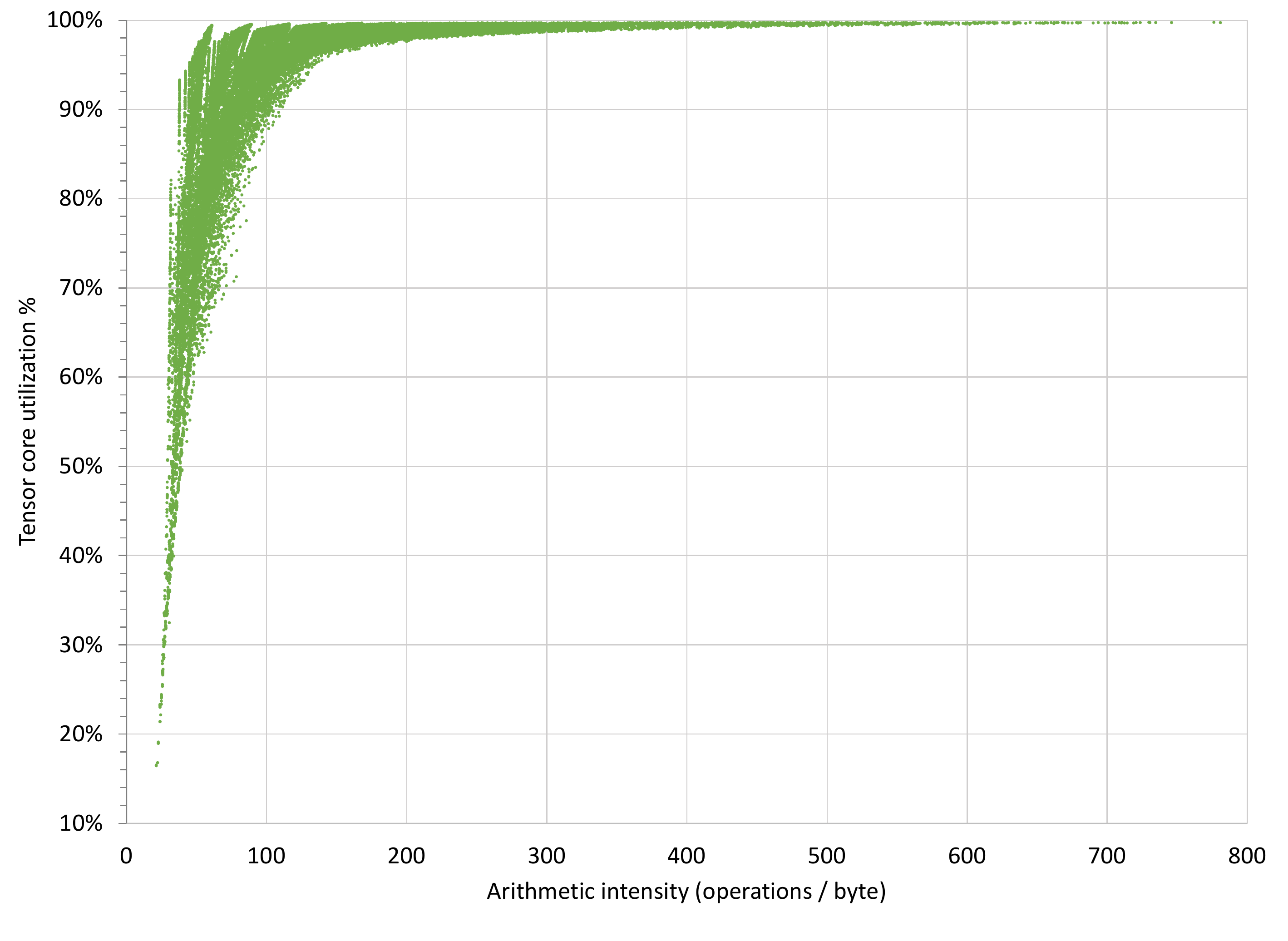}
        \caption{\emph{Stream-K} 
        ($\text{blocking factors} = $ 64$\times$64$\times$16)} \label{fig:dgemm_roofline_streamk}
    \end{subfigure}
    \caption[FP64 GEMM performance landscape]{FP64 GEMM ``roofline'' performance utilization landscapes on NVIDIA A100 across 32K problem shapes and sizes.} \label{fig:dgemm_comparison}
\end{figure}

\begin{table}
    \begin{tabularx}{\columnwidth}{rXXXX}
        \toprule
        & \thead{vs.\ \\ CUTLASS\\ {$64\times64\times16$}} 
        & \thead{vs.\ \\ cuBLAS} 
        & \thead{vs.\ \\ cuBLAS\\ {$>150$ ops/B}} 
        & \thead{vs.\ \\ CUTLASS \\ oracle} \\
        \midrule
        \textbf{Average} & $1.23\times$ & $1.06\times$ & $1.03\times$ & $1.05\times$ \\
        \textbf{StdDev} & $0.45$ & $0.10$ & $0.03$ & $0.09$ \\
        \textbf{Min} & $0.77\times$ & $0.68\times$ & $0.99\times$ & $0.70\times$ \\
        \textbf{Max} & $5.63\times$ & $2.55\times$ & $1.24\times$ & $1.64\times$ \\
        \bottomrule
    \end{tabularx}
    \caption{\emph{Stream-K} FP64 Relative Performance} 
    \label{tab:streamk_speedup_fp64}

    \begin{tabularx}{\columnwidth}{rXXXX}
        \toprule
        & \thead{vs.\ \\ CUTLASS\\ {$128\times128\times32$}} 
        & \thead{vs.\ \\ cuBLAS} 
        & \thead{vs.\ \\ cuBLAS\\ {$>150$ ops/B}} 
        & \thead{vs.\ \\ CUTLASS \\ oracle} \\
        \midrule
        \textbf{Average} & $1.63\times$ & $1.13\times$ & $1.15\times$ & $1.12\times$ \\
        \textbf{StdDev} & $1.46$ & $0.45$ & $0.12$ & $0.37$ \\
        \textbf{Min} & $0.80\times$ & $0.64\times$ & $0.98\times$ & $0.61\times$ \\
        \textbf{Max} & $14.7\times$ & $6.74\times$ & $1.85\times$ & $4.63\times$ \\
        \bottomrule
    \end{tabularx}
    \caption{\emph{Stream-K} FP16$\rightarrow$32 Relative Performance} 
    \label{tab:streamk_speedup_fp16}
\end{table}

The ``roofline'' plots of Figure~\ref{fig:dgemm_roofline_cutlass} and 
Figure~\ref{fig:hgemm_roofline_cutlass} highlight the spread of 
performance produced by the singleton \emph{data-parallel} CUTLASS kernels.  
They plot the percentage of FP64 and FP16$\rightarrow$32 processor utilization 
as a function of computational intensity. Ideally, a GEMM implementation's 
performance response would manifest as a narrow band that adheres tightly to 
the machine's bandwidth- and compute-bound performance ceilings. Here, the 
\emph{data-parallel} kernels exhibit a fairly large dynamic range for any given 
regime of arithmetic intensity.  In contrast, the performance responses from 
the equivalent \emph{Stream-K} kernels in Figure~\ref{fig:dgemm_roofline_streamk} 
and Figure~\ref{fig:hgemm_roofline_streamk} are much tighter. These 
observations are corroborated by Table~\ref{tab:streamk_speedup_fp64} 
and Table~\ref{tab:streamk_speedup_fp16}, which show the \emph{Stream-K} 
kernels outperforming their \emph{data-parallel} FP64 and FP16$\rightarrow$32 
equivalents by an average of 1.23$\times$ and 1.63$\times$, respectively.  
For extreme strong-scaling scenarios where $m \times n$ is small and $k$ is 
large, our \emph{Stream-K} kernels demonstrate up to 5.63$\times$ and 
14.7$\times$ speedup, respectively.  

The second columns of Table~\ref{tab:streamk_speedup_fp64} and 
Table~\ref{tab:streamk_speedup_fp16} compare our \emph{Stream-K} performance 
with that of cuBLAS\@. On average, our FP64 and FP16$\rightarrow$32
\emph{Stream-K} GEMM kernels respectively deliver 6\% and 13\% greater 
throughput than their corresponding cuBLAS ensembles, with peak improvement of 2.55$\times$ and 6.74$\times$. This is a significant improvement over the breadth of 32K GEMM problem shapes and sizes with 20$\times$ \emph{less} executable code (a single kernel for each precision) than NVIDIA's vendor GEMM library, cuBLAS\@.


Furthermore, the contrast between the FP64 and FP16$\rightarrow$32
cuBLAS performance responses (Figure~\ref{fig:dgemm_roofline_cublas} and
Figure~\ref{fig:hgemm_roofline_cublas}) versus those of our hypothetical CUTLASS
oracle ensembles (Figure~\ref{fig:dgemm_roofline_oracle} and 
Figure~\ref{fig:hgemm_roofline_oracle}) reveal the difficulties of designing
kernel selection heuristics that deliver consistently good performance.  
Despite having access to the same blocking factor specializations, cuBLAS 
exhibits substantially wider dynamic ranges than the idealized
\emph{data-parallel} CUTLASS oracle. The performance spreads of our
\emph{Stream-K} kernels are narrower still, achieving up to 4.6$\times$
the idealized oracle performance and underscoring their ability to achieve
utilization levels that are simply not possible from tile-centric
work decompositions.

\begin{figure}
    \begin{subfigure}[t]{0.49\textwidth}
        \centering
        \includegraphics[width=\columnwidth]{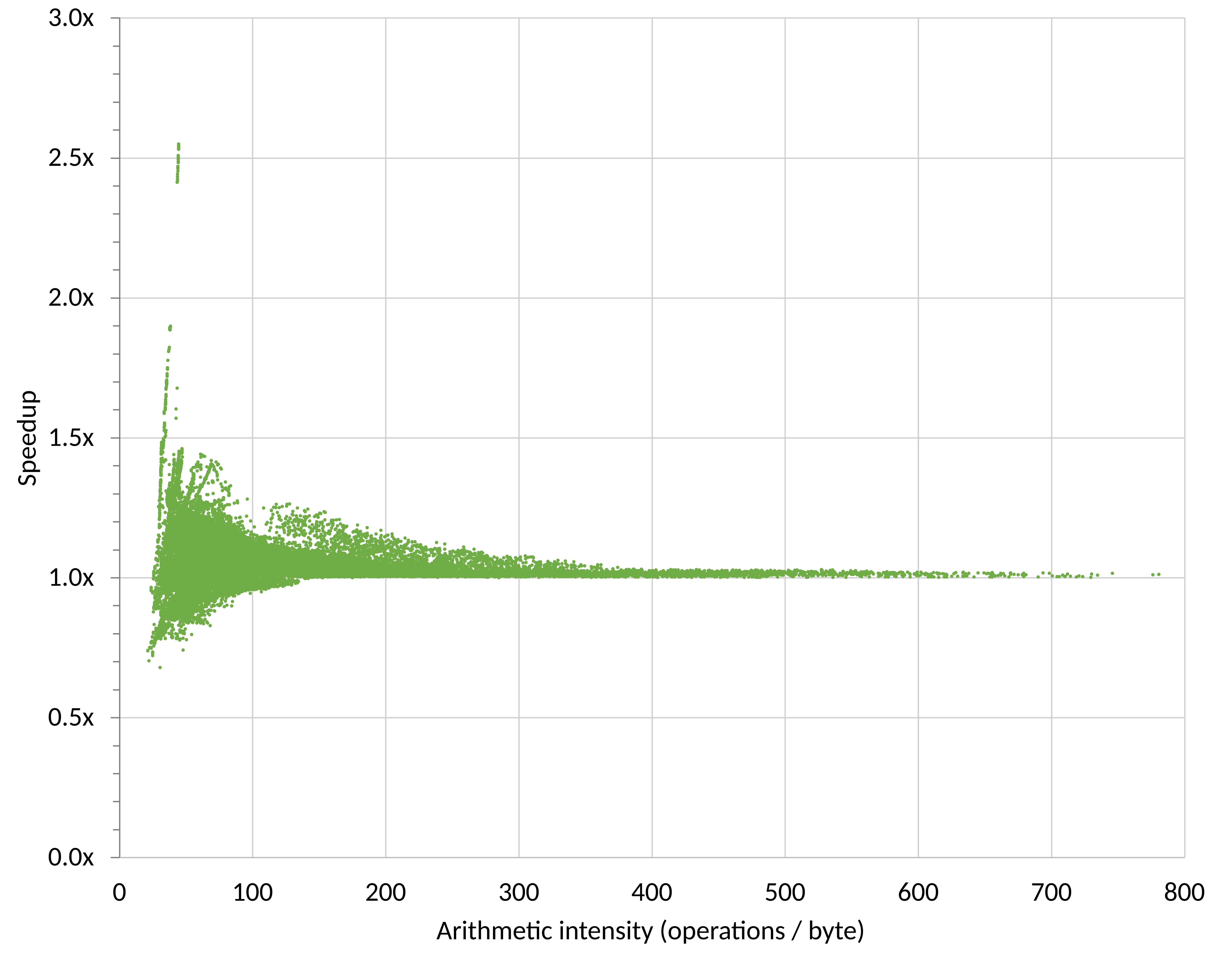}
        \caption{FP64 \emph{Stream-K} speedup vs.\ cuBLAS.} \label{fig:dgemm_speedup}
    \end{subfigure}
    \hfill%
    \begin{subfigure}[t]{0.49\textwidth}
        \centering
        \includegraphics[width=\columnwidth]{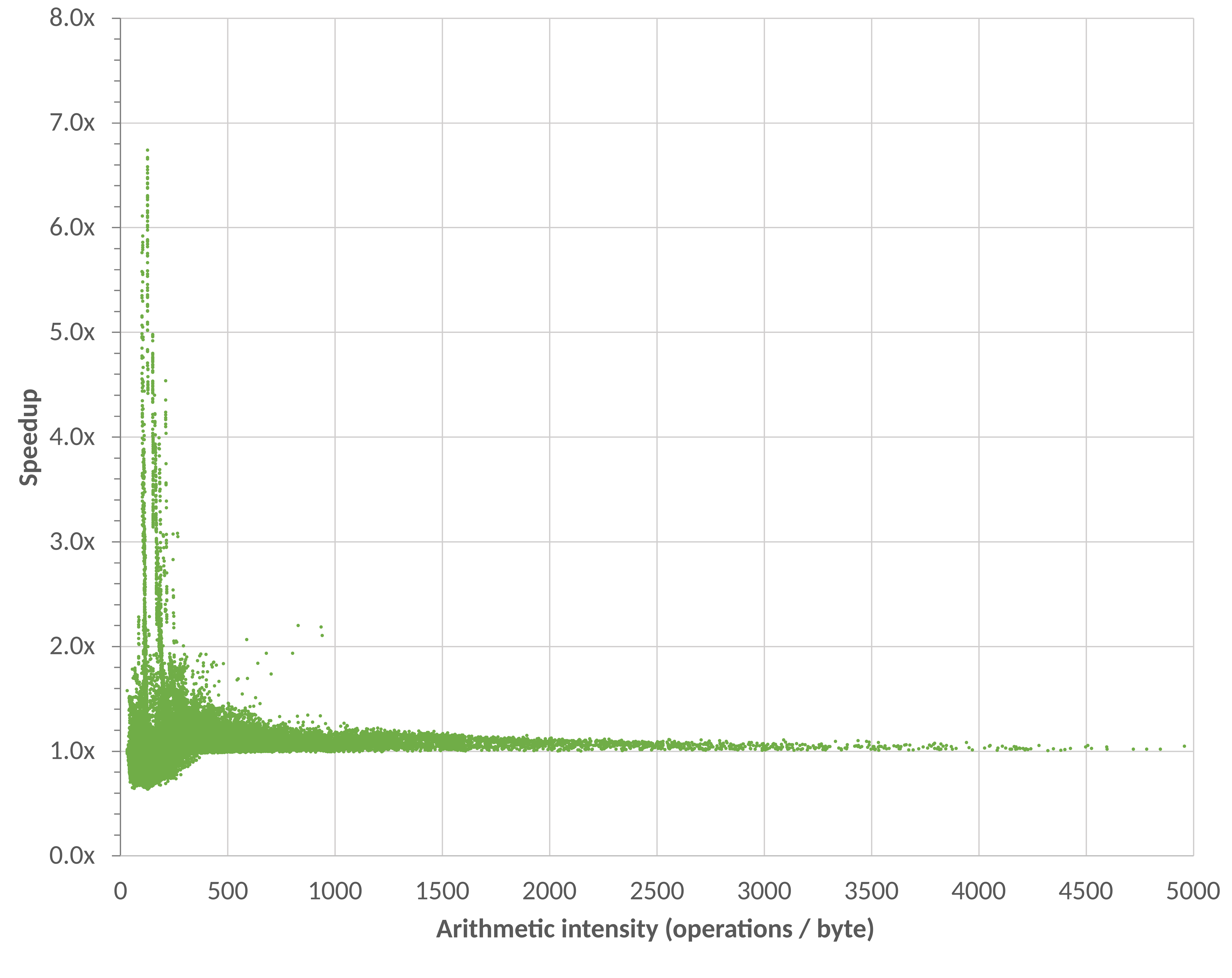}
        \caption{FP16$\rightarrow$32 \emph{Stream-K} speedup vs.\ cuBLAS.} \label{fig:hgemm_speedup}
    \end{subfigure}
        \caption[\emph{Stream-K} speedup vs.\ cuBLAS]{\emph{Stream-K} speedup vs.\ cuBLAS.} \label{fig:cublas_speedup}
\end{figure}

Finally, we observe regimes of small, bandwidth-bound problem shapes
where our largish blocking factors do not compete well against cuBLAS. 
However, if we restrict our scope to the domain of compute-bound problems 
(i.e., FP64 problems having compute intensity $>$~150~ops/byte and 
FP16$\rightarrow$32 problems $>$~400~ops/byte), Figure~\ref{fig:dgemm_speedup} 
and Figure~\ref{fig:hgemm_speedup} demonstrate that our singleton 
\emph{Stream-K} kernels achieve unilaterally higher performance than the 
cuBLAS ensembles.
The ``noisy'' relative performance in the regimes below these thresholds is not
surprising, as \emph{Stream-K} is attempting to make memory-bound
computations run faster by adding more memory workload. This suggests a
few avenues for future work, namely separate cost-modeling for the
memory-bound regime and/or the bundling of a second \emph{Stream-K} kernel 
having smaller tile size into a two-kernel ensemble.

%% file: ch_streamk_v2/chapters/conclusion.tex
\section{Conclusion}

We presented \emph{Stream-K}, a novel parallel workload decomposition
technique for scheduling general matrix multiplication (GEMM) and
similar computations on wide architectures such as GPUs. Unlike other
tile-splitting techniques, the MAC-loop iteration is our unit of
workload quantization across processor cores. This affords excellent
strong scaling and workload balancing because its cost is (1) a constant
with respect to the problem shape, and (2) substantially smaller than
that of an entire output tile.

Furthermore, \emph{Stream-K} produces an $O(p)$ number of splitting
seams that are bound by the number of processor cores. Consequently, the
overheads of strong scaling and workload balancing scale with processor
width rather than problem size. This is a welcome feature for many
applications that cannot afford to allocate large amounts of temporary storage
equivalent to the problem output.

Finally, we evaluated our \emph{Stream-K} approach across a broad spectrum 
of GEMM shapes and sizes.  We showed that a single blocking configuration of
\emph{Stream-K} can (1) achieve levels of absolute performance that match 
and/or exceed that of NVIDIA's cuBLAS library, even when the latter is 
operating at near-peak processor utilization, and (2) do so with much 
higher levels of performance consistency. Additionally, \emph{Stream-K} 
is an attractive option for library construction and maintenance, as 
it presents an opportunity to reduce distribution sizes by an order 
of magnitude and removes the need for complex handcoded heuristics or
machine learning models for kernel selection without compromising performance.

For future works, we identify cache-aware, tile-access patterns such as 
Morton Order, an avenue for optimization. We also believe that 
\emph{Stream-K} decomposition could provide a similar improved performance 
response for other GEMM-like workloads that struggle with the same 
quantization inefficiencies.

%% file: tex/conclusion.tex
\chapter{Conclusion}
\label{chapter:conclusion}

This dissertation identifies load balancing as the key component to achieve high-performance for both regular and irregular computations on the GPU. In efforts to address challenges associated with load balancing, we propose an abstraction that provides a new way to program irregular applications. Expressed using our simple, yet powerful abstraction, we are able to extend previously application-specific load-balancing techniques to other workloads and parallel domains. We also highlight the need to consider resource imbalance for regular fine-grained problems, and show that our general \emph{Stream-K} schedule rids of any complicated heuristics and provides state-of-the art performance on an already well-optimized problem like General Matrix-Multiplication. Built on the insights from this work, we propose the following future research directions.

\section{Future Research Directions}
\label{sec:future-research-directions}

\subsection{Multi-GPU Load Balancing} 
In Chapter~\ref{ch:loadbalance} we built an abstraction for load-balancing centered around \emph{atoms}, \emph{tiles} and \emph{sets} as fundamental building blocks targetted for single GPU implementations. Some of the underlying load-balancing algorithms can easily extend to support multi-GPU problems. This is going to be the key advancement going forward, as sparse-irregular problems in the real world can be extremely large (for example, graphs with billions of vertices and edges). These large problems do not fit in a single GPU's memory, and to accelerate computation on these problems, the abstraction must be extended to support multi-GPU and multi-node systems. 

Our insights in this dissertation inspire a new research direction, where specific modular algorithms (such as merge-path~\ref{alg:work-oriented} SpMV) are well suited for a multi GPU enviroment and can be easily exposed with the existing building blocks with an additional levels of hierarchy for \emph{sets} of \emph{sets}, where each \emph{set} is large enough to fit in a single GPU's memory. We call workloads such as merge-path SpMV, isomorphic problems. Isomorphism here implies that regardless of where the computation occurs, a thread, block, or a separate GPU, the structure of the algorithm stays the same. 

A much more challenging research direction is further extending the multi GPU programming model to then support amorphous workloads, where different levels of compute hierarchy will require different kinds of algorithms to complete a workload. For example, the Radix Sorting algorithm scales nicely for a single GPU compute hierarchy as implemented in the CUB library~\cite{CUB:2016}, but it may not be the best choice at a multi-GPU boundary as communicating bit-level information among different GPUs may not be enough to fully utilize the memory bandwidth. Therefore, we may require different sorting algorithms at different parts of the compute hierarchy---and similarly, may require different load-balancing schedules at the compute boundaries. This will make for an interesting challenge to explore abstractions and APIs for hybrid load-balancing schemes with heuristics to context switch depending on the amount of work and the compute boundary one is working at.

\subsection{Heuristics using Roofline Model}

\begin{figure}
  \centering
  \includegraphics[width=\columnwidth]{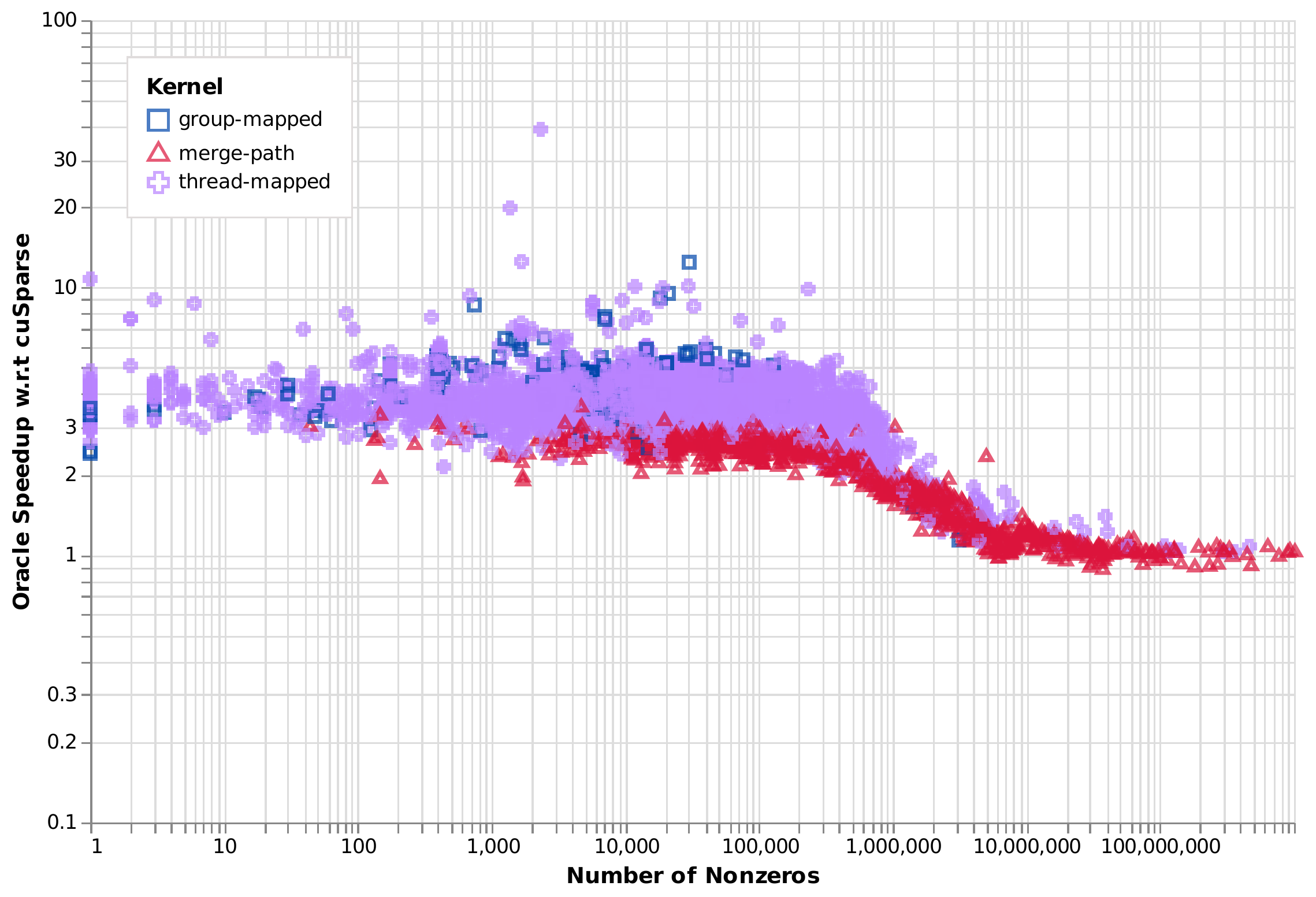}
  \caption[Oracle SpMV vs.\ cuSparse's SpMV]{Speedup of an oracle heuristics SpMV vs.\ cuSparse's SpMV across SuiteSparse. Selects the best load-balancing schedule from the available ones for a given problem using exhaustive search.}
  \label{fig:spmv-speedup-oracle}
\end{figure}

One optimization we highlighted in our results was the use of simple heuristics to determine which load-balancing algorithm to use. However, our results show that our simple heuristics are far from an oracle schedule (see Figure~\ref{fig:spmv-speedup-oracle}) that will pick the right load-balancing algorithm for all sparse datasets and workloads. One direction for research is to look at the traditional roofline analytical model and expand it to capture the irregularity within sparse-linear algebra operations like SpMV or SpMM. Using this model, we can then build more robust heuristics to select the among the different supported load-balancing schedules.

\subsection{Standardization of Sparse-Matrix Formats}
In Chapter~\ref{ch:loadbalance} we highlight the diversity and importance of sparse-matrix formats for sparse-linear algebra and graph analytics. Our APIs for the load-balancing abstraction were designed to accomodate multiple formats for different workloads at both high- and low-level of the abstraction. Based on these insights, we propose the standardization of sparse-matrix formats as new \cpp{} containers for both sequential and parallel programming use cases. The research challenge here is to find a general and high-performant way to build and use these containers for various different domains (graphs, sparse-linear algebra, machine learning, etc.) One potential way to build a general, high-performant sparse-matrix container will be to use specialized \cpp{} iterators to define the ``sparse-view'' of the container. For example, instead of a sparse-matrix accessed using the three arrays of a CSR (offsets, indices and values), a sparse-matrix will be accessed using an iterator over the matrix that provides the tuple: row, column, value (see Listing~\ref{lst:std_sp_matrix}).

\begin{listing}
    \caption[An example of generic SpMV.]{Standardized sparse matrix accessed within a generic SpMV. The underlying sparse format can be changed to a CSC, COO, DIA, but the high-level loop remains the same.}
    \label{lst:std_sp_matrix}
    \begin{minted}[
        obeytabs=true,
        tabsize=4,
        linenos=true,
        numbersep=-1pt]{c++}
    // Implements a "format" unaware SpMV.
    template<typename type_t>
    void spmv(
      matrix_t<type_t, std::size_t, csr_view_t>& A,
      std::vector<type_t>& x,
      std::vector<type_t>& y) {
      // Internally a csr type,
      // a sparse matrix simply accessed by 
      // specialized iterator that return row, 
      // column, value.
      for (auto& [i, j, v] : A) {
        y[i] += v * x[j];
      }
    }
    \end{minted}
  \end{listing}

\subsection{Programming Model for Locality} 
We believe that along with efficient load-balancing algorithms for irregular problems, there is also a need for further research in an orthogonal model for better leveraging locality within these problems. A dynamic tiling approach for grouping similar data for better caching and a light-weight reordering approach for reordering sparse-matrices are two possible directions that can be exploited within this programming model. The challenge is the tradeoffs that arises when discussing between the impact load-balancing might have on locality and vice-versa, and this needs to be better quantified/researched.

\subsection{\emph{Stream-K} Abstraction} 
In our work in Chapter~\ref{ch:streamk}, we presented a work-centric approach for parallel decomposition targeted at GEMM. Impactful future-work will be using a further generalized version of \emph{Stream-K} that can distill any regular computation (such as tensor algebra, convolutions and other machine learning workloads) to ``amount of work'' each thread group needs to perform, and a loop that consumes the assigned work within the thread group until all work is completed (using a persistent-CTA).
The goal of this future work will be to make the performance benefits attained from using \emph{Stream-K} more accessible to a wide breadth of regular problems that suffer from wave quantization and require a sophisticated heuristics driven approach (e.g. kernels within cuDNN).

\subsection{Beyond Today's CUDA}
One thrust we see going forward is looking at new CUDA programming model additions and support that will better allow for addressing scheduling fine-grained parallel computations on the GPUs. 

\paragraph{CUDA and C++} 
Our load-balancing abstraction and APIs are fully realized with the help of tools available within modern \cpp{}. However, current CUDA (major version 11) has limited support for the latest \cpp{} standard. We look forward to future CUDA/\cpp{} with native support for the \cpp{}20 Ranges library that extends and generalizes the standard algorithms and iterator libraries making composable APIs even more powerful and efficient.

\paragraph{Reconfigurable CUDA Grids} 
A natural way to think about the load-balancing or scheduling problem within CUDA is through CUDA's dynamic parallelism, where one can launch CUDA kernels from within a kernel. Irregular problems can then be load-balanced by first a kernel looking at the amount of work available for a given problem and then launching a CUDA kernel with the exact amount of resources required to process that part of the problem. However, dynamic parallelism within CUDA is not performant due to the work required by the driver and the APIs to configure a fresh kernel launch per CUDA thread. One approach to addressing poor dynamic parallelism performance is by utilizing CUDA's compute hierarchy. A single grid can be oversubscribed with the number of threads and/or resources, and then reconfigured to fit the problem size; effectively ``freeing'' up resources for other threads to consume. Since no new work is being spawned, a reconfiguration on-the-fly can be achieved at a much cheaper cost than launching new kernels from within a kernel. We propose this as an interesting direction to improve CUDA for irregular parallel problems.
